# Evaluating Hive and Spark SQL with BigBench

*Technical Report No. 2015-2*
*January 11, 2016*

Todor Ivanov and Max-Georg Beer

Frankfurt Big Data Lab
Chair for Databases and Information Systems
Institute for Informatics and Mathematics
Goethe University Frankfurt
Robert-Mayer-Str. 10,
60325, Bockenheim
Frankfurt am Main, Germany

www.bigdata.uni-frankfurt.de

## Table of Contents

## 1. Introduction

The objective of this work was to utilize BigBench [1] as a Big Data benchmark and evaluate and compare two processing engines: MapReduce [2] and Spark [3]. MapReduce is the established engine for processing data on Hadoop. Spark is a popular alternative engine that promises faster processing times than the established MapReduce engine. BigBench was chosen for this comparison because it is the first end-to-end analytics Big Data benchmark and it is currently under public review as TPCx-BB [4]. One of our goals was to evaluate the benchmark by performing various scalability tests and validate that it is able to stress test the processing engines. First, we analyzed the steps necessary to execute the available MapReduce implementation of BigBench [1] on Spark. Then, all the 30 BigBench queries were executed on MapReduce/Hive with different scale factors in order to see how the performance changes with the increase of the data size. Next, the group of HiveQL queries were executed on Spark SQL and compared with their respective Hive runtimes.

This report gives a detailed overview on how to setup an experimental Hadoop cluster and execute BigBench on both Hive and Spark SQL. It provides the absolute times for all experiments preformed for different scale factors as well as query results which can be used to validate correct benchmark execution. Additionally, multiple issues and workarounds were encountered and solved during our work. An evaluation of the resource utilization (CPU, memory, disk and network usage) of a subset of representative BigBench queries is presented to illustrate the behavior of the different query groups on both processing engines.

Last but not least it is important to mention that larger parts of this report are taken from the master thesis of Max-Georg Beer, entitled "Evaluation of BigBench on Apache Spark Compared to MapReduce" [5].

The rest of the report is structured as follows: Section 2 provides a brief description of the technologies involved in our study. Brief summary of the BigBench benchmark is presented in Section 3. Section 4 evaluates the steps needed to complete in order to execute BigBench on Spark. An overview of the hardware and software setup used for the experiments is given in Section 5. The performed experiments together with the evaluation of the results are presented in Section 6. Section 7 depicts a comparison between the cluster resource utilization during the execution of representative BigBench queries. Finally, Section 8 concludes with lessons learned.

## 2. Background

**Big Data** has emerged as a new term not only in IT, but also in numerous other industries such as healthcare, manufacturing, transportation, retail and public sector administration [6][7] where it quickly became relevant. There is still no single definition which adequately describes all Big Data aspects [8], but the "*V*" characteristics (*Volume*, *Variety*, *Velocity*, *Veracity* and more) are among the widely used one. Exactly these new Big Data characteristics challenge the capabilities of the traditional data management and analytical systems [8][9]. These challenges also motivate the researchers and industry to develop new types of systems such as Hadoop and NoSQL databases [10].

**Apache Hadoop** [11] is a software framework for distributed storing and processing of large data sets across computer clusters using the map and reduce programming model. The architecture allows scaling up from a single server to thousands of machines. At the same time Hadoop delivers high-availability by detecting and handling failures at the application layer. The use of data replication guarantees the data reliability and fast access. The core Hadoop components are the Hadoop Distributed File System (HDFS) [12][13] and the MapReduce framework [2].

HDFS has a master/slave architecture with a *NameNode* as a master and multiple *DataNodes* as slaves. The *NameNode* is responsible for storing and managing all file structures, metadata, transactional operations and logs of the file system. The *DataNodes* store the actual data in the form of files. Each file is split into blocks of a preconfigured size. Every block is copied and stored on multiple *DataNodes*. The number of block copies depends on the *Replication Factor*.

MapReduce is a software framework that provides general programming interfaces for writing applications that process vast amounts of data in parallel, using a distributed file system, running on the cluster nodes. The MapReduce unit of work is called *job* and consists of input data and a MapReduce program. Each job is divided into *map* and *reduce* tasks. The map task takes a split, which is a part of the input data, and processes it according to the user-defined map function from the MapReduce program. The reduce task gathers the output data of the map tasks and merges them according to the user-defined reduce function. The number of reducers is specified by the user and does not depend on input splits or number of map tasks. The parallel application execution is achieved by running map tasks on each node to process the local data and then send the result to a reduce task which produces the final output.

Hadoop implements the MapReduce (version 1) model by using two types of processes – *JobTracker* and *TaskTracker*. The *JobTracker* coordinates all jobs in Hadoop and schedules tasks to the *TaskTrackers* on every cluster node. The *TaskTracker* runs tasks assigned by the *JobTracker*.

Multiple other applications were developed on top of the Hadoop core components, also known as the Hadoop ecosystem, to make it more ease to use and applicable to variety of industries. Example for such applications are Hive [14], Pig [15], Mahout [16], HBase [17], Sqoop [18] and many more.

**YARN** (Yet Another Resource Negotiator) [19] is the next generation Apache Hadoop platform, which introduces new architecture by decoupling the programming model from the resource management infrastructure and delegating many scheduling-related functions to per-application components. This new design [19] offers some improvements over the older platform:

- Scalability
- Multi-tenancy
- Serviceability
- Locality awareness
- High Cluster Utilization
- Reliability/Availability
- Secure and auditable operation
- Support for programming model diversity
- Flexible Resource Model
- Backward compatibility

The major difference is that the functionality of the JobTracker is split into two new daemons – *ResourceManager* (RM) and *ApplicationMaster* (AM). The RM is a global service, managing all the resources and jobs in the platform. It consists of a scheduler and the *ApplicationManager*. The scheduler is responsible for allocation of resources to the various running applications based on their resource requirements. The *ApplicationManager* is responsible for accepting jobs-submissions and negotiating resources from the scheduler. Additionally, there is a *NodeManager* (NM) agent that runs on each worker. It is responsible for allocating and monitoring of node resources (CPU, memory, disk and network) usage and reports back to the *ResourceManager* (scheduler). An instance of the *ApplicationMaster* runs per-application on each node and negotiates the appropriate resource container from the scheduler. It is important to mention that the new MapReduce 2.0 maintains API compatibility with the older stable versions of Hadoop and therefore, MapReduce jobs can run unchanged.

**Hive** [10][17] is a data warehouse infrastructure built on top of Hadoop. Hive was originally developed by Facebook and supports the analysis of large data sets stored on HDFS by queries in a SQL-like declarative query language. This SQL-like language is called *HiveQL* and is based on the SQL language, but does not strictly follow the SQL-92 standard. For example, the additional feature *Use Defined Functions* (UDF) of *HiveQL* allows to filter data by custom Java or Python scripts. Plugging in custom scripts makes the implementation of in *HiveQL* natively unsupported statements possible.

When a *HiveQL* statement is submitted through the Hive command-line interface, the compiler of Hive translates the statement into jobs that are submitted to the MapReduce engine [14]. This allows users to analyze large data sets without actually having to apply the MapReduce programming model themselves. The MapReduce programming model is very low-level and requires developers to write custom programs, whereas Hive can be used by analysts with SQL skills.

Before data stored on HDFS can be analyzed by Hive, Hive's *Metastore* has to be created. The *Metastore* is the central repository for Hive's metadata and stores all information about the available databases, the available tables, the available table columns, table columns' types etc. The *Metastore* is stored on a traditional RDBMS like MySQL.

When a table is created with *HiveQL*, the user can define the format of the file that is stored on HDFS and which contains the actual data of the table [21]. Besides the default text file format, more compressed formats like *ORC* and *Parquet* are available. The applied file format affects the performance of Hive.

**Apache Spark** [22] is a processing engine that promises to perform much faster than Hadoop's MapReduce engine. This performance advantage of Spark is achieved in part by its heavy reliance on in-memory computing. In contrast to that, MapReduce is strongly based on disk. Spark was originally created in 2009 by the AMPLab at UC Berkeley and was developed to run independent of Hadoop. Instead, Spark is a generic framework for a wide variety of distributed storage systems including Hadoop.

The Spark project consists of several components [22]. The *Spark Core* is the general execution engine that provides APIs for programming languages like Java, Scala and Python and enables an easy development of Spark programs. All the other Spark components are built on top of the *Spark Core*. These components are *Spark SQL* for analyzing structured data, *Spark Streaming* for analyzing streaming data, the machine learning framework *MLlib* and the graph processing framework *GraphX*.

*Spark SQL* [23] integrates relational processing into Spark and allows users to intermix relational and procedural processing techniques. Besides the general support for structured data processing, *Spark SQL* supports SQL-like statements. These statements can be executed through a command-line interface similar to Hive's command-line interface. Moreover, *Spark SQL* is pretty compatible to run unmodified *HiveQL* queries and to use the *Hive Metastore* [24]. In summary

*Spark SQL* relates to Spark in the same way as Hive relates to MapReduce: an interface to execute SQL-like statements on the respective processing engine.

The general programming model of *Spark Core* and therefore the fundamentals for all the other Spark components can be summarized as follows [3]. To write a program running on Spark, the developer has to write the so called driver program that implements the program flow and launches various operations in parallel.

Spark provides the two main abstractions Resilient Distributed Datasets (RDD) and parallel operations. A RDD is a read-only, partitioned collection of elements.

The separate partitions of the RDD are distributed across a set of machines and can be stored in a persistent storage as well as in-memory. Persisting and caching the RDD in memory allows very efficient operations.

Besides allowing Spark's driver program to run its operations on the RDD in parallel on various machines, a RDD can automatically recover from machine failures.

**Cloudera Hadoop Distribution (CDH)** [25] is a 100% Apache-licensed open source Hadoop distribution offered by Cloudera. It includes the core Apache Hadoop elements - Hadoop Distributed File System (HDFS) and MapReduce (YARN), as well as several additional projects from the Apache Hadoop Ecosystem. All components are tightly integrated to enable ease of use and managed by a central application - Cloudera Manager [26].

## 3. BigBench

BigBench [26][27] is a proposal for an end-to-end analytics benchmark suite for Big Data systems. To fit the needs of a Big Data benchmark and to allow the performance comparison of different Big Data systems, BigBench focuses on the three Big Data characteristics volume, variety and velocity. It provides a specification describing a data model and workloads of a non-system-specific end-to-end analytics benchmark. Additionally, a data generator is available to create data for the data model.

Since the BigBench specification is general and technology agnostic, it should be implemented specifically for each Big Data system. The initial implementation of BigBench was made for the Teradata Aster platform [29]. It was done in the Aster’s SQL-MR syntax served - additionally to a description in the English language - as an initial specification of BigBench's workloads. Meanwhile, BigBench is implemented for Hadoop [1], using the MapReduce engine and other components like Hive, Mahout and OpenNLP from the Hadoop Ecosystem.

To summarize, BigBench covers the data model, depicted in Figure 1, the data generator and the specification of the workloads. Figure 1 shows how BigBench implements the variety property of Big Data. This is done by categorizing the data model into three parts: *structured*, *semi-structured* and *unstructured* data. A fictional product retailer is used as the underlying business model [27]. The business model and a large portion of the data model's structured part is derived from the TPC-DS benchmark [30]. The structured part was extended with a table for the prices of the retailer's competitors, the semi-structured part was added represented by a table with website logs and the unstructured part was added by a table showing product reviews.

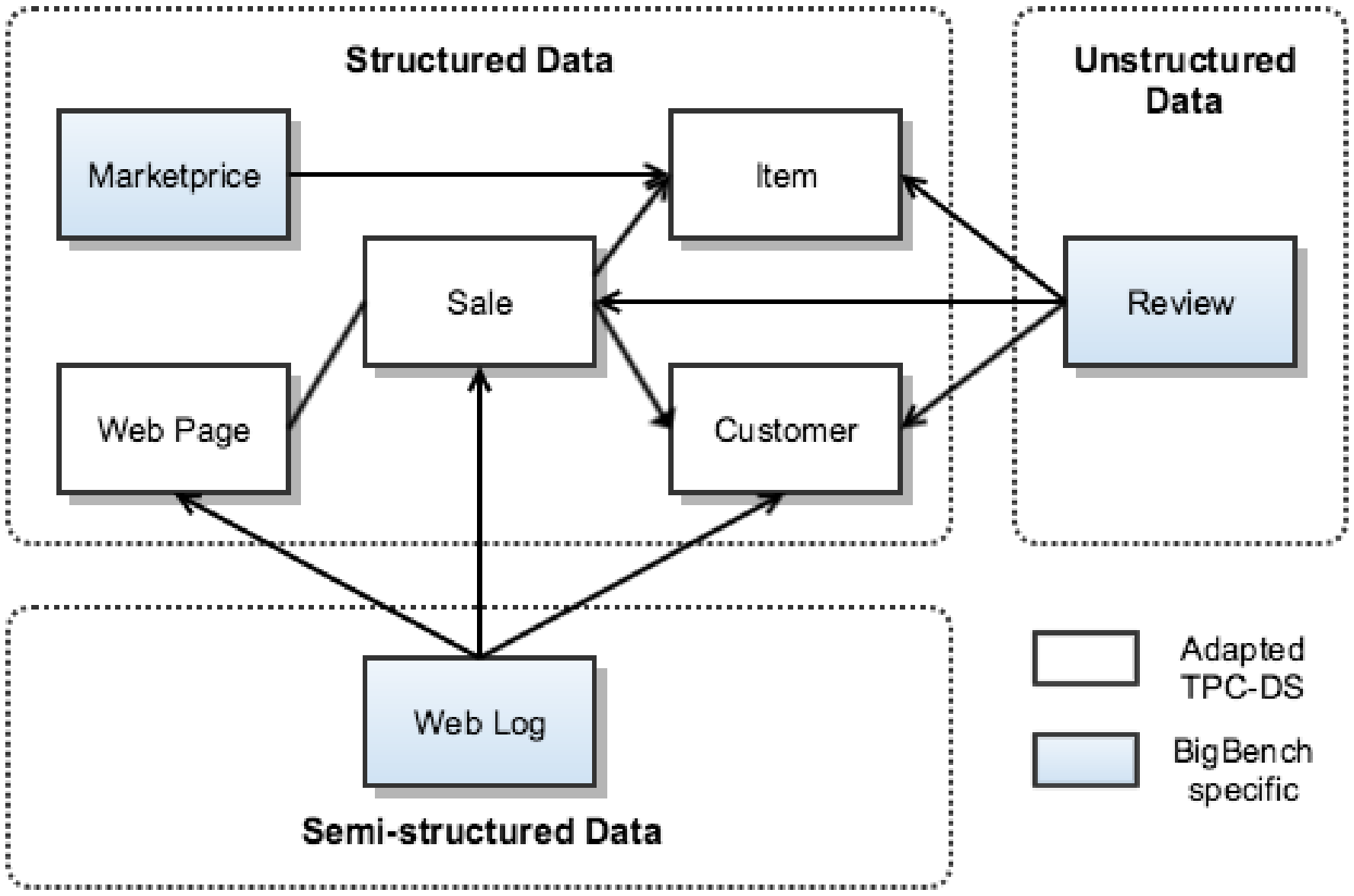

Figure 1: BigBench Schema [31]

The data generator is based on an extension of PDGF [32] and allows generating data in accordance with BigBench's data model, including the structured, semi-structured and unstructured parts. The data generator can scale the amount of data based on a scale factor. Due to parallel processing of the data generator, it runs efficiently for large scale factors. In this way, the Big Data volume property is implemented in BigBench. Additionally, the velocity property of Big Data is implemented by a periodic refresh scheme that constantly adds new data to the different tables of the data model.

The workloads are a major part of BigBench. The workloads are represented by 30 queries, which are defined as questions about the BigBench's underlying business model. Ten of these queries are taken from the TPC-DS benchmark's workload. The other 20 queries were defined based on the five major areas of Big Data analytics identified in the McKinsey report on Big Data use cases and opportunities [6]. These areas are marketing, merchandising, operations, supply chain and new business models. However, besides these business areas it was made sure that the following three technical dimensions are also included in the queries:

a) The three different data types (structured, semi-structured and unstructured type)
b) The two paradigms of processing (declarative and procedural MR)
c) Different algorithms of analytic processing (classifications, clustering, regression etc.)

A list of the BigBench queries grouped by the technologies their implementation is based on can be found in Table 1.

| Query Types | Queries | Number of Queries |
|---|---|---|
| Pure **HiveQL** | Q6, Q7, Q9, Q11, Q12, Q13, Q14, Q15, Q16, Q17, Q21, Q22, Q23, Q24 | 14 |
| **Java MapReduce** with HiveQL | Q1, Q2 | 2 |
| **Python Streaming** MR with HiveQL | Q3, Q4, Q8, Q29, Q30 | 5 |
| **Mahout** (Java MR) with HiveQL | Q5, Q20, Q25, Q26, Q28 | 5 |
| **OpenNLP** (Java MR) with HiveQL | Q10, Q18, Q19, Q27 | 4 |

Table 1: BigBench Queries

The combination of factoring in relevant business areas as well as technical dimensions within the scope of the Big Data characteristics makes BigBench to a Big Data analytics benchmark suite. Besides the objective of becoming an industry standard as TPCx-BB [4], BigBench will be extended to incorporate additional use cases in the future [28].

# 4. BigBench on Spark

A major focus of this work is to evaluate and run BigBench on Spark. Because Spark SQL supports HiveQL, the queries of the type "*Pure HiveQL*" were successfully ported to Spark and executed. However, to provide a comprehensive evaluation the additional BigBench queries will also be considered in this chapter.
The validation references described in the subsection Query Validation Reference significantly supported the evaluation. With their help, the verification of successful query executions was quite easy. The first section of this chapter presents workarounds that had to be applied at the beginning of our research. At that time, Spark SQL was at an earlier stage and did not support some of the syntactical expressions. During the project, many issues were solved by developers of the Spark project and the described workarounds became obsolete. Below, the final outcomes of the evaluation of running BigBench on Spark are described and all necessary porting tasks are listed.

## 4.1. Workarounds

Since the start of our research, further development on Spark solved several issues. However, before these improvements on Spark were available, workarounds for those issues had to be developed. In the following part, two major problems are examined to give an example of our work and an idea of the current state of Spark's component Spark SQL. The issues are described as follows: First, the actual issue is described. Then, the temporarily implemented workaround is explained. Finally, a reference to the reported ticket in the official issue tracker of the Spark project is given.

**Variables substitution**

The Hive variable substitution mechanism allows using variables within the queries. The so called *hiveconf* variables can be set by passing them with the *hiveconf* parameter to the client program or by setting them directly with the set command in the query. Furthermore, values of ordinary environment variables can be accessed within queries. Depending on whether it is a *hiveconf* variable or an environment variable, the variable can be retrieved by using the syntax *${hiveconf:variable_name}* or *${env:variable_name}* [33].
The available BigBench implementation for MapReduce uses this mechanism intensively. Initially, Spark SQL did not support this mechanism. Because this mechanism was used intensively as well as to avoid big changes on the BigBench implementation, the variable substitution concept was retained. The approach of the workaround was to retrieve and substitute the variables before the queries were passed to the Spark SQL client program. By doing so, no variables were within the queries and the actual variable substitution mechanism was obsolete. The procedure implemented in the script-based solution, which was executed before the query was passed to the Spark SQL client program, can be described as follows:

1) Searching for the variable syntax *${hiveconf: variable_name}* and *${env:variable_name}* in the query.
2) Retrieving these variables to obtain their values.
3) Replacing each variable with its received value.
4) Passing the query with replaced variables to the Spark SQL client program.

This workaround became obsolete with the resolution of the issue [SPARK-5202] *HiveContext doesn't support the Variables Substitution*[1] in the Spark project.

**User-Defined Functions (UDFs) with multiple fields as output**

When an UDF output has multiple fields, it was not possible to assign an alias for each individual field. The following example shows the desired, but unsupported expression.

```
SELECT extract_sentiment(pr_item_sk,pr_review_content) AS (pr_item_sk, review_sentence,
sentiment, sentiment_word) FROM product_reviews;
```

Because this expression was syntactically not accepted by Spark SQL, the subsequent workaround was used to solve this issue.

```
SELECT `result._c0` AS pr_item_sk, `result._c1` AS review_sentence, `result._c2` AS sentiment,
`result._c3` AS sentiment_word
FROM (
  SELECT extract_sentiment(pr.pr_item_sk,pr.pr_review_content) AS return
  FROM product_reviews pr
) result;
```

This workaround became obsolete with the resolution of the issue [SPARK-5237] *UDTF don't work with multi-alias of multi-columns as output on Spark SQL*[2] in the Spark project.

## 4.2. Porting Issues

This section documents the final outcomes of running the BigBench queries on Spark. Table 2 gives an overview of all the different porting tasks that have been identified together with the affected queries attached to each task.

| Issue | Affected Queries |
|---|---|
| External scripts in Spark SQL | Q1, Q2, Q3, Q4, Q8, Q10, Q18, Q19, Q27, Q29, Q30 |
| Different expression of null values | Q3, Q8, Q29, Q30 |
| Scripts implemented for MapReduce | Q1, Q2 |
| External libraries | Q5, Q10, Q18, Q19, Q20, Q25, Q26, Q27, Q28 |
| Query specific settings | Q3, Q4, Q7, Q8, Q16, Q21, Q22, Q23, Q24, Q29, Q30 |
| Type definition for return values | Q1, Q2, Q3, Q4, Q8 |

Table 2: Porting tasks and queries that are affected by them

Subsequently, all different porting tasks are explained in more detail.

[1] https://issues.apache.org/jira/browse/SPARK-5202
[2] https://issues.apache.org/jira/browse/SPARK-5237

**External scripts in Spark SQL**

Calling external scripts within queries executed with Spark SQL requires passing of the respective script file paths to the Spark SQL client program. This ensures that these scripts are distributed to all of the Spark executors. This is relevant for all queries containing user-defined functions (UDFs) or custom reduce scripts. Depending on whether these are represented as Java programs (JAR files) or Python scripts (PY files), the parameter to be used differs.
To make Python scripts available on the executors, the *files* parameter should be used. This places the scripts in the working directory of each executor. Affected by this issue are the BigBench queries Q1, Q2, Q10, Q18, Q19 and Q27. The usage of the *files* parameter is shown by the following generalized command. The ***$SPARK_ROOT*** variable represents the path to the root of the local Spark repository.

```
$SPARK_ROOT/bin/spark-sql --files $PY_FILE_PATH
```

To make Java programs available, the *jars* parameter should be used. Besides distributing the files to the Spark executors, this ensures that the programs will be included to the Java Classpath on each executor. Affected by this issue are the BigBench queries Q3, Q4, Q8, Q29 and Q30. Using the *jars* parameter is shown by the following generalized command.

```
$SPARK_ROOT/bin/spark-sql --jars $JAR_FILE_PATH
```

**Different expression of null values**

It became apparent that in Hive and Spark SQL, specific calculations lead to different results. Examples for such different calculation results can be found in Table 3.

| Query | Hive Result | Spark SQL Result |
|---|---|---|
| SELECT CAST(1 as double) / CAST(0 as double) FROM table; | NULL | Infinity |
| SELECT CAST(-1 as double) / CAST(0 as double) FROM table; | NULL | -Infinity |
| SELECT CAST(0 as double) / CAST(0 as double) FROM table; | NULL | NaN |

Table 3: Hive and Spark SQL differences

Furthermore, *Hive* and *Spark SQL* show different expression of null values in the context of external scripts. This different expression has impact on the row counts of several BigBench query result tables. Conditions that check if the value of a row field is equal/unequal to null, lead to different results. Null values are expressed in ***Hive*** as ***\N*** and in ***Spark SQL*** as ***null***. Subsequently, a generalized Python code example illustrates the required adjustments to ensure correct query execution when using Spark SQL.
When executing with Hive, the following condition is valid to check if a row field is unequal to null.

```
if rowField != '\N' :
        # do something
```

When executing with Spark SQL, the condition must be adjusted as follows.

```
if rowField != 'null' :
        # do something
```

Affected by this issue are external scripts of the BigBench queries Q3, Q8, Q29 and Q30.

**Scripts implemented for MapReduce**

External scripts that are specifically implemented for the MapReduce framework are not usable when running BigBench on Spark. Those scripts have to be rewritten to run with the Spark framework. This task requires understanding the respective MapReduce code and transforming it to code compatible with the Spark framework. Performing this task requires certain knowledge in the mentioned technologies. The affected BigBench queries are Q1 and Q2.

**External libraries**

The implementation of BigBench for MapReduce utilizes a small number of external libraries. It uses Apache OpenNLP for processing natural language text and Apache Mahout for performing machine learning tasks. These libraries, which are implemented to run on MapReduce, have to be replaced. In case of Apache Mahout, this means waiting for the release that runs on Spark or choosing a different machine learning library that is already running on Spark like MLlib [34]. This issue affects all queries utilizing the functionality of libraries such as Apache OpenNLP (Java MR) and Mahout (Java MR) (see Table 1).

**Query specific settings**

Contrary to Hive, Spark SQL does not dynamically determine some of the settings during query execution. The need for manually defining settings for specific queries and scale factors became obvious in the case of queries with exhaustive join operations and queries with streaming functionality. The higher the scale factor the more relevant were those settings in terms of query runtime.

Open tickets in the Spark issue tracker like [SPARK-2211] *Join Optimization*[3] and [SPARK-5791] *show poor performance when multiple table do join operation*[4] document the missing join optimization functionality in Spark, which causes the need of tweaking settings specifically for individual queries. In the official Spark documentation [35] the unsupported functionality of dynamically determining the number of partitions is described. It became apparent that setting this value properly was especially relevant for queries with streaming functionality. Ryza [36] gives a formula that roughly estimate this value. However, despite utilizing the formula, it is not a simple task to determine this setting.

Due to the complexity and the fact that the configuration of such specific settings has to be individually processed for each query and scale factor, it does not seem to be a practical approach. With further development, Spark will probably improve its abilities of dynamic settings determination and query optimization.

Affected by this issue concerning determination of query specific settings are the BigBench queries Q7, Q16, Q21, Q22, Q23, Q24 with exhaustive join operations and the BigBench queries Q3, Q4, Q8, Q29, Q30 with streaming functionality.

[3] https://issues.apache.org/jira/browse/SPARK-2211
[4] https://issues.apache.org/jira/browse/SPARK-5791

**Type definition for return values**

HiveQL supports an operation to integrate custom reduce scripts in the query data stream. Records output by these scripts have a certain number of fields. By default these fields are of the type string. However, it is possible to cast each field to a specified data type. Typecasting the fields of reduce script outputs is used in several BigBench queries. In case of Spark SQL this type casting has not worked properly and caused wrong query execution. Removing the type cast definition can solve this issue. However, Hive allows typecasting the return values of functions. Affected by this issue are the BigBench queries Q1, Q2, Q3, Q4 and Q8. This is shown in the following example:

```
SELECT result_field_one, result_field_two
FROM (
  FROM (
    SELECT
      wcs_user_sk       AS user,
      wcs_click_date_sk AS lastviewed_date,
    FROM source_table
  ) my_return_table
  REDUCE
  my_return_table.user,
  my_return_table.lastviewed_date,
  USING 'python reduce_script.py'
  AS (result_field_one BIGINT, result_field_two BIGINT));
```

When executing on Spark SQL, typecasting return values should be prevented.

```
SELECT result_field_one, result_field_two
FROM (
  FROM (
    SELECT
      wcs_user_sk       AS user,
      wcs_click_date_sk AS lastviewed_date,
    FROM source_table
  ) my_return_table
  REDUCE
  my_return_table.user,
  my_return_table.lastviewed_date,
  USING 'python reduce_script.py'
  AS (result_field_one, result_field_two));
```

## 5. Experimental Setup

This section presents the hardware and software setup of the cluster as well as the exact configuration of the Hadoop and BigBench components as used in our experiments.

### 5.1. Hardware

The experiments were performed on a cluster consisting of 4 nodes connected directly through 1GBit Netgear switch, as shown on Figure 2. All 4 nodes are Dell PowerEdge T420 servers. The master node is equipped with 2x Intel Xeon E5-2420 (1.9GHz) CPUs each with 6 cores, 32GB of RAM and 1TB (SATA, 3.5 in, 7.2K RPM, 64MB Cache) hard drive. The worker nodes are equipped with 1x Intel Xeon E5-2420 (2.20GHz) CPU with 6 cores, 32GB of RAM and 4x 1TB (SATA, 3.5 in, 7.2K RPM, 64MB Cache) hard drives. More detailed specification of the node servers is provided in the Appendix (Table 19 and Table 20).

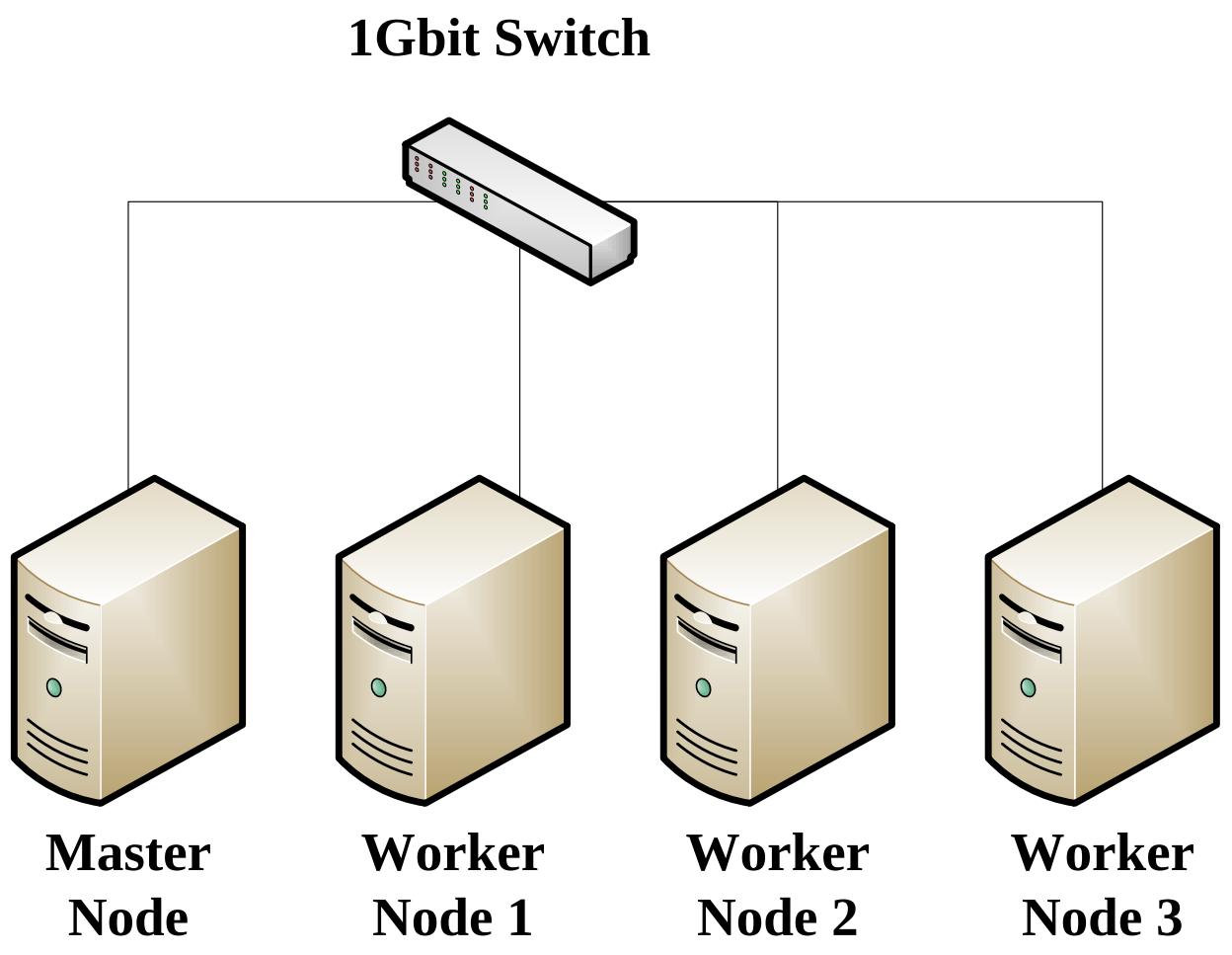

Figure 2: Cluster Setup

| Setup Description | Summary |
|---|---|
| *Total Nodes:* | 4 x Dell PowerEdge T420 |
| *Total Processors/ Cores/Threads :* | 5 CPUs/ 30 Cores/ 60 Threads |
| *Total Memory:* | 128 GB |
| *Total Number of Disks:* | 13 x 1TB,SATA, 3.5 in, 7.2K RPM, 64MB Cache |
| *Total Storage Capacity:* | 13 TB |
| *Network:* | 1 GBit Ethernet |

Table 4: Summary of Total System Resources

Table 4 summarizes the total cluster resources that are used in the calculation of the benchmark ratios in the next sections.

### 5.2. Software

This section describes the software setup of the cluster. The exact software versions that were used are listed in Table 5. Ubuntu Server LTS was installed on all 4 nodes, allocating the entire first disk. The number of open files per user was changed from the default value of 1024 to 65000 as suggested by the TPCx-HS benchmark and Cloudera guidelines [37]. Additionally, the OS ***swappiness option*** was turned permanently off (*vm.swappiness = 0*). The remaining three disks, on all worker nodes, were formatted as ***ext4*** partitions and permanently mounted with options ***noatime*** and ***nodiratime***. Then the partitions were configured to be used by HDFS through the Cloudera Manager. Each 1TB disk provides in total 916.8GB of effective HDFS space, which means that all three workers (3 x 916.8GB = 8251.2GB = 8.0578TB) have in total around 8TB of effective HDFS space.

| Software | Version |
|---|---|
| *Ubuntu Server 64 Bit* | 14.04.1 LTS, Trusty Tahr, Linux 3.13.0-32-generic |
| *Java (TM) SE Runtime Environment* | 1.6.0_31-b04, 1.7.0_72-b14 |
| *Java HotSpot (TM) 64-Bit Server VM* | 20.6-b01, mixed mode 24.72-b04, mixed mode |
| *OpenJDK Runtime Environment* | 7u71-2.5.3-0ubuntu0.14.04.1 |
| *OpenJDK 64-Bit Server VM* | 24.65-b04, mixed mode |
| *Cloudera Hadoop Distribution* | 5.2.0-1.cdh5.2.0.p0.36 |
| *BigBench* | [38] |
| *Spark* | 1.4.0-SNAPSHOT (March 27th 2015) |

Table 5: Software Stack of the System under Test

Cloudera CDH 5.2, with default configurations, was used for all experiments. Table 6 summarizes the software services running on each node. Due to the resource limitation (only 3 worker nodes) of our experimental setup, the cluster was configured to work with replication factor of 2. This means that our cluster can store at most 4TB of data on HDFS.

| Server | Disk Drive | Software Services |
|---|---|---|
| *Master Node* | Disk 1/ sda1 | Operating System, Root, Swap, Cloudera Manager Services, Name Node, SecondaryName Node, Hive Metastore, Hive Server2, Oozie Server, Spark History Server, Sqoop 2 Server, YARN Job History Server, Resource Manager, Zookeeper Server |
| *Worker Nodes 1-3* | Disk 1/ sda1 | Operating System, Root, Swap, Data Node, YARN Node Manager |
| | Disk 2/ sdb1 | Data Node |
| | Disk 3/ sdc1 | Data Node |
| | Disk 4/ sdd1 | Data Node |

Table 6: Software Services per Node

## 5.3. Cluster Configuration

Besides making modifications on the BigBench implementation as described previously, configuration parameters for the different components of the cluster have to be properly set so that BigBench queries run stable (also with higher scale factors). Determining these configuration parameters is not connected to the particular case of running the BigBench benchmark. Instead, this is part of the general complexity of Big Data systems and is essential to their proper operation. As a basic principle when setting the configuration parameters, we tried to follow the rule that these should not differ from their default values unless adjusting is needed to ensure correct cluster operation. This principle avoids tuning of special cases to guarantee meaningful benchmarking results. However, utilizing all the available cluster resources and running BigBench with higher scale factors demonstrated the need for adjusting some of the parameters. Furthermore, some configuration parameters of Spark were not set by default and had to be defined accordingly. The process of determining the configuration parameters can be described as follows and was executed for each individual BigBench query:

1) Identifying errors and abnormal runtime in BigBench query execution.
2) Figuring out which configuration parameters cause the problem.
3) Trying to find problem-solving values for configuration parameters.
4) Validating the configuration parameter values by re-executing the BigBench query: Parameter values are determined successfully when errors are fixed and abnormal runtime is solved.

Components of the cluster that were actually affected by adjusted configuration parameters are YARN, Spark, MapReduce and Hive. It should be noted that changing the configuration parameters of YARN has an impact on Spark as well as MapReduce because both processing engines are dependent on the resource manager YARN. Hereafter, the changed configuration parameters of the particular cluster components are documented and explained.

**YARN**

To adjust the configuration of the resource manager YARN in order to fit the experimental cluster and to ensure efficient resource utilization, two configuration parameters were adjusted initially. The amount of memory that can be allocated for YARN ResourceContainers per node (***yarn.nodemanager.resource.memory-mb = 28672***) and the maximum allocation for every YARN ResourceContainer request were set to 28 GB (***yarn.scheduler.maximum-allocation-mb = 28672***). Later, following the recommendations published by Ryza [36], the amount of memory that can be allocated for YARN ResourceContainers was changed to 31 GB per node (***yarn.nodemanager.resource.memory-mb = 31744***). As described in Hortonworks' manual [39], the maximum allocation for every YARN ResourceContainer request was set to be exactly the same as the amount of memory that can be allocated for YARN ResourceContainers. In short, this parameter defines the largest ResourceContainers size YARN will allow. It was also set to 31 GB (***yarn.scheduler.maximum-allocation-mb = 31744***). Following the recommendations published by Ryza [36], the number of CPU cores that can be allocated for YARN ResourceContainers was changed to 11 per node (***yarn.nodemanager.resource.cpu-vcores = 11***). The final configuration gives YARN plenty of resources, but still leaves 1 GB of memory and 1 CPU core to the operating system. All of the above YARN configuration adjustments were made in the respective ***yarn-site.xml*** configuration file.

**Spark**

Since the Spark version shipped with CDH 5.2.0 was not used, the Spark configuration that comes with CDH was deactivated. Many configuration parameters can be set by passing them to the Spark client program. Besides passing ***--master yarn*** to run YARN in client mode, the configuration parameters ***--num-executors***, ***--executor-cores*** and ***--executor-memory*** should be passed with proper values. Initially, finding proper values for the above mentioned configuration parameters was done by performing spot-check tests. The different configuration parameter values of the performed tests and their runtime for two randomly chosen BigBench queries can be found in Table 7. The test results prompted us to set the configuration parameters to the values used in configuration 4 (***--num-executors 12***, ***--executor-cores 2***, ***--executor-memory 8G***).

| # | num-executors | executor-memory | executor-cores | Time (min) | Time (min) |
|---|---|---|---|---|---|
| | | | | Q7 | Q24 |
| 1 | 3 | 26G | 12 | 2.98 | 4.15 |
| 2 | 3 | 26G | 10 | 3.02 | 4.12 |
| 3 | 6 | 16G | 4 | 3.05 | 4.05 |
| 4 | 12 | 8G | 2 | 2.73 | 3.55 |

Table 7: Runtime for different Spark configurations

However, the recommendations published by Ryza [36], give a more methodical guideline regarding the Spark configuration parameters.
The sample cluster in the guide configured 3 executors on each DataNode except the one operating the ApplicationMaster, which has only 2 executors. Due to different hardware resources, there are maximum 3 executors on every DataNode. Because of the configuration parameter values ***--executor-cores*** and ***--executor-memory,*** on every DataNode there will be available resources for the ApplicationMaster. Consequently, this results in total of 9 executors (-***-num-executors 9***).
Every DataNode in the experimental cluster has 12 virtual CPU cores. Since one core is left for the operating system and Hadoop daemons, there are 11 virtual cores available for the executors. Dividing the number of cores by the 3 executors per node results in 3 cores per executor (--***executor-cores 3***). Therefore, 9 virtual cores per node are used for executors, 1 core is left for the operating system and 2 spare cores are available. These two cores are the ones available for the ApplicationMaster.
The amount of memory per executor can be determined by the following calculation:

$$approx_{_em} = \frac{\text{yarn. nodemanager. resource. memory} - \text{mb}}{\text{num} - \text{executors}} = \frac{31\,744}{3} = 10\,581$$

The variable approx_em stores the amount of memory which is theoretically available for each executor. However, the Java Virtual Machine (JVM) overhead has to be considered and included into the calculation. This can be done by subtracting the value of the property ***spark.yarn.executor.memoryOverhead*** from the calculated approx_em value. If the property ***spark.yarn.executor.memoryOverhead*** is not explicitly set by the user, its default value is calculated by ***max (384, 0.07 * executor-memory)***. Listed below is the calculation done in order to determine the memory per executor:

$$\begin{aligned}
&\text{executor} - \text{memory} = \\
&= \text{approx}_{\text{em}} - \text{spark. yarn. executor. memoryOverhead} = \\
&= \text{approx}_{\text{em}}\text{em} - \max(384, 0.07 * \text{approx}_{\text{em}}) = \\
&= 10581 - \max(384, 0.07 * 10581) = \\
&= 9840
\end{aligned}$$

The resulting integral value 9840 MB is adjusted downward to 9 GB (***--executor-memory = 9G***).
In addition to the above configurations, which have to be passed as parameter when calling the client program, the default serializer used for object serialization was also changed (***spark.serializer = org.apache.spark.serializer.KryoSerializer***). The faster Kryo serializer was chosen over the default serializer as recommended by various sources [36], [35]. The serializer option was adjusted in the respective ***spark-defaults.conf*** configuration file.

**MapReduce**

Specifically for the BigBench queries, which include Java MapReduce programs (Q1 and Q2), configuration parameters had to be adjusted to ensure accurate execution. Execution errors were caused by not allowing enough memory for the map and reduce tasks. Also the allowed Java heap size of the map and reduce tasks [40] had to be increased. To find proper values for these parameters, values were raised incrementally until errors were eliminated. This resulted in the following adjusted parameters: ***mapreduce.map.java.opts.max.heap = 2GB***, ***mapreduce.reduce.java.opts.max.heap = 2GB***, ***mapreduce.map.memory.mb = 3GB***, ***mapreduce.reduce.memory.mb = 3GB***. These settings were changed in the respective ***mapred-site.xml*** configuration file.

**Hive**

When executing BigBench's query Q9 with the default configuration, Hive encountered an out of memory error. Initially, this issue was solved by deactivating MapJoins for this particular query. The MapJoin feature allows loading a table in memory, so that a very fast table scan can be performed [41]. As a consequence, performing a MapJoin requires more memory resources. In our case this caused out of memory errors, which could be resolve by simply deactivating this feature. Deactivating was done by just setting ***hive.auto.convert.join = false*** in the file ***engines/hive/queries/q09/q09.sql*** of the BigBench repository.
Even though deactivating MapJoins solves the problem, it entails a significant performance decline. A better solution is the increase of the heap size of the local Hadoop JVM to prevent the out of memory error. In our case the heap size was increased to 2 GB. This was done by adding the parameters ***-Xms2147483648*** and ***-Xmx2147483648*** to the environment variable ***HADOOP_CLIENT_OPTS*** in the responsible ***hive-env.sh*** file.

**Configuration validation**

During the progress of determining proper parameter values, multiple validations were performed. Especially after applying the guidelines published by Ryza [36] and after choosing the better solution for the MapJoin issue described in section 4.2, the values were validated against the one previously used. It should be noted that the previous configuration can be also seen as a viable configuration. However, the following validation results should verify Ryzas' guidelines [36] and demonstrate the performance difference between the two configurations. Table 8 lists the different parameters for the ***default***, ***initial*** and ***final*** configurations as used in our cluster configuration. Figure 3 illustrates the effect on queries' runtime when changing the ***initial*** cluster configuration to the ***final*** cluster configuration.

| Component | Parameter | Default Configuration | Initial Configuration | Final Configuration |
|---|---|---|---|---|
| YARN | yarn.nodemanager.resource.memory-mb | 8GB | 28GB | ***31GB*** |
| | yarn.scheduler.maximum-allocation-mb | 8GB | 28GB | ***31GB*** |
| | yarn.nodemanager.resource.cpu-vcores | 8 | 8 | ***11*** |
| Spark | master | local | yarn | ***yarn*** |
| | num-executors | 2 | 12 | ***9*** |
| | executor-cores | 1 | 2 | ***3*** |
| | executor-memory | 1GB | 8GB | ***9GB*** |
| | spark.serializer | org.apache.spark. serializer.JavaSerializer | org.apache.spark. serializer.JavaSerializer | ***org.apache.spark. serializer.KryoSerializer*** |
| MapReduce | mapreduce.map.java.opts.max.heap | 788MB | 2GB | ***2GB*** |
| | mapreduce.reduce.java.opts.max.heap | 788MB | 2GB | ***2GB*** |
| | mapreduce.map.memory.mb | 1GB | 3GB | ***3GB*** |
| | mapreduce.reduce.memory.mb | 1GB | 3GB | ***3GB*** |
| Hive | hive.auto.convert.join (Q9 only) | true | false | ***true*** |
| | Client Java Heap Size | 256MB | 256MB | ***2GB*** |

Table 8: Initial and final configuration

Considering the differences in the runtimes of the individual queries depicted in Figure 3, no big difference can be seen when running them on MapReduce except for query Q9. The reason for this was that the maximum client Java heap size was raised to 2GB. However, it seems that no other query except query Q9 was running into that limit, so this change did not have any impact on the runtimes. As mentioned in the above Hive section, not turning off MapJoins for query Q9, but raising the maximum client Java heap size instead, significantly improved its runtime. In case of running the queries with Apache Spark, the runtime of 8 queries became faster whereas 4 queries became slower.

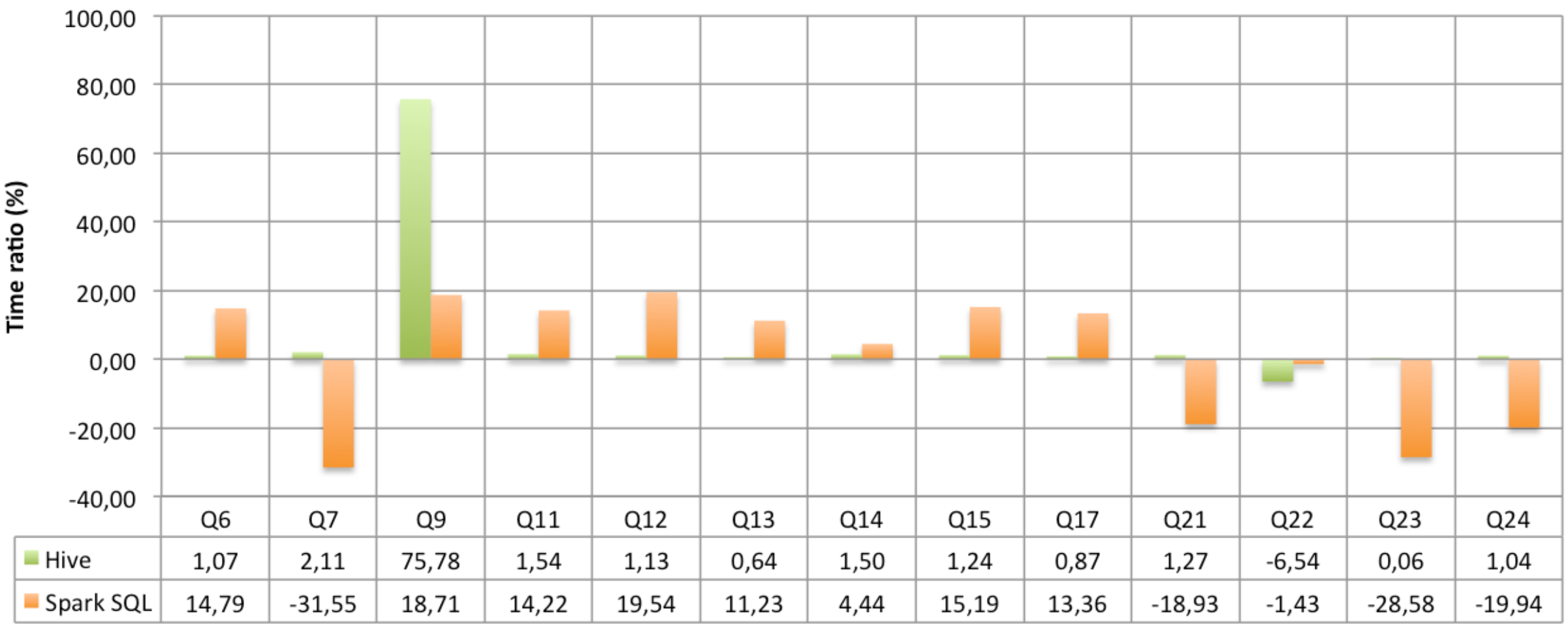

Figure 3: Differences in runtime for different cluster configurations

In summary it can be said that the ***initial*** configuration that was determined through testing can be considered a decent configuration because it showed slightly slower runtimes compared to the ***final*** configuration. Therefore, the ***final*** configuration following the best practices was chosen for the real benchmarking experiments. Investigating the performance of different configurations in advance allowed us to validate the ***final*** configuration. This was sufficient for our benchmark purposes since our goal was not to find the optimal cluster configuration.

## 6. Experimental Results

This section presents the query execution results. Experiments were performed with all the 30 BigBench queries on MapReduce/Hive and the group of 14 pure HiveQL queries on Spark SQL for four different scale factors (100GB, 300GB, 600GB and 1TB).

### 6.1. BigBench on MapReduce

Table 9 summarizes the absolute runtimes of the 30 BigBench queries on MapReduce/Hive for 100GB scale factor. There are three columns depicting the times for each run in minutes, a column with the average execution time from the three runs and two columns with the standard deviation in minutes and in %. The cells with yellow mark the queries with standard deviation higher than 2%.

| Query | Run1 (min) | Run2 (min) | Run3 (min) | Average Time (min) | Standard Deviation (min) | Standard Deviation % |
|---|---|---|---|---|---|---|
| Q1 | 3.75 | 3.77 | 3.73 | 3.75 | 0.02 | 0.44 |
| Q2 | 8.40 | 8.27 | 8.03 | 8.23 | 0.19 | 2.25 |
| Q3 | 10.20 | 10.22 | 9.55 | 9.99 | 0.38 | 3.81 |
| Q4 | 72.58 | 72.98 | 68.55 | 71.37 | 2.45 | 3.44 |
| Q5 | 27.85 | 28.18 | 27.07 | 27.70 | 0.57 | 2.07 |
| Q6 | 6.43 | 6.37 | 6.27 | 6.36 | 0.08 | 1.32 |
| Q7 | 9.18 | 9.10 | 8.92 | 9.07 | 0.14 | 1.50 |
| Q8 | 8.57 | 8.60 | 8.60 | 8.59 | 0.02 | 0.22 |
| Q9 | 3.12 | 3.08 | 3.18 | 3.13 | 0.05 | 1.63 |
| Q10 | 15.58 | 15.50 | 15.25 | 15.44 | 0.17 | 1.12 |
| Q11 | 2.90 | 2.87 | 2.88 | 2.88 | 0.02 | 0.58 |
| Q12 | 7.18 | 6.97 | 6.97 | 7.04 | 0.13 | 1.78 |
| Q13 | 8.43 | 8.28 | 8.43 | 8.38 | 0.09 | 1.03 |
| Q14 | 3.12 | 3.12 | 3.27 | 3.17 | 0.09 | 2.73 |
| Q15 | 2.03 | 2.05 | 2.03 | 2.04 | 0.01 | 0.47 |
| Q16 | 5.88 | 5.90 | 5.57 | 5.78 | 0.19 | 3.25 |
| Q17 | 7.55 | 7.53 | 7.72 | 7.60 | 0.10 | 1.33 |
| Q18 | 8.47 | 8.57 | 8.57 | 8.53 | 0.06 | 0.68 |
| Q19 | 6.53 | 6.53 | 6.60 | 6.56 | 0.04 | 0.59 |
| Q20 | 8.50 | 8.62 | 8.02 | 8.38 | 0.32 | 3.80 |
| Q21 | 4.58 | 4.53 | 4.63 | 4.58 | 0.05 | 1.09 |
| Q22 | 16.53 | 16.92 | 16.48 | 16.64 | 0.24 | 1.42 |
| Q23 | 18.18 | 18.05 | 18.37 | 18.20 | 0.16 | 0.87 |
| Q24 | 4.82 | 4.80 | 4.77 | 4.79 | 0.03 | 0.53 |
| Q25 | 6.25 | 6.22 | 6.22 | 6.23 | 0.02 | 0.31 |
| Q26 | 5.32 | 5.18 | 5.08 | 5.19 | 0.12 | 2.25 |
| Q27 | 0.93 | 0.87 | 0.93 | 0.91 | 0.04 | 4.22 |
| Q28 | 18.30 | 18.38 | 18.40 | 18.36 | 0.05 | 0.29 |
| Q29 | 5.27 | 5.28 | 4.97 | 5.17 | 0.18 | 3.45 |
| Q30 | 19.68 | 19.77 | 18.98 | 19.48 | 0.43 | 2.21 |

Table 9: MapReduce Executions for all BigBench Queries with SF 100 (100GB data)

Similarly, Table 10 illustrates the three runtimes for the BigBench queries on MapReduce/Hive with 300GB scale factor. Again reported are the average time from three runs and the standard deviation in minutes and in %. For this scale factor all of the queries show standard deviation within 2%, which is an indicator for stable performance.

| Query | Run1 (min) | Run2 (min) | Run3 (min) | Average Time (min) | Standard Deviation (min) | Standard Deviation % |
|---|---|---|---|---|---|---|
| Q1 | 5.57 | 5.53 | 5.47 | 5.52 | 0.05 | 0.92 |
| Q2 | 21.15 | 21.03 | 21.03 | 21.07 | 0.07 | 0.32 |
| Q3 | 26.28 | 26.37 | 26.30 | 26.32 | 0.04 | 0.17 |
| Q4 | 221.53 | 221.83 | 220.58 | 221.32 | 0.65 | 0.29 |
| Q5 | 76.68 | 76.52 | 76.48 | 76.56 | 0.11 | 0.14 |
| Q6 | 10.75 | 10.60 | 10.73 | 10.69 | 0.08 | 0.77 |
| Q7 | 17.02 | 16.88 | 16.87 | 16.92 | 0.08 | 0.49 |
| Q8 | 17.78 | 17.62 | 17.82 | 17.74 | 0.11 | 0.60 |
| Q9 | 6.60 | 6.52 | 6.57 | 6.56 | 0.04 | 0.64 |
| Q10 | 19.70 | 19.82 | 19.48 | 19.67 | 0.17 | 0.86 |
| Q11 | 4.60 | 4.62 | 4.60 | 4.61 | 0.01 | 0.21 |
| Q12 | 11.68 | 11.60 | 11.52 | 11.60 | 0.08 | 0.72 |
| Q13 | 12.97 | 13.08 | 12.95 | 13.00 | 0.07 | 0.56 |
| Q14 | 5.47 | 5.53 | 5.45 | 5.48 | 0.04 | 0.80 |
| Q15 | 3.02 | 2.97 | 3.05 | 3.01 | 0.04 | 1.39 |
| Q16 | 14.72 | 14.90 | 14.88 | 14.83 | 0.10 | 0.68 |
| Q17 | 10.95 | 10.85 | 10.93 | 10.91 | 0.05 | 0.49 |
| Q18 | 11.12 | 11.05 | 10.88 | 11.02 | 0.12 | 1.09 |
| Q19 | 7.20 | 7.22 | 7.25 | 7.22 | 0.03 | 0.35 |
| Q20 | 20.22 | 20.42 | 20.23 | 20.29 | 0.11 | 0.55 |
| Q21 | 6.90 | 6.85 | 6.92 | 6.89 | 0.03 | 0.50 |
| Q22 | 19.35 | 19.85 | 19.08 | 19.43 | 0.39 | 2.00 |
| Q23 | 20.12 | 20.92 | 20.48 | 20.51 | 0.40 | 1.95 |
| Q24 | 7.00 | 7.05 | 7.00 | 7.02 | 0.03 | 0.41 |
| Q25 | 11.18 | 11.20 | 11.23 | 11.21 | 0.03 | 0.23 |
| Q26 | 8.58 | 8.57 | 8.55 | 8.57 | 0.02 | 0.19 |
| Q27 | 0.63 | 0.63 | 0.62 | 0.63 | 0.01 | 1.53 |
| Q28 | 21.27 | 21.25 | 21.22 | 21.24 | 0.03 | 0.12 |
| Q29 | 11.68 | 11.72 | 11.80 | 11.73 | 0.06 | 0.51 |
| Q30 | 57.73 | 57.60 | 57.72 | 57.68 | 0.07 | 0.13 |

Table 10: MapReduce Executions for all BigBench Queries with SF 300 (300GB data)

Table 11 depicts the three runtimes for the BigBench queries on MapReduce/Hive with 600GB scale factor. Reported are the average runtime from the three runs and the standard deviation in minutes and in %. For this scale factor all of the queries show standard deviation within 2%, except for Q22, which has standard deviation of around 5.6%.

| Query | Run1 (min) | Run2 (min) | Run3 (min) | Average Time (min) | Standard Deviation (min) | Standard Deviation % |
|---|---|---|---|---|---|---|
| Q1 | 8.10 | 8.10 | 8.13 | 8.11 | 0.02 | 0.24 |
| Q2 | 40.35 | 40.02 | 39.97 | 40.11 | 0.21 | 0.52 |
| Q3 | 53.60 | 53.52 | 53.23 | 53.45 | 0.19 | 0.36 |
| Q4 | 510.25 | 502.00 | 493.67 | 501.97 | 8.29 | 1.65 |
| Q5 | 155.18 | 155.52 | 156.35 | 155.68 | 0.60 | 0.39 |
| Q6 | 16.65 | 16.78 | 16.75 | 16.73 | 0.07 | 0.41 |
| Q7 | 29.43 | 29.47 | 29.63 | 29.51 | 0.11 | 0.36 |
| Q8 | 32.45 | 32.47 | 32.47 | 32.46 | 0.01 | 0.03 |
| Q9 | 11.45 | 11.58 | 11.47 | 11.50 | 0.07 | 0.63 |
| Q10 | 24.07 | 24.53 | 24.28 | 24.29 | 0.23 | 0.96 |
| Q11 | 7.47 | 7.48 | 7.42 | 7.46 | 0.03 | 0.47 |
| Q12 | 18.83 | 18.62 | 18.55 | 18.67 | 0.15 | 0.79 |
| Q13 | 20.22 | 20.20 | 20.27 | 20.23 | 0.03 | 0.17 |
| Q14 | 9.00 | 8.98 | 8.98 | 8.99 | 0.01 | 0.11 |
| Q15 | 4.45 | 4.48 | 4.47 | 4.47 | 0.02 | 0.37 |
| Q16 | 28.83 | 29.27 | 29.28 | 29.13 | 0.26 | 0.88 |
| Q17 | 14.60 | 14.63 | 14.57 | 14.60 | 0.03 | 0.23 |
| Q18 | 14.48 | 14.32 | 14.53 | 14.44 | 0.11 | 0.79 |
| Q19 | 7.55 | 7.67 | 7.52 | 7.58 | 0.08 | 1.04 |
| Q20 | 39.27 | 39.30 | 39.40 | 39.32 | 0.07 | 0.18 |
| Q21 | 10.22 | 10.25 | 10.20 | 10.22 | 0.03 | 0.25 |
| Q22 | 18.72 | 19.80 | 20.93 | 19.82 | 1.11 | 5.59 |
| Q23 | 23.10 | 23.08 | 23.47 | 23.22 | 0.22 | 0.93 |
| Q24 | 10.27 | 10.30 | 10.33 | 10.30 | 0.03 | 0.32 |
| Q25 | 19.88 | 20.02 | 20.07 | 19.99 | 0.09 | 0.47 |
| Q26 | 15.05 | 15.00 | 15.20 | 15.08 | 0.10 | 0.69 |
| Q27 | 0.98 | 0.98 | 0.97 | 0.98 | 0.01 | 0.98 |
| Q28 | 24.77 | 24.73 | 24.82 | 24.77 | 0.04 | 0.17 |
| Q29 | 22.78 | 22.73 | 22.82 | 22.78 | 0.04 | 0.18 |
| Q30 | 119.38 | 120.27 | 119.93 | 119.86 | 0.45 | 0.37 |

Table 11: MapReduce Executions for all BigBench Queries with SF 600 (600GB data)

Table 9 summarizes the absolute runtimes of the 30 BigBench queries on MapReduce/Hive for 1000GB/1TB scale factor. There are three columns depicting the times for each run in minutes, a column with the average execution time from the three runs and two columns with the standard deviation in minutes and in %. Similar to the smaller scale factors, only Q22 has a slightly higher than 2% standard deviation and is marked in yellow.

| Query | Run1 (min) | Run2 (min) | Run3 (min) | Average Time (min) | Standard Deviation (min) | Standard Deviation % |
|---|---|---|---|---|---|---|
| Q1 | 10.48 | 10.45 | 10.63 | 10.52 | 0.10 | 0.93 |
| Q2 | 68.12 | 66.47 | 66.90 | 67.16 | 0.86 | 1.27 |
| Q3 | 89.30 | 91.48 | 90.87 | 90.55 | 1.13 | 1.24 |
| Q4 | 927.67 | 918.05 | 940.33 | 928.68 | 11.18 | 1.20 |
| Q5 | 272.53 | 268.67 | 264.27 | 268.49 | 4.14 | 1.54 |
| Q6 | 25.28 | 25.40 | 25.67 | 25.42 | 0.20 | 0.77 |
| Q7 | 46.40 | 46.47 | 46.33 | 46.33 | 0.07 | 0.14 |
| Q8 | 53.30 | 53.78 | 53.93 | 53.67 | 0.33 | 0.62 |
| Q9 | 17.62 | 17.87 | 17.68 | 17.72 | 0.13 | 0.73 |
| Q10 | 22.92 | 22.62 | 22.67 | 22.73 | 0.16 | 0.71 |
| Q11 | 11.20 | 11.23 | 11.23 | 11.24 | 0.02 | 0.17 |
| Q12 | 29.93 | 29.88 | 30.05 | 29.86 | 0.09 | 0.29 |
| Q13 | 30.30 | 30.30 | 30.07 | 30.18 | 0.13 | 0.45 |
| Q14 | 13.88 | 13.83 | 13.87 | 13.84 | 0.03 | 0.18 |
| Q15 | 6.35 | 6.38 | 6.38 | 6.37 | 0.02 | 0.30 |
| Q16 | 48.77 | 48.63 | 48.87 | 48.85 | 0.12 | 0.24 |
| Q17 | 18.53 | 18.63 | 18.62 | 18.57 | 0.05 | 0.29 |
| Q18 | 27.60 | 27.75 | 27.45 | 27.60 | 0.15 | 0.54 |
| Q19 | 8.18 | 8.15 | 8.13 | 8.16 | 0.03 | 0.31 |
| Q20 | 64.83 | 64.92 | 64.77 | 64.84 | 0.08 | 0.12 |
| Q21 | 14.90 | 14.93 | 14.98 | 14.92 | 0.04 | 0.28 |
| Q22 | 29.78 | 31.03 | 30.67 | 29.84 | 0.64 | 2.15 |
| Q23 | 25.05 | 25.23 | 25.12 | 25.16 | 0.09 | 0.37 |
| Q24 | 14.75 | 14.77 | 14.83 | 14.75 | 0.04 | 0.30 |
| Q25 | 31.65 | 31.65 | 31.55 | 31.62 | 0.06 | 0.18 |
| Q26 | 22.92 | 22.85 | 23.07 | 22.94 | 0.11 | 0.48 |
| Q27 | 0.70 | 0.68 | 0.68 | 0.69 | 0.01 | 1.40 |
| Q28 | 28.87 | 28.87 | 29.05 | 28.93 | 0.11 | 0.37 |
| Q29 | 37.05 | 37.37 | 37.35 | 37.21 | 0.18 | 0.48 |
| Q30 | 199.30 | 203.10 | 200.50 | 200.97 | 1.94 | 0.97 |

Table 12: MapReduce Executions for all BigBench Queries with SF 1000 (1TB data)

## 6.2. BigBench on Spark SQL

This part presents the group of 14 pure HiveQL BigBench queries executed on Spark SQL with different scale factors.

Table 13 summarizes the absolute runtimes of the 14 queries run on Spark SQL for 100GB scale factor. Reported are the absolute times from the three runs, the average runtime in minutes and the standard deviation in minutes and in %. The yellow cells indicate the queries with standard deviation higher or equal to 2% and possibly unstable behavior.

| Query | Run1 (min) | Run2 (min) | Run3 (min) | Average Time (min) | Standard Deviation (min) | Standard Deviation % |
|---|---|---|---|---|---|---|
| Q6 | 2.53 | 2.60 | 2.50 | 2.54 | 0.05 | 2.00 |
| Q7 | 2.53 | 2.53 | 2.55 | 2.54 | 0.01 | 0.38 |
| Q9 | 1.25 | 1.25 | 1.22 | 1.24 | 0.02 | 1.55 |
| Q11 | 1.17 | 1.15 | 1.15 | 1.16 | 0.01 | 0.83 |
| Q12 | 1.95 | 1.98 | 1.95 | 1.96 | 0.02 | 0.98 |
| Q13 | 2.43 | 2.42 | 2.43 | 2.43 | 0.01 | 0.40 |
| Q14 | 1.25 | 1.23 | 1.25 | 1.24 | 0.01 | 0.77 |
| Q15 | 1.40 | 1.40 | 1.40 | 1.40 | 0.00 | 0.00 |
| Q16 | 3.40 | 3.38 | 3.43 | 3.41 | 0.03 | 0.75 |
| Q17 | 1.55 | 1.55 | 1.57 | 1.56 | 0.01 | 0.62 |
| Q21 | 2.70 | 2.68 | 2.65 | 2.68 | 0.03 | 0.95 |
| Q22 | 31.75 | 45.50 | 32.73 | 36.66 | 7.67 | 20.92 |
| Q23 | 16.08 | 17.45 | 16.52 | 16.68 | 0.70 | 4.19 |
| Q24 | 3.32 | 3.33 | 3.33 | 3.33 | 0.01 | 0.29 |

Table 13: Spark SQL Executions for the group of 14 HiveQL BigBench Queries with SF 100 (100GB data)

Analogous Table 14 presents the execution times with 300GB scale factor. The reported columns are the same and the yellow cells indicate standard deviation higher than 2%.

| Query | Run1 (min) | Run2 (min) | Run3 (min) | Average Time (min) | Standard Deviation (min) | Standard Deviation % |
|---|---|---|---|---|---|---|
| Q6 | 3.42 | 3.50 | 3.63 | 3.52 | 0.11 | 3.11 |
| Q7 | 6.03 | 6.17 | 5.93 | 6.04 | 0.12 | 1.94 |
| Q9 | 1.70 | 1.73 | 1.68 | 1.71 | 0.03 | 1.49 |
| Q11 | 1.37 | 1.38 | 1.38 | 1.38 | 0.01 | 0.70 |
| Q12 | 3.05 | 3.05 | 3.08 | 3.06 | 0.02 | 0.63 |
| Q13 | 3.58 | 3.60 | 3.60 | 3.59 | 0.01 | 0.27 |
| Q14 | 1.55 | 1.58 | 1.55 | 1.56 | 0.02 | 1.23 |
| Q15 | 1.58 | 1.58 | 1.60 | 1.59 | 0.01 | 0.61 |
| Q16 | 7.85 | 8.00 | 7.80 | 7.88 | 0.10 | 1.32 |
| Q17 | 2.13 | 2.22 | 2.22 | 2.19 | 0.05 | 2.20 |
| Q21 | 10.12 | 11.13 | 10.67 | 10.64 | 0.51 | 4.78 |
| Q22 | 54.90 | 62.10 | 65.07 | 60.69 | 5.23 | 8.61 |
| Q23 | 25.57 | 26.60 | 28.88 | 27.02 | 1.70 | 6.28 |
| Q24 | 15.22 | 15.23 | 15.35 | 15.27 | 0.07 | 0.48 |

Table 14: Spark SQL Executions for the group of 14 HiveQL BigBench Queries with SF 300 (300GB data)

Table 15 shows the absolute Spark SQL runtimes with 600GB scale factor. The higher standard deviations are marked with yellow.

| Query | Run1 (min) | Run2 (min) | Run3 (min) | Average Time (min) | Standard Deviation (min) | Standard Deviation % |
|---|---|---|---|---|---|---|
| Q6 | 4.80 | 4.82 | 4.88 | 4.83 | 0.04 | 0.91 |
| Q7 | 24.67 | 21.07 | 18.67 | 21.47 | 3.02 | 14.07 |
| Q9 | 2.32 | 2.30 | 2.30 | 2.31 | 0.01 | 0.42 |
| Q11 | 1.67 | 1.70 | 1.68 | 1.68 | 0.02 | 0.99 |
| Q12 | 4.92 | 4.92 | 4.93 | 4.92 | 0.01 | 0.20 |
| Q13 | 5.57 | 5.55 | 5.60 | 5.57 | 0.03 | 0.46 |
| Q14 | 2.10 | 2.12 | 2.08 | 2.10 | 0.02 | 0.79 |
| Q15 | 1.93 | 1.92 | 1.93 | 1.93 | 0.01 | 0.50 |
| Q16 | 23.78 | 23.40 | 22.78 | 23.32 | 0.50 | 2.16 |
| Q17 | 2.90 | 2.90 | 2.92 | 2.91 | 0.01 | 0.33 |
| Q21 | 28.32 | 27.38 | 25.83 | 27.18 | 1.25 | 4.62 |
| Q22 | 96.00 | 78.55 | 92.22 | 88.92 | 9.18 | 10.32 |
| Q23 | 57.77 | 53.78 | 44.78 | 52.11 | 6.65 | 12.76 |
| Q24 | 41.62 | 46.08 | 38.87 | 42.19 | 3.64 | 8.63 |

Table 15: Spark SQL Executions for the group of 14 HiveQL BigBench Queries with SF 600 (600GB data)

Finally, Table 16 summerizes the query times for the largest 1000GB scale factor. Most of the queries show standard deviation higher than 2% which is marked with yellow.

| Query | Run1 (min) | Run2 (min) | Run3 (min) | Average Time (min) | Standard Deviation (min) | Standard Deviation % |
|---|---|---|---|---|---|---|
| Q6 | 6.68 | 6.72 | 6.75 | 6.70 | 0.03 | 0.50 |
| Q7 | 39.68 | 42.73 | 42.67 | 41.07 | 1.74 | 4.24 |
| Q9 | 2.78 | 2.90 | 2.78 | 2.82 | 0.07 | 2.42 |
| Q11 | 2.08 | 2.07 | 2.08 | 2.07 | 0.01 | 0.46 |
| Q12 | 7.55 | 7.62 | 7.52 | 7.56 | 0.05 | 0.67 |
| Q13 | 8.03 | 7.95 | 7.97 | 7.98 | 0.04 | 0.55 |
| Q14 | 2.95 | 2.82 | 2.87 | 2.83 | 0.07 | 2.38 |
| Q15 | 2.37 | 2.35 | 2.35 | 2.36 | 0.01 | 0.41 |
| Q16 | 42.73 | 45.63 | 42.83 | 43.65 | 1.65 | 3.77 |
| Q17 | 3.55 | 3.47 | 3.52 | 3.55 | 0.04 | 1.18 |
| Q21 | 45.30 | 51.73 | 49.37 | 48.08 | 3.25 | 6.77 |
| Q22 | 110.27 | 114.78 | 138.92 | 122.68 | 15.40 | 12.56 |
| Q23 | 69.40 | 74.78 | 71.57 | 69.01 | 2.71 | 3.92 |
| Q24 | 83.32 | 77.20 | 76.02 | 77.05 | 3.92 | 5.09 |

Table 16: Spark SQL Executions for the group of 14 HiveQL BigBench Queries with SF 1000 (1TB data)

## 6.3. Query Validation Reference

This section provides the tables with exact values that were used in the process of porting and evaluation of the BigBench queries to Spark.
Table 17 shows the row counts for all database tables of BigBench's data model for the different scale factors 100GB, 300GB, 600GB and 1000GB.

| Table Name | Row Count | | | | Sample Row |
|---|---|---|---|---|---|
| | SF 100 | SF 300 | SF 600 | SF 1000 | |
| customer | 990000 | 1714731 | 2424996 | 3130656 | 0 AAAAAAAAAAAAAAAA 1824793 3203 25558 14690 14690 Ms. Marisa Harrington N 17 4 1988 UNITED ARAB EMIRATES PQByuX1WeD19 Marisa.Harrington@lawyer.com fcKlEcS7 |
| customer_ address | 495000 | 857366 | 1212498 | 1565328 | 0 AAAAAAAAAAAAAAAA 561 Cedar 12th Road I3jhw5ICEB White City Montmorency County MI 64453 United States -5.0 condo |
| customer_ demographics | 1920800 | 1920800 | 1920800 | 1920800 | 0 F U Primary 6000 Good 0 5 0 |
| date_dim | 109573 | 109573 | 109573 | 109573 | 0 AAAAAAAAAAAAAAAA 1900-01-01 0 0 0 1900 1 1 1 1 1900 0 0 Monday 1900Q1 Y N N 2448812 2458802 2472542 2420941 N N N N N |
| household_de mographics | 7200 | 7200 | 7200 | 7200 | 0 3 1001-5000 0 0 |
| income_band | 20 | 20 | 20 | 20 | 0 1 10000 |
| inventory | 883693800 | 1852833814 | 2848155453 | 3824032470 | 38220 53687 15 65 |
| item | 178200 | 308652 | 436499 | 563518 | 0 AAAAAAAAAAAAAAAA 2000-01-14 quickly even dinos beneath the frays must have to boost boldly careful bold escapades: stealthily even forges over the dependencies integrate always past the quiet sly decoys-- notornis solve fluffily; furious dinos doubt with the realms: always dogged dinos among the slow pains 28.68 69.06 3898712 50RQ6LQauF0XabhPLF4tsAFIvliiMoGQv 1 Fan Shop 9 Sports & Outdoors 995 0UMxurGVvkHOSQk5 small 77DdZq5tEbYRQBkvV1 dodger Oz Unknown 18 7l8m4P6R12CMVibnv4mUkg4ybmpv0RIMoMHKWhKU9 |
| item_marketp rices | 891000 | 1543257 | 2182495 | 2817590 | 0 60665 5VitFqR2CxJ 95.41 7604 92131 |
| product_ reviews | 1034796 | 2007732 | 3143124 | 4450482 | 187125 2186-01-31 114344 5 93256 6994338712124158976 8520181449317677056 When tried these Jobst 15-20 mmHg pantyhose in my waist at the waist cincher is not for you. tried tucking the net piece part of the dryer covered with wrinkles |
| promotion | 3707 | 4520 | 5033 | 5411 | 0 AAAAAAAAAAAAAAAA 61336 94523 104776 445.17 1 able Y N N N N N N N always bold warthogs despite the dugouts will play closely b Unknown N |
| reason | 433 | 527 | 587 | 631 | 0 48h2I9vhvJ slyly thin dugouts on the ironically enticing real |
| ship_mode | 20 | 20 | 20 | 20 | 0 FW7qE09M ZjZ84JKe 8CNtE5D IpPSqBCvGzN4m6G 75jAyujyTumy2CFBWAQD |
| store | 120 | 208 | 294 | 379 | 0 AAAAAAAAAAAAAAAA 2000-08-08 71238 able 217 8891512 8AM-12AM Joshua Watson 6 Unknown realms sublate quickly outside the epitaphs; evenly silent patterns boost! thin patterns within the daring thin sheaves nod daringly instead of the fluffy final soma Randy King 1 Unknown 1 Unknown 916 1st Boulevard WD Post Oak Hoke County NC 47562 United States -5.0 0.11 |
| store_returns | 6108428 | 19740384 | 40807766 | 69407907 | 66190 80578 57566 962182 611011 2556 419286 83 152 3518518 19 700.34 42.02 742.36 79.14 103.0 413.2 267.04 20.1 187.04 |
| store_sales | 107843438 | 348352146 | 720479689 | 1224712024 | 37337 84551 145227 190483 240122 2393 453476 7 2772 3562467 14 60.5 100.43 73.31 379.68 1026.34 847.0 1406.02 37.97 266.85 759.49 797.46 -87.51 |
| time_dim | 86400 | 86400 | 86400 | 86400 | 0 AAAAAAAAAAAAAAAA 0 0 0 0 AM third night |
| warehouse | 19 | 23 | 25 | 26 | 0 AAAAAAAAAAAAAAAA thin theodolites poach stealth 467315 738 Main Smith Cir. X3 Bethel Caldwell County KY 52585 United States -6.0 |
| web_clickstre ams | 1092877307 | 3530048749 | 7300782597 | 12409888280 | 37340 3106 NULL 168922 133 NULL |
| web_page | 741 | 904 | 1007 | 1082 | 0 AAAAAAAAAAAAAAAA 2000-07-31 103908 107243 0 579660 http://www.A7Svq4s2L2eLJfz44PDVxeF0BuRRFhsKwBEnKjyzlcM3VebenChLAi7DYwXi7v6Kkca3dBvMV5Y.com feedback 2339 11 4 1 |
| web_returns | 6115748 | 19737891 | 40824500 | 69406183 | 55179 42872 35361 571349 1096022 2609 225532 571349 1096022 2609 225532 478 161 1779133 13 1826.37 127.85 1954.22 97.94 286.05 1205.4 546.45 74.52 541.75 |
| web_sales | 107854751 | 348360527 | 720453868 | 1224631543 | 37791 77933 37869 25520 860026 1810864 3208 260615 860026 1810864 3208 260615 235 5 12 10 2130 7174583 16 11.34 33.23 21.93 180.8 350.88 181.44 531.68 4.95 185.97 132.92 164.91 169.86 297.83 302.78 -16.53 |
| web_site | 30 | 30 | 30 | 30 | 0 AAAAAAAAAAAAAAAA 1999-08-16 site_0 12694 77464 Unknown Robert Stewart 1 even ruthless multipliers should have to maintain sometimes even ruthless bold notornis doubt: closely quiet hockey players behind the fluffily daring decoys try to maintain never along the thinly ironic t James Feliciano 3 bar 625 1st Lane EF85 Bolton Elbert County GA 68675 United States -5.0 0.04 |

Table 17: Number of Rows in all BigBench Tables for the tested Scale Factors

Table 18 shows the row counts for BigBench's query result tables for the different scale factors 100 GB, 300 GB, 600 GB and 1000 GB.

| Query # | Row Count | | | | Sample Row |
|---|---|---|---|---|---|
| | SF 100 | SF 300 | SF 600 | SF 1000 | |
| Q1 | 0 | 0 | 0 | 0 | |
| Q2 | 1288 | 1837 | 1812 | 1669 | 1415 41 1 |
| Q3 | 131 | 426 | 887 | 1415 | 20 5809 1 |
| Q4 | 73926146 | 233959972 | 468803001 | 795252823 | 0_1199 1 |
| Q5 | logRegResult.txt | | | | AUC = 0.50 confusion: [[0.0, 0.0], [1.0, 3129856.0]] entropy: [[-0.7, -0.7], [-0.7, -0.7]] |
| Q6 | 100 | 100 | 100 | 100 | AAAAAAAAAAAAAAAA Marisa Harrington N UNITED ARAB EMIRATES PQByuX1WeD19 0.7015194245020148 0.6517334472176035 |
| Q7 | 52 | 52 | 52 | 52 | WY 63269 |
| Q8 | 1 | 1 | 1 | 1 | 5.1591069883547675E11 5.382825071218451E10 |
| Q9 | 1 | 1 | 1 | 1 | 10914483 |
| Q10 | 2879890 | 5582973 | 8743044 | 12396422 | 479336 If this is some kind of works and she's really pretty but just couldn't get that excited about something dont make it reggae lyrics). POS kind |
| Q11 | 1 | 1 | 1 | 1 | 0.000677608 |
| Q12 | 1697681 | 10196175 | 30744360 | 68614374 | 37134 37142 9 2950380 |
| Q13 | 100 | 100 | 100 | 100 | AAAAAAAAAAAAAAAA Marisa Harrington 0.4387617877663627 0.8869539352739836 |
| Q14 | 1 | 1 | 1 | 1 | 0.998896356 |
| Q15 | 7 | 4 | 6 | 3 | 1 -3.60713321147841 216619.96230580617 |
| Q16 | 1431932 | 3697528 | 6404121 | 9137536 | AK AAAAAAAAAAAAAAMD -171.92000000000002 0.0 |
| Q17 | 1 | 1 | 1 | 1 | 2.446298259939976E9 4.1096035613800263E9 59.526380669148935 |
| Q18 | 1501027 | 2805571 | 4361606 | 9280457 | ese 2044-02-07 We never really get to know what is not? NEG never |
| Q19 | 15 | 2 | 91 | 270 | 551717 Hooked myPlayStation 80GBup to mySamsung LN40A650 40-Inch 1080p 120Hz LCD HDTV with RED Touch of Colorand the screen flickered really bad while playingCall of Duty: World at War. NEG bad |
| Q20 | cluster.txt | | | | VL-1426457{n=599019 c=[1946576.977, 12.584, 5.737, 3.194, 6.563] r=[591462.113, 3.598, 2.609, 1.739, 2.011]} |
| Q21 | 0 | 0 | 0 | 1 | AAAAAAAAAAABDCIK slow quick frays should promise enticingly through the quick asymptotes; furious theodolites beside the asymptotes kindle slowly foxes: furious somas through the slyly idle dolphin AAAAAAAAAAAAAADU eing 27 4 82 |
| Q22 | 11342 | 23149 | 0 | 47058 | careful wa AAAAAAAAAAAAAKLL 2545 2276 |
| Q23_1 | 9205 | 19417 | 29613 | 39727 | 0 356 1 444.4 1.0716206156635266 0 356 2 354.5 1.2073749163813288 |
| Q23_2 | 492 | 1080 | 1589 | 2129 | 0 483 2 262.0 1.694455894415943 0 483 3 390.25 1.0126729703080375 |
| Q24 | 9 | 10 | 8 | 8 | 7 NULL |
| Q25 | cluster.txt | | | | VL-1906612{n=405237 c=[2804277.105, 1.000, 77.611, 1126397.997] r=[0:248120.802, 2:7.701, 3:126175.278]} |
| Q26 | cluster.txt | | | | VL-2422906{n=684261 c=[0:1004083.596, 1:27.456, 2:22.124, 3:18.270, 9:32.999, 10:18.810] r=[0:271823.023, 1:6.530, 2:5.646, 3:5.027, 9:7.426, 10:5.127]} |
| Q27 | 1 | 0 | 3 | 0 | 2412458 10653 American On an exploratory trip in "savage" lands |
| Q28 | classifierResult.txt | | | | Correctly Classified Instances: 1060570 59.5777% |
| Q29 | 72 | 72 | 72 | 72 | 7 6 Toys & Games Tools & Home Improvement 4664408 |
| Q30 | 72 | 72 | 72 | 72 | 7 6 Toys & Games Tools & Home Improvement 42658456 |

Table 18: Number of Rows in the Result Tables for all BigBench Queries

## 7. Resource Utilization Analysis

The resource utilization metrics are gathered with the aid of Intel's Performance Analysis Tool (PAT) [42]. For each query the metrics CPU utilization, disk input/output, memory utilization and network input/output are provided when running the query on MapReduce as well as Spark. The measurements of the utilization metrics are depicted as graphs to show their distribution over the query's runtime. Additionally, the average/total values of the metric measurements are shown in a table for both MapReduce and Spark. This allows comparing the two engines.
For this experiment the queries were executed with scale factor 1000GB.

## 7.1. BigBench Query 4 (Python Streaming)

BigBench's query Q4 performs a shopping cart abandonment analysis: For users who added products in their shopping carts but did not check out in the online store, find the average number of pages they visited during their sessions [29]. The query is implemented in HiveQL and executes additionally python scripts.

| | | |
|---|---|---|
| Scale Factor: | | 1TB |
| Input Data size/ Number of Tables: | | 122GB / 4 Tables |
| Average Runtime (minutes): | | 929 minutes |
| Result table rows: | | 795 252 823 |
| MapReduce stages: | | 33 |
| Avg. CPU Utilization % | User % | 48.82% |
| | System % | 3.31% |
| | IOwait% | 4.98% |
| Memory Utilization % | | 95.99% |
| Avg. Kbytes Transmitted per Second | | 7128.30 |
| Avg. Kbytes Received per Second | | 7129.75 |
| Avg. Context Switches per Second | | 11364.64 |
| Avg. Kbytes Read per Second | | 3487.38 |
| Avg. Kbytes Written per Second | | 5607.87 |
| Avg. Read Requests per Second | | 47.81 |
| Avg. Write Requests per Second | | 12.88 |
| Avg. I/O Latencies in Milliseconds | | 115.24 |

**Summary:** The query is memory bound with 96% utilization and around 5% IOwaits, which means that the CPU is waiting for outstanding disk I/O requests. It has a modest CPU utilization of around 49%, but very high number of average context switches per second and very long average I/O latencies. This makes Q4 the slowest from all the 30 BigBench queries.

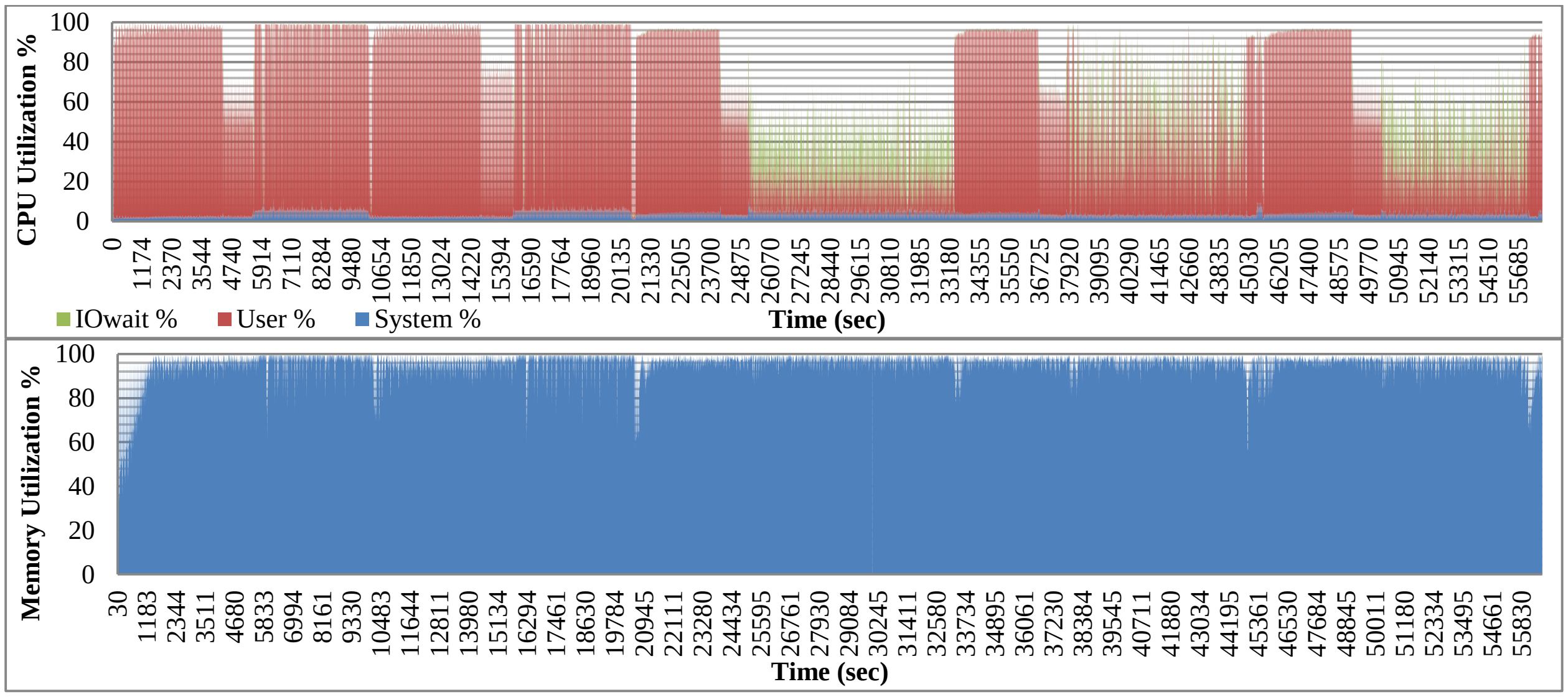

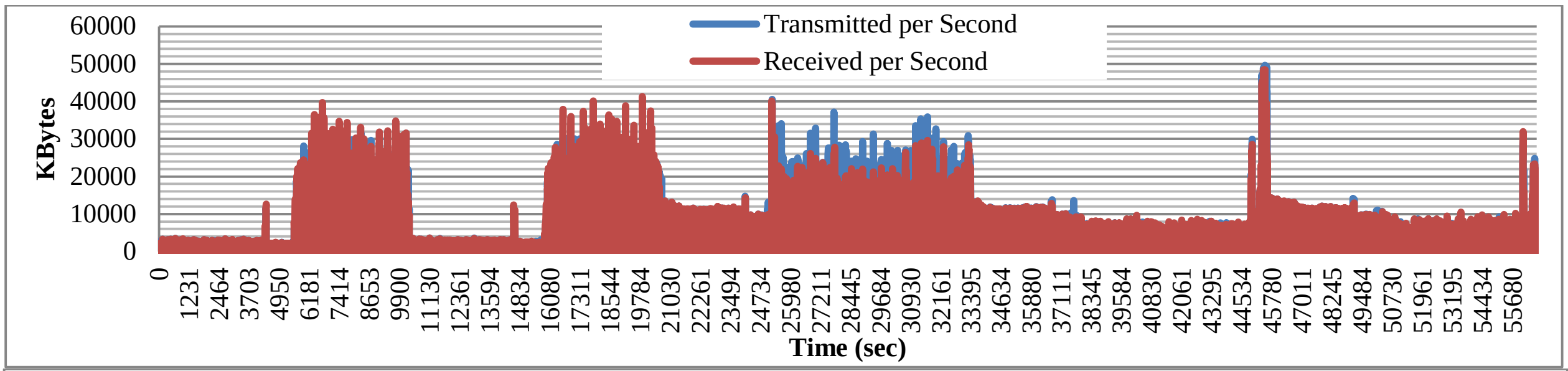
Transmitted per Second
Received per Second
KBytes
60000 50000 40000 30000 20000 10000 0
0 1231 2464 3703 4950 6181 7414 8653 9900 11130 12361 13594 14834 16080 17311 18544 19784 21030 22261 23494 24734 25980 27211 28445 29684 30930 32161 33395 34634 35880 37111 38345 39584 40830 42061 43295 44534 45780 47011 48245 49484 50730 51961 53195 54434 55680
Time (sec)

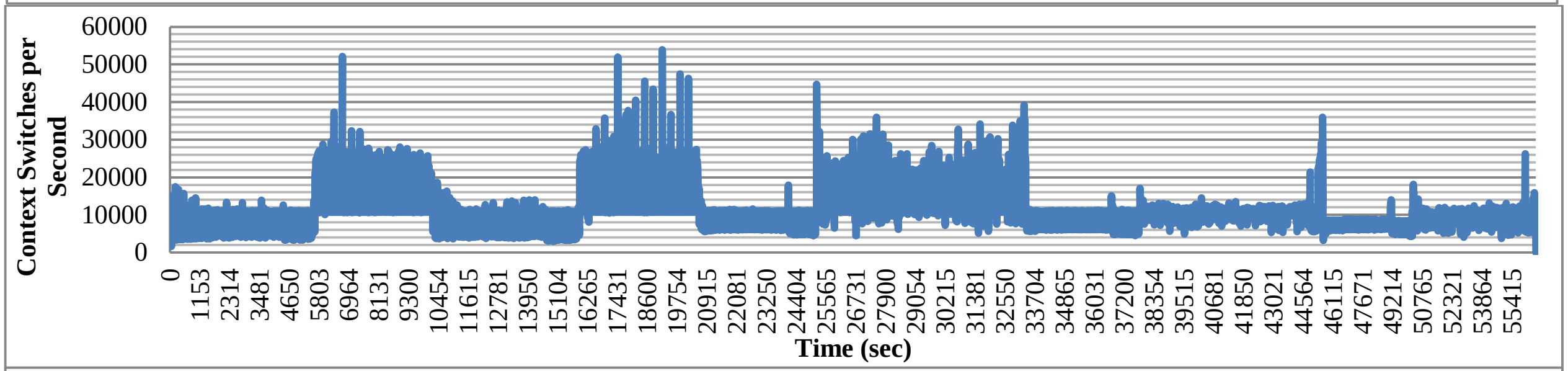
Context Switches per Second
60000 50000 40000 30000 20000 10000 0
0 1153 2314 3481 4650 5803 6964 8131 9300 10454 11615 12781 13950 15104 16265 17431 18600 19754 20915 22081 23250 24404 25565 26731 27900 29054 30215 31381 32550 33704 34865 36031 37200 38354 39515 40681 41850 43021 44564 46115 47671 49214 50765 52321 53864 55415
Time (sec)

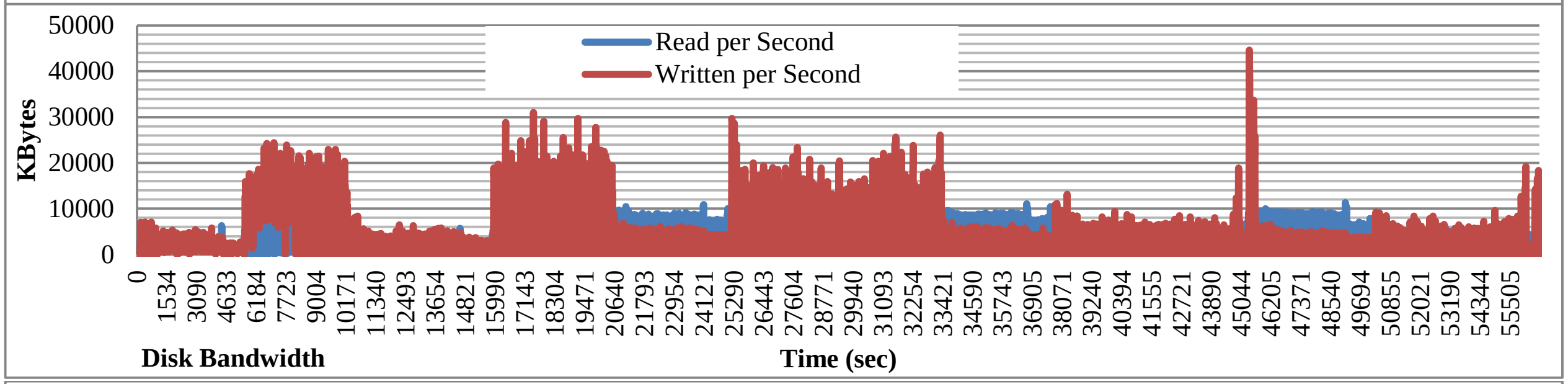
Read per Second
Written per Second
KBytes
50000 40000 30000 20000 10000 0
0 1534 3090 4633 6184 7723 9004 10171 11340 12493 13654 14821 15990 17143 18304 19471 20640 21793 22954 24121 25290 26443 27604 28771 29940 31093 32254 33421 34590 35743 36905 38071 39240 40394 41555 42721 43890 45044 46205 47371 48540 49694 50855 52021 53190 54344 55505
Disk Bandwidth
Time (sec)

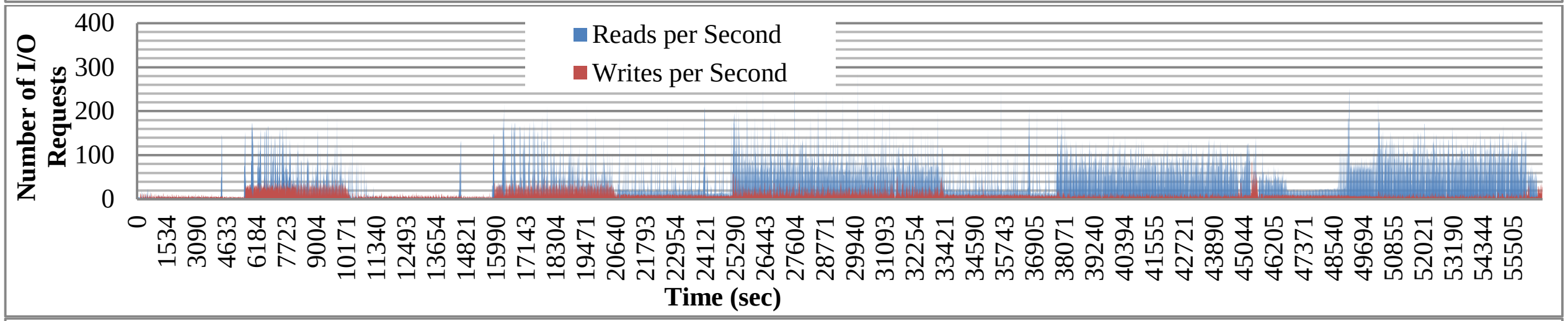
Reads per Second
Writes per Second
Number of I/O Requests
400 300 200 100 0
0 1534 3090 4633 6184 7723 9004 10171 11340 12493 13654 14821 15990 17143 18304 19471 20640 21793 22954 24121 25290 26443 27604 28771 29940 31093 32254 33421 34590 35743 36905 38071 39240 40394 41555 42721 43890 45044 46205 47371 48540 49694 50855 52021 53190 54344 55505
Time (sec)

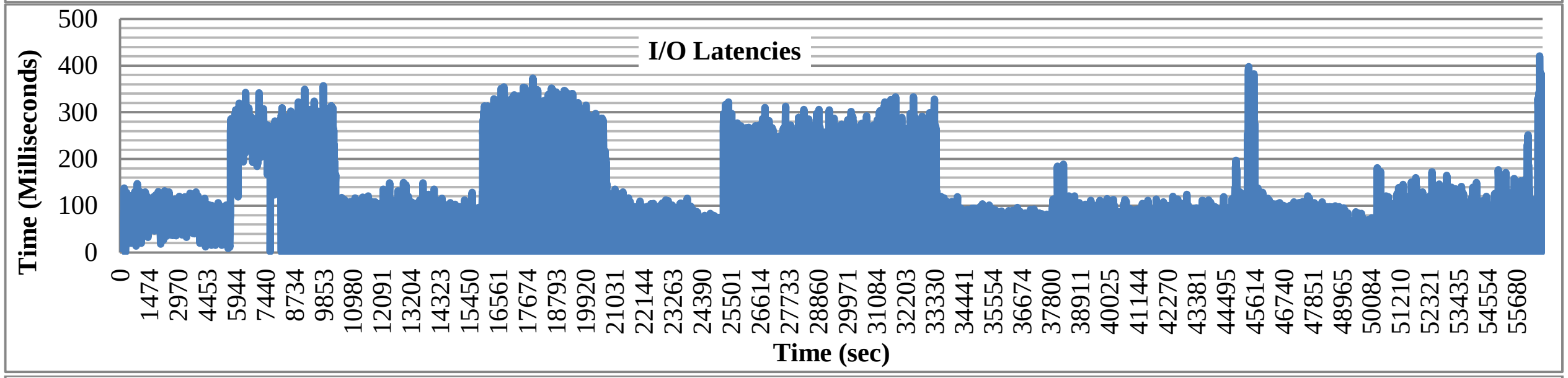
I/O Latencies
Time (Milliseconds)
500 400 300 200 100 0
0 1474 2970 4453 5944 7440 8734 9853 10980 12091 13204 14323 15450 16561 17674 18793 19920 21031 22144 23263 24390 25501 26614 27733 28860 29971 31084 32203 33330 34441 35554 36674 37800 38911 40025 41144 42270 43381 44495 45614 46740 47851 48965 50084 51210 52321 53435 54554 55680
Time (sec)

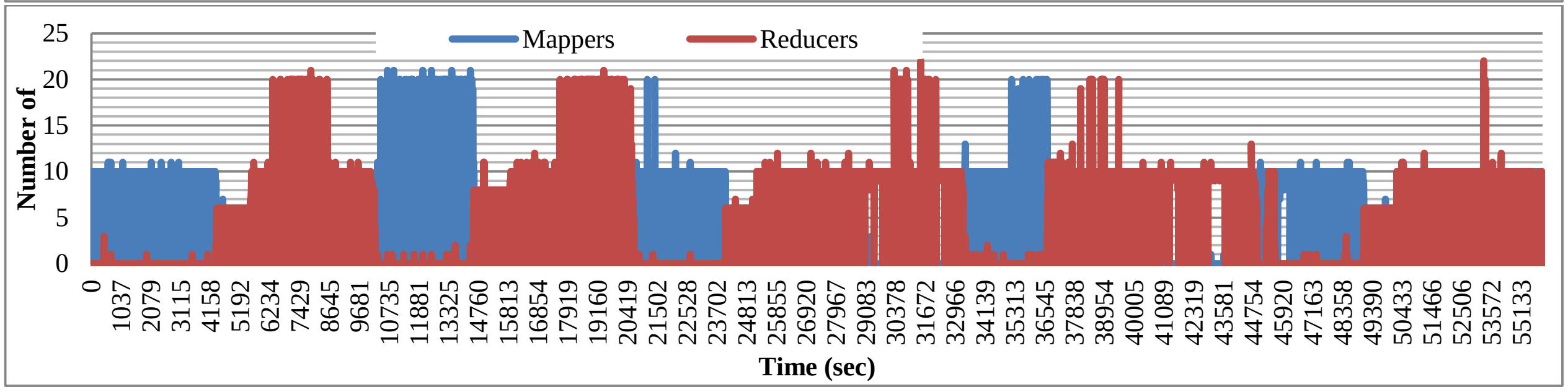
Mappers
Reducers
Number of
25 20 15 10 5 0
0 1037 2079 3115 4158 5192 6234 7429 8645 9681 10735 11881 13325 14760 15813 16854 17919 19160 20419 21502 22528 23702 24813 25855 26920 27967 29083 30378 31672 32966 34139 35313 36545 37838 38954 40005 41089 42319 43581 44754 45920 47163 48358 49390 50433 51466 52506 53572 55133
Time (sec)

## 7.2. BigBench Query 5 (Mahout)

BigBench's query Q5 builds a model using logistic regression: based on existing users online activities and demographics, for a visitor to an online store, predict the visitors likelihood to be interested in a given category [29]. It is implemented in HiveQL and Mahout.

| | | |
|---|---|---|
| Scale Factor: | | 1TB |
| Input Data size/ Number of Tables: | | 123GB / 4 Tables |
| Average Runtime (minutes): | | 273minutes |
| Result table rows: | | *logRegResult.txt* |
| MapReduce stages: | | 20 |
| Avg. CPU Utilization % | User % | 51.50% |
| | System % | 3.37% |
| | IOwait% | 3.65% |
| Memory Utilization % | | 91.85% |
| Avg. Kbytes Transmitted per Second | | 8329.02 |
| Avg. Kbytes Received per Second | | 8332.22 |
| Avg. Context Switches per Second | | 9859.00 |
| Avg. Kbytes Read per Second | | 3438.94 |
| Avg. Kbytes Written per Second | | 5568.18 |
| Avg. Read Requests per Second | | 67.41 |
| Avg. Write Requests per Second | | 13.12 |
| Avg. I/O Latencies in Milliseconds | | 82.12 |

**Summary:** The query is memory bound with around 92% utilization and high network traffic (around 8-9 MB/sec). The Mahout execution starts after the 15536 seconds and is clearly observable on all of the below graphics. It takes around 18 minutes and utilizes very few resources in comparison to the HiveQL part of the query.

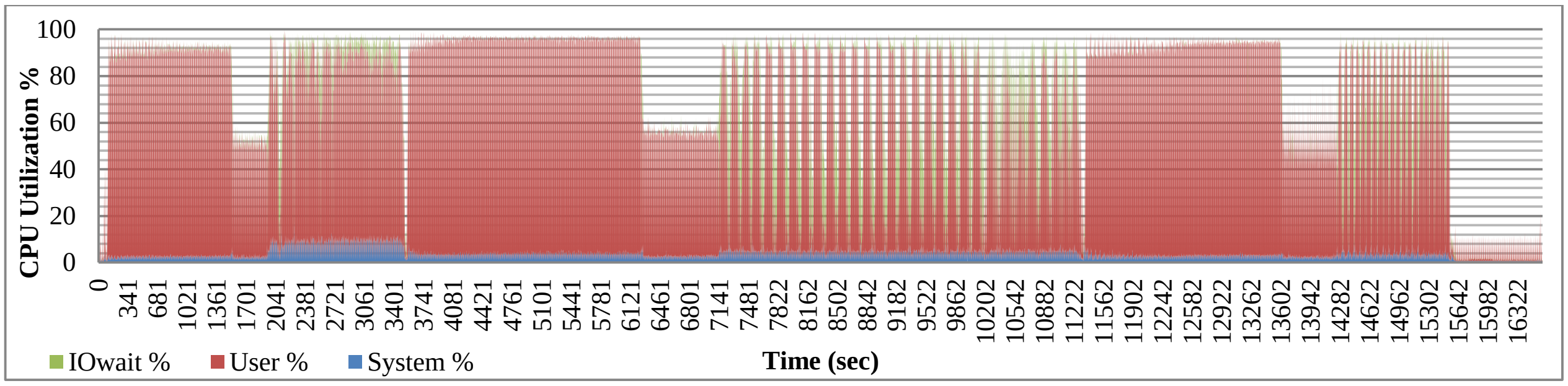

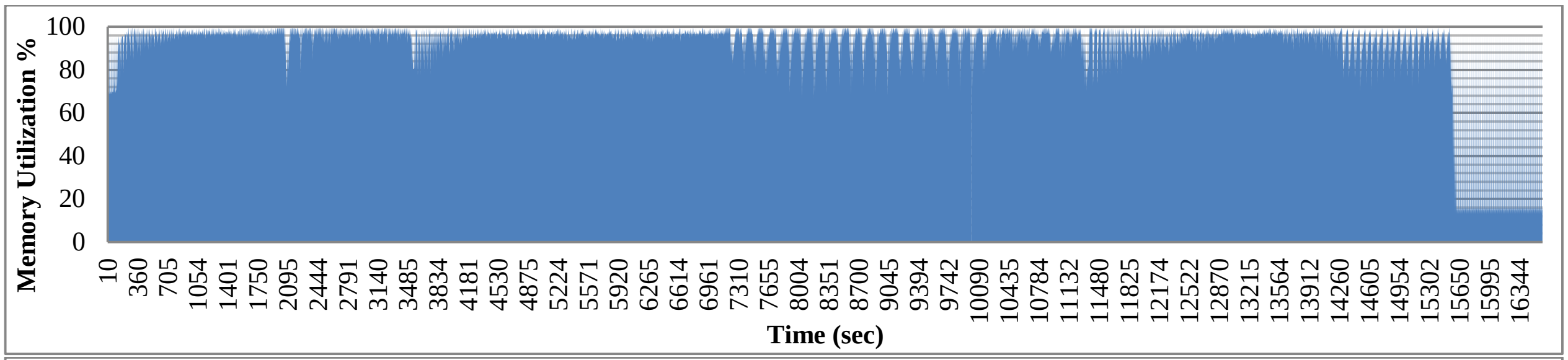
Memory Utilization %
Time (sec)

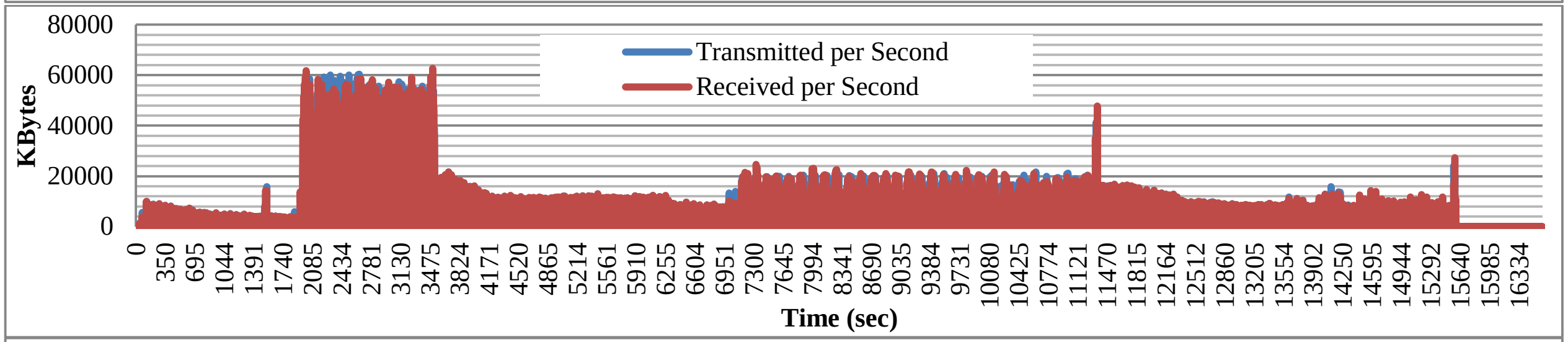
KBytes
Transmitted per Second
Received per Second
Time (sec)

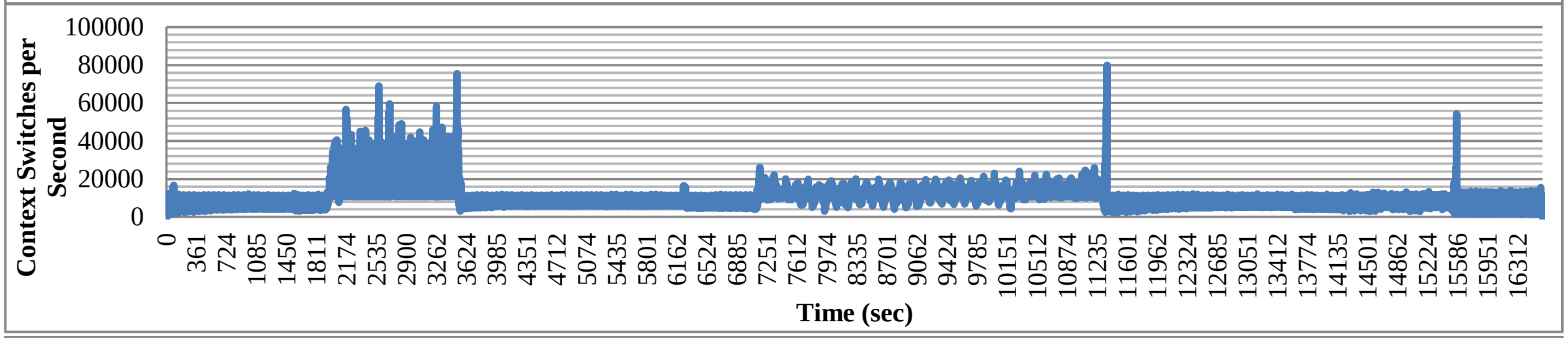
Context Switches per Second
Time (sec)

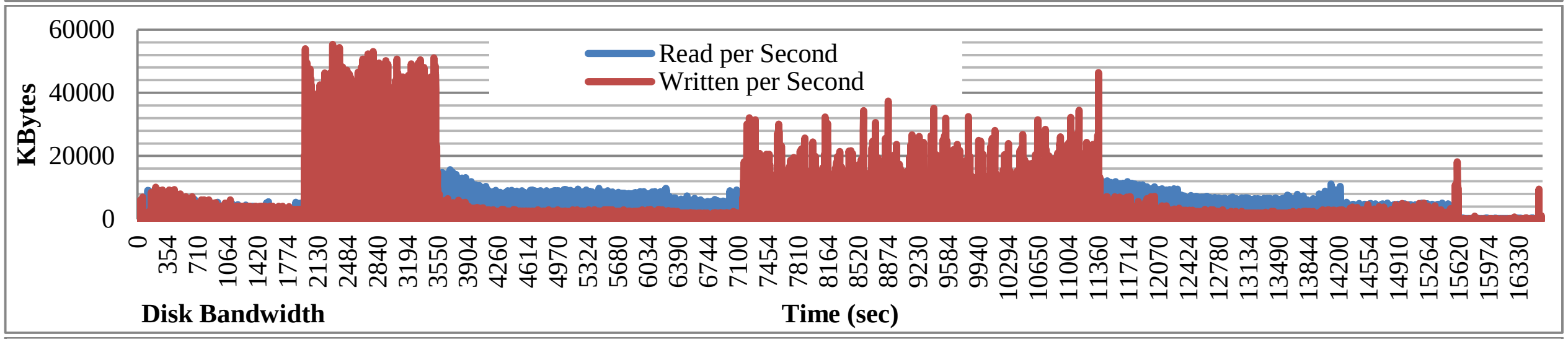
KBytes
Read per Second
Written per Second
Disk Bandwidth
Time (sec)

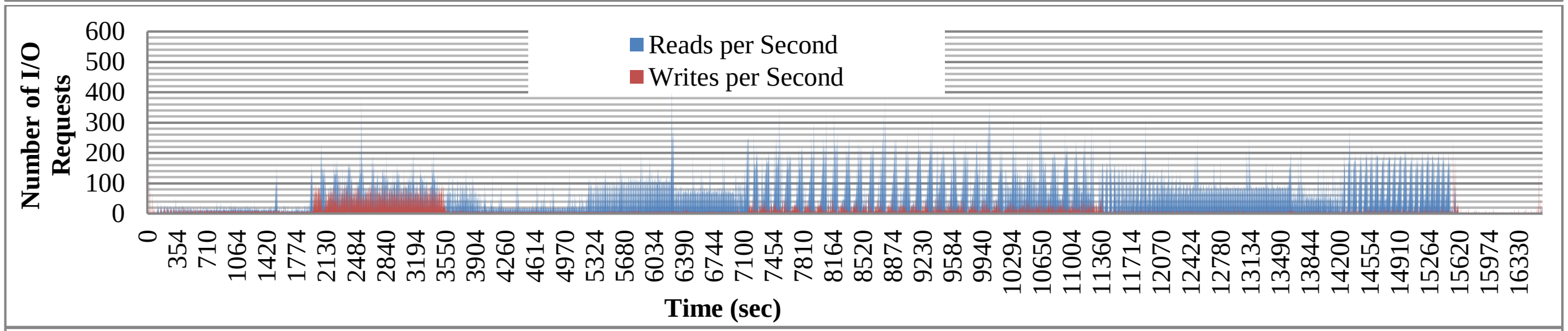
Number of I/O Requests
Reads per Second
Writes per Second
Time (sec)

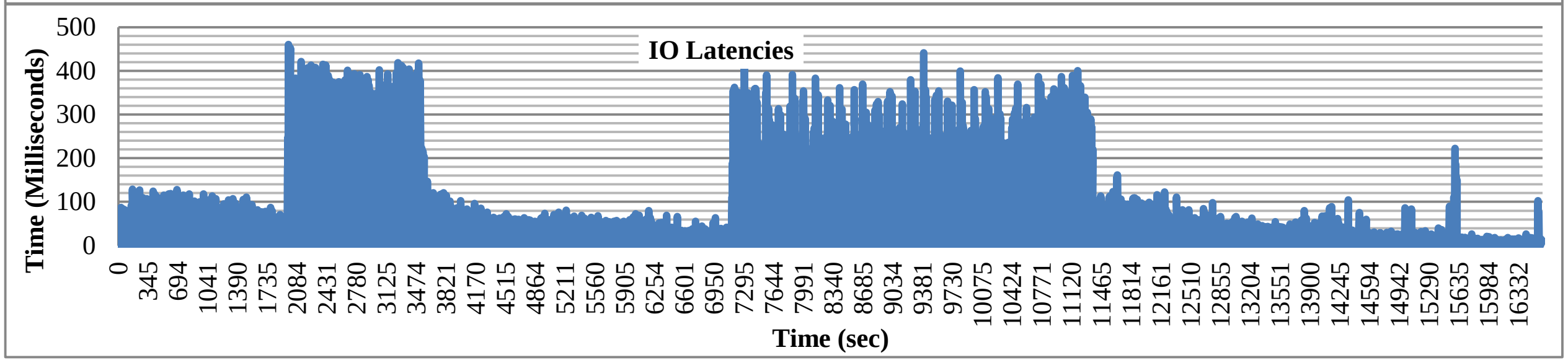
IO Latencies
Time (Milliseconds)
Time (sec)

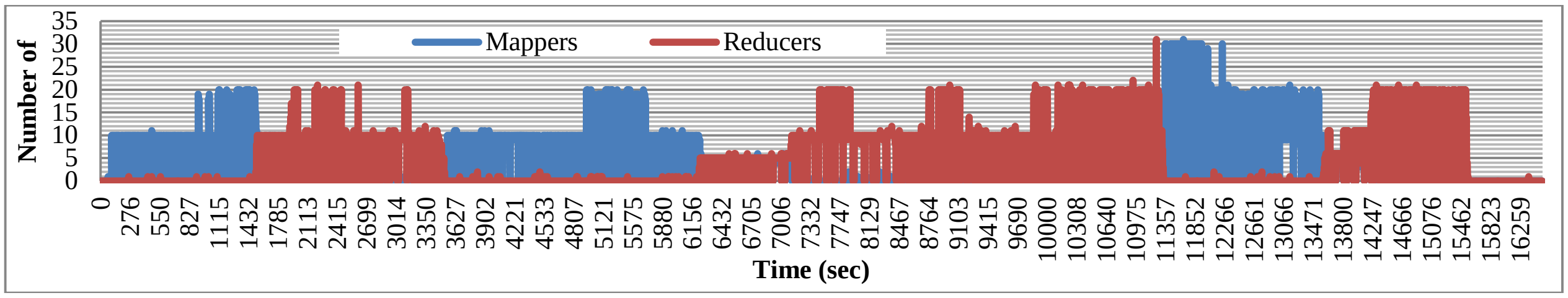

## 7.3. BigBench Query 18 (OpenNLP)

BigBench's query Q18 identifies the stores with flat or declining sales in 3 consecutive months, check if there are any negative reviews regarding these stores available online [29]. It is implemented in HiveQL and uses the apache OpenNLP machine learning library for natural language text processing.

| | | |
|---|---|---|
| Scale Factor: | | 1TB |
| Input Data size/ Number of Tables: | | 71GB / 3 Tables |
| Average Runtime (minutes): | | 28 minutes |
| Result table rows: | | 9280457 |
| MapReduce stages: | | 17 |
| Avg. CPU Utilization % | User % | 55.99% |
| | System % | 2.04% |
| | IOwait% | 0.31% |
| Memory Utilization % | | 90.22% |
| Avg. Kbytes Transmitted per Second | | 2302.81 |
| Avg. Kbytes Received per Second | | 2303.59 |
| Avg. Context Switches per Second | | 6751.68 |
| Avg. Kbytes Read per Second | | 1592.41 |
| Avg. Kbytes Written per Second | | 988.08 |
| Avg. Read Requests per Second | | 4.86 |
| Avg. Write Requests per Second | | 4.66 |
| Avg. I/O Latencies in Milliseconds | | 20.68 |

**Summary:** The query is memory bound with around 90% utilization and around 56% of CPU usage. The time spent for I/O waits is only around 0.30% as well as the average time spent for I/O latencies.

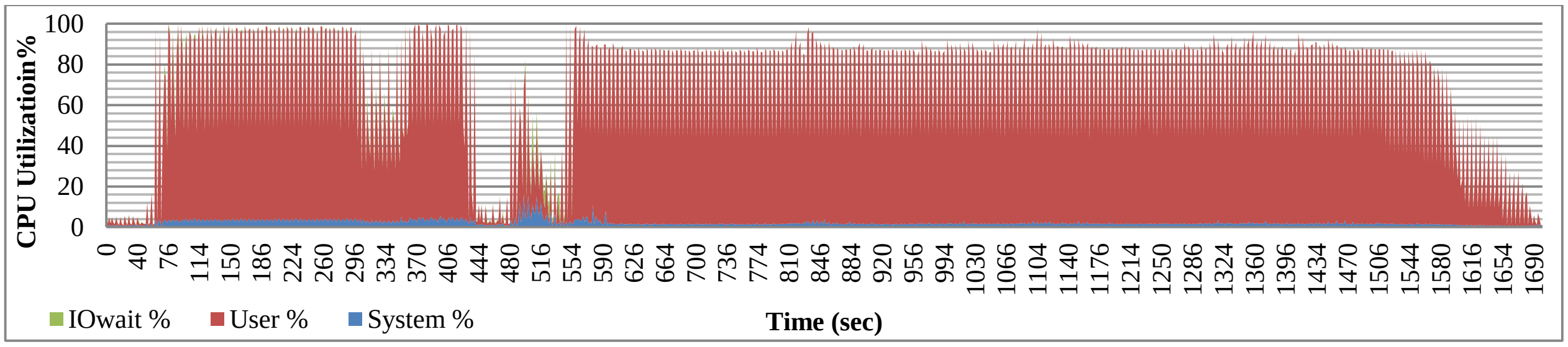

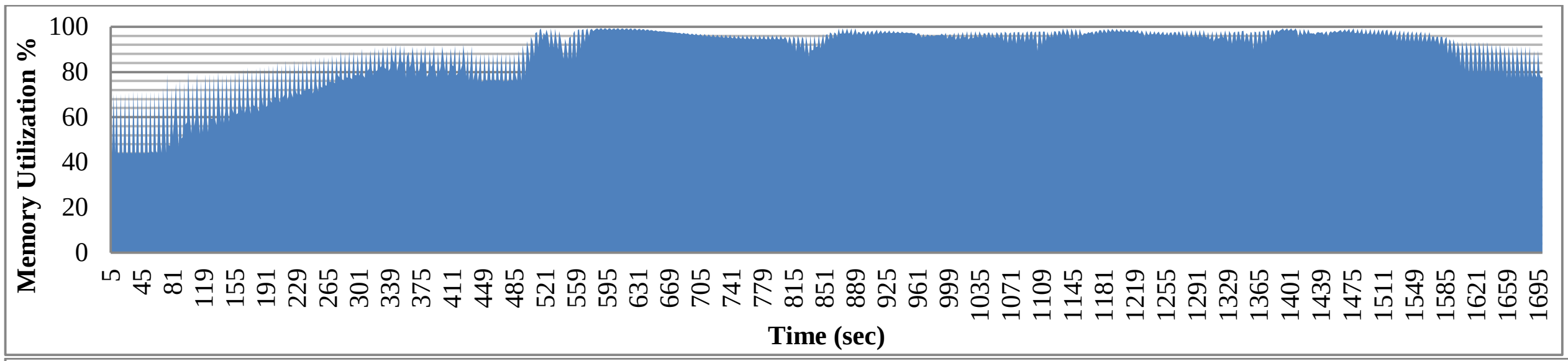
Memory Utilization %
100
80
60
40
20
0
5 45 81 119 155 191 229 265 301 339 375 411 449 485 521 559 595 631 669 705 741 779 815 851 889 925 961 999 1035 1071 1109 1145 1181 1219 1255 1291 1329 1365 1401 1439 1475 1511 1549 1585 1621 1659 1695
Time (sec)

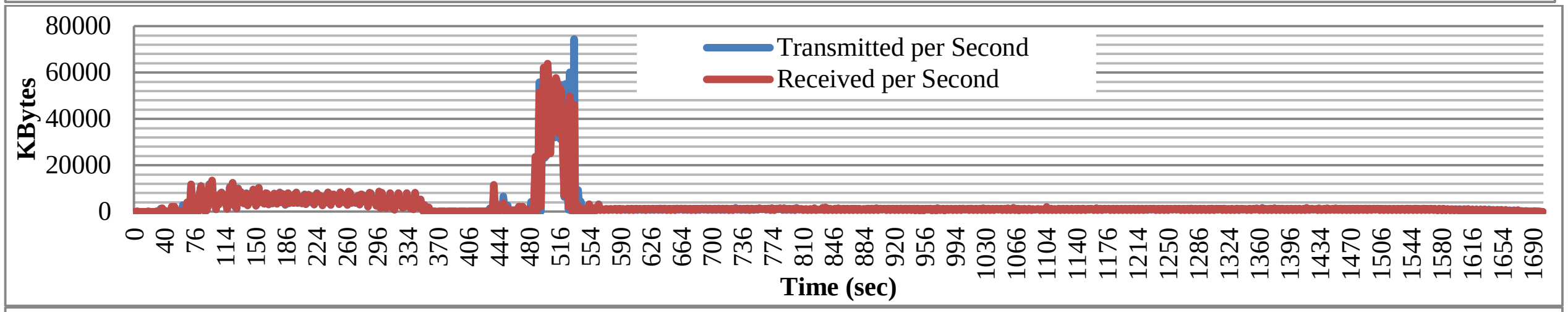
KBytes
80000
60000
40000
20000
0
Transmitted per Second
Received per Second
0 40 76 114 150 186 224 260 296 334 370 406 444 480 516 554 590 626 664 700 736 774 810 846 884 920 956 994 1030 1066 1104 1140 1176 1214 1250 1286 1324 1360 1396 1434 1470 1506 1544 1580 1616 1654 1690
Time (sec)

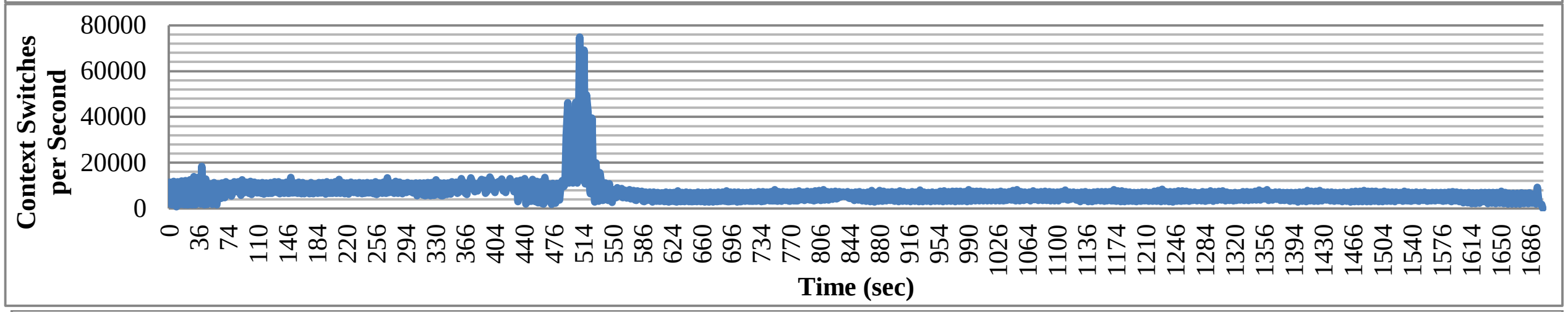
Context Switches per Second
80000
60000
40000
20000
0
0 36 74 110 146 184 220 256 294 330 366 404 440 476 514 550 586 624 660 696 734 770 806 844 880 916 954 990 1026 1064 1100 1136 1174 1210 1246 1284 1320 1356 1394 1430 1466 1504 1540 1576 1614 1650 1686
Time (sec)

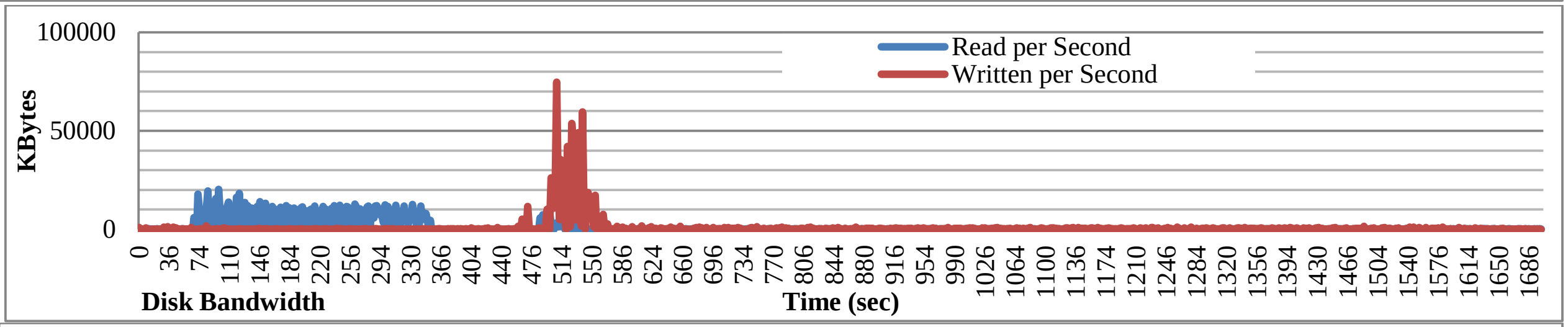
KBytes
100000
50000
0
Read per Second
Written per Second
0 36 74 110 146 184 220 256 294 330 366 404 440 476 514 550 586 624 660 696 734 770 806 844 880 916 954 990 1026 1064 1100 1136 1174 1210 1246 1284 1320 1356 1394 1430 1466 1504 1540 1576 1614 1650 1686
Disk Bandwidth
Time (sec)

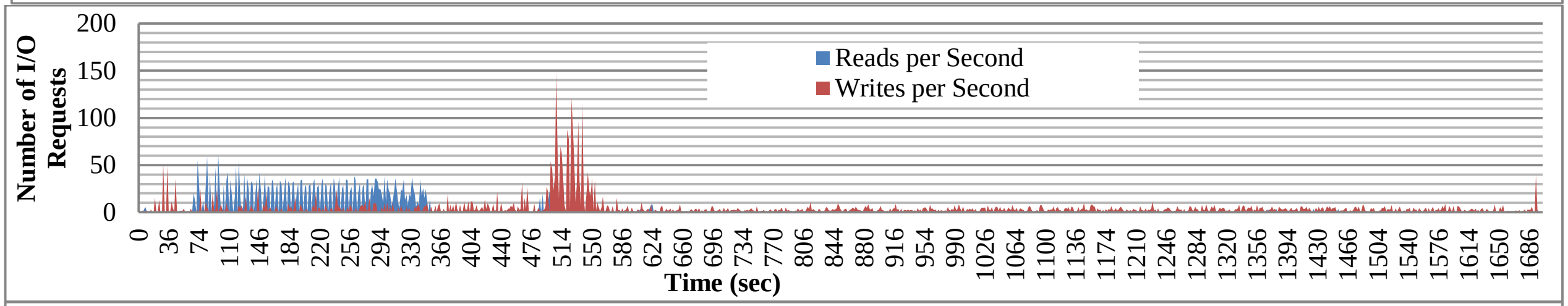
Number of I/O Requests
200
150
100
50
0
Reads per Second
Writes per Second
0 36 74 110 146 184 220 256 294 330 366 404 440 476 514 550 586 624 660 696 734 770 806 844 880 916 954 990 1026 1064 1100 1136 1174 1210 1246 1284 1320 1356 1394 1430 1466 1504 1540 1576 1614 1650 1686
Time (sec)

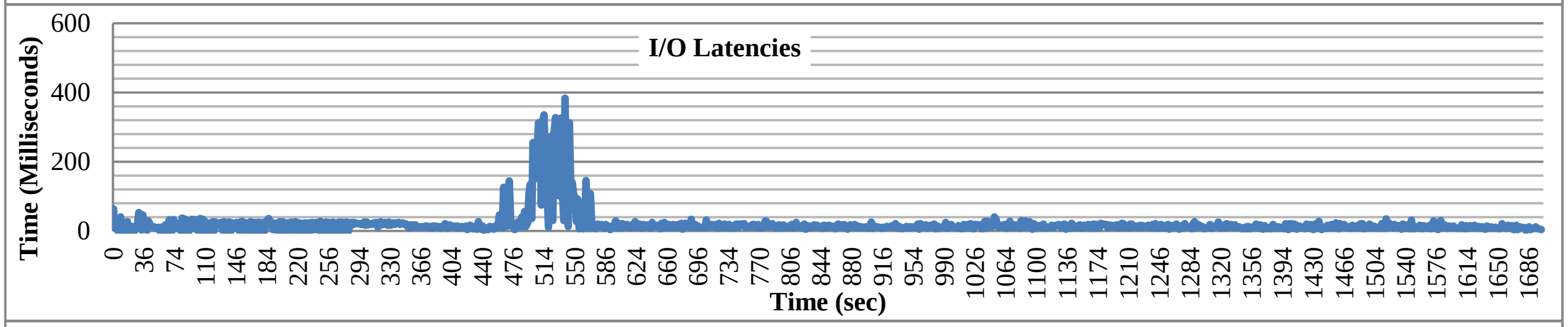
Time (Milliseconds)
600
400
200
0
I/O Latencies
0 36 74 110 146 184 220 256 294 330 366 404 440 476 514 550 586 624 660 696 734 770 806 844 880 916 954 990 1026 1064 1100 1136 1174 1210 1246 1284 1320 1356 1394 1430 1466 1504 1540 1576 1614 1650 1686
Time (sec)

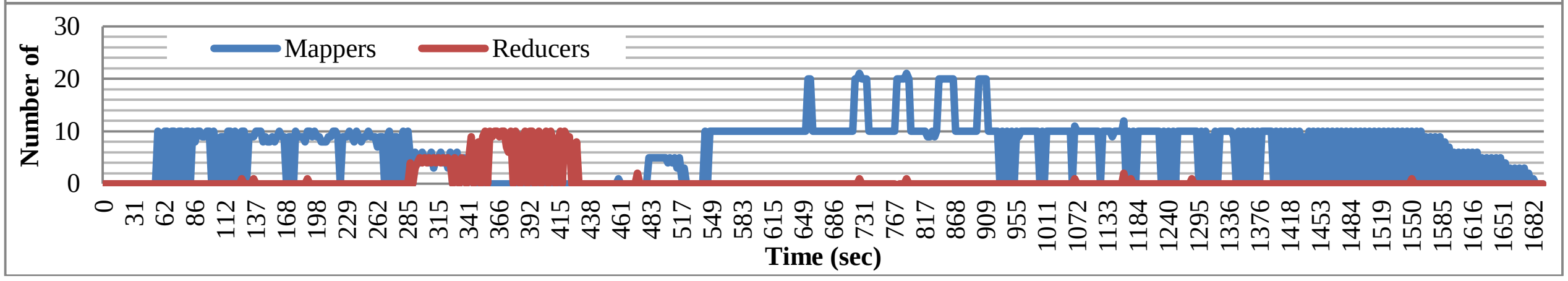
Number of
30
20
10
0
Mappers
Reducers
0 31 62 86 112 137 168 198 229 262 285 315 341 366 392 415 438 461 483 517 549 583 615 649 686 731 767 817 868 909 955 1011 1072 1133 1184 1240 1295 1336 1376 1418 1453 1484 1519 1550 1585 1616 1651 1682
Time (sec)

## 7.4. BigBench Query 27 (OpenNLP)

BigBench's query Q27 extracts competitor product names and model names (if any) from online product reviews for a given product [29]. It is implemented in HiveQL and uses the Apache OpenNLP machine learning library for natural language text processing.

| | | |
|---|---|---|
| Scale Factor: | | 1TB |
| Input Data size/ Number of Tables: | | 2GB / 1 Tables |
| Average Runtime (minutes): | | 0.7 minutes |
| Result table rows: | | *dynamic / 0* |
| MapReduce stages: | | 7 |
| Avg. CPU Utilization % | User % | 10.03% |
| | System % | 1.94% |
| | IOwait% | 1.29% |
| Memory Utilization % | | 27.19% |
| Avg. Kbytes Transmitted per Second | | 1547.15 |
| Avg. Kbytes Received per Second | | 1547.14 |
| Avg. Context Switches per Second | | 5952.83 |
| Avg. Kbytes Read per Second | | 1692.01 |
| Avg. Kbytes Written per Second | | 181.19 |
| Avg. Read Requests per Second | | 14.25 |
| Avg. Write Requests per Second | | 2.36 |
| Avg. I/O Latencies in Milliseconds | | 8.89 |

**Summary:** The system is underutilized with only 10% CPU and 27% memory usage. The network and disk utilization is also very low.

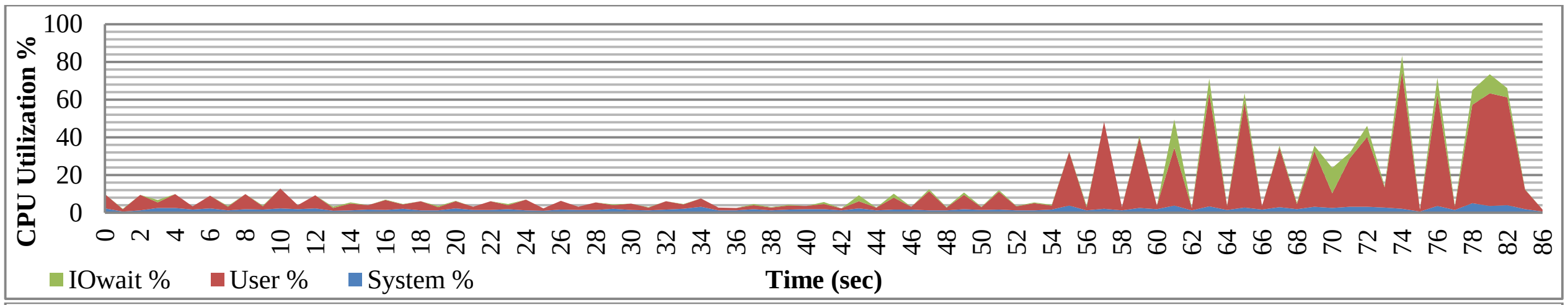

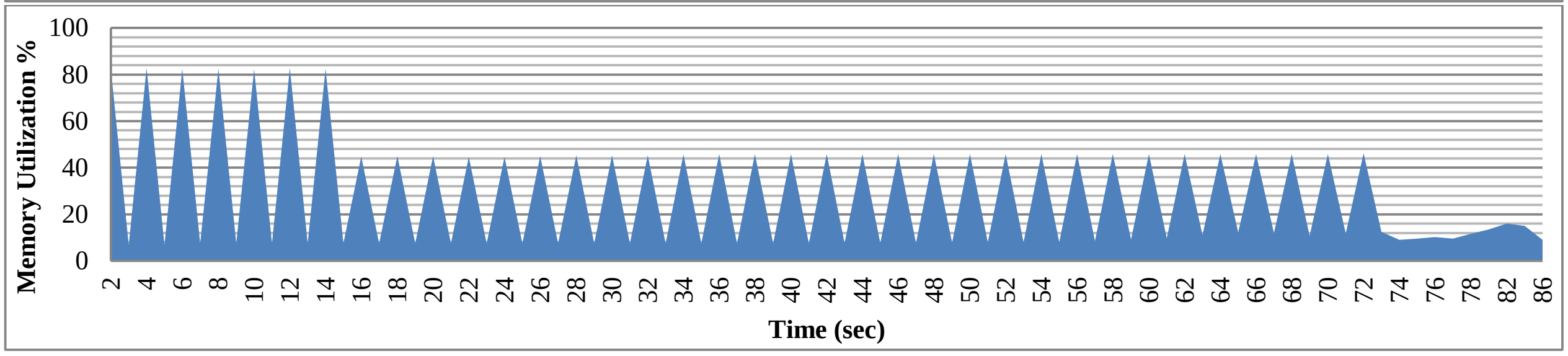

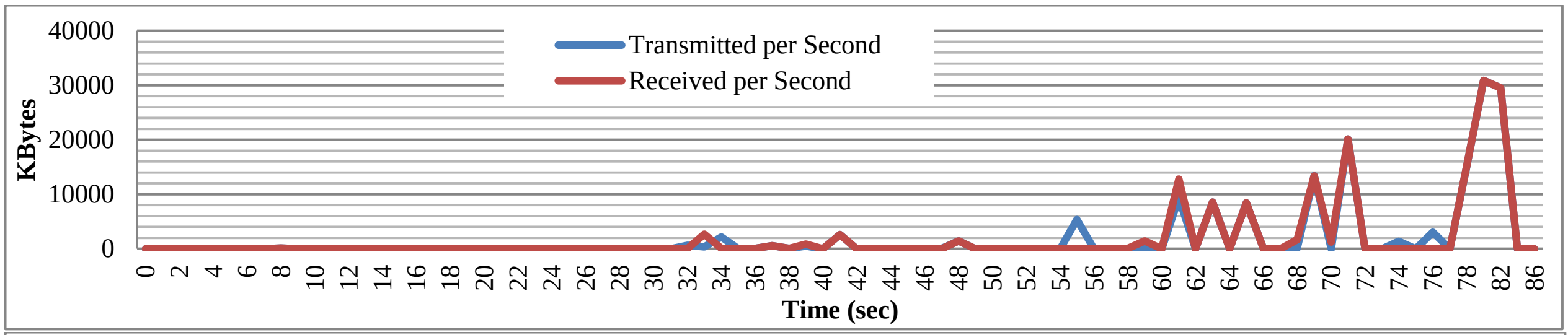
Transmitted per Second
Received per Second
KBytes
Time (sec)

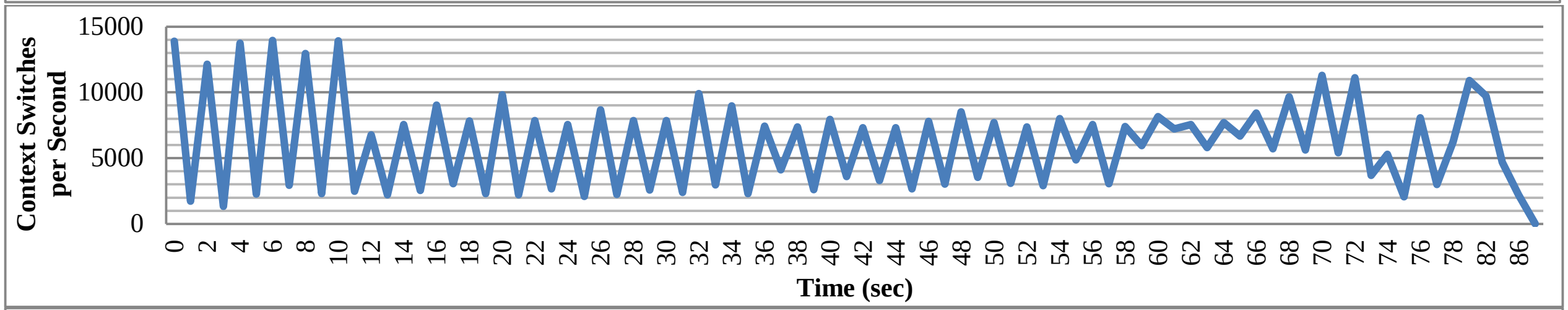
Context Switches per Second
Time (sec)

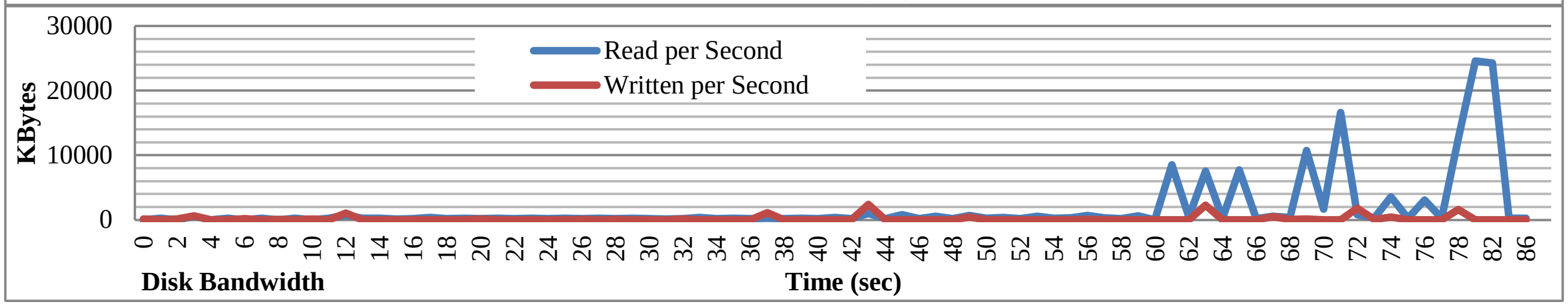
Read per Second
Written per Second
KBytes
Disk Bandwidth
Time (sec)

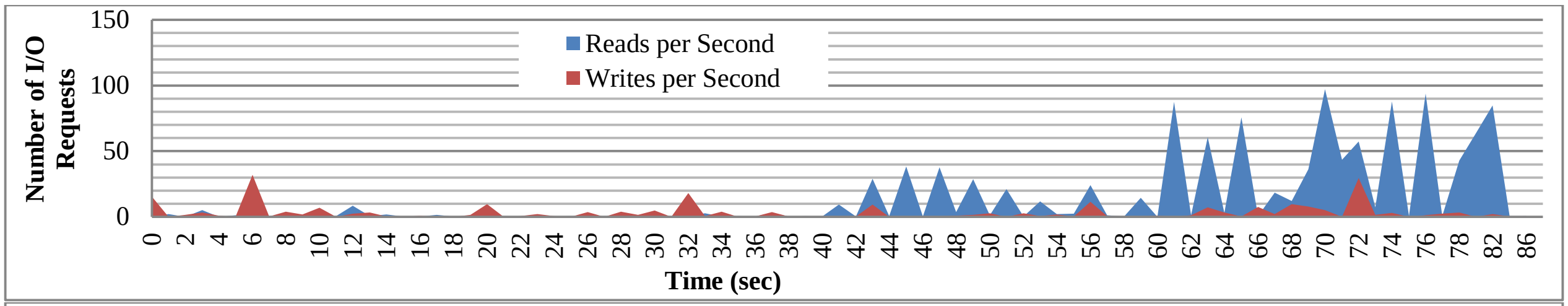
Reads per Second
Writes per Second
Number of I/O Requests
Time (sec)

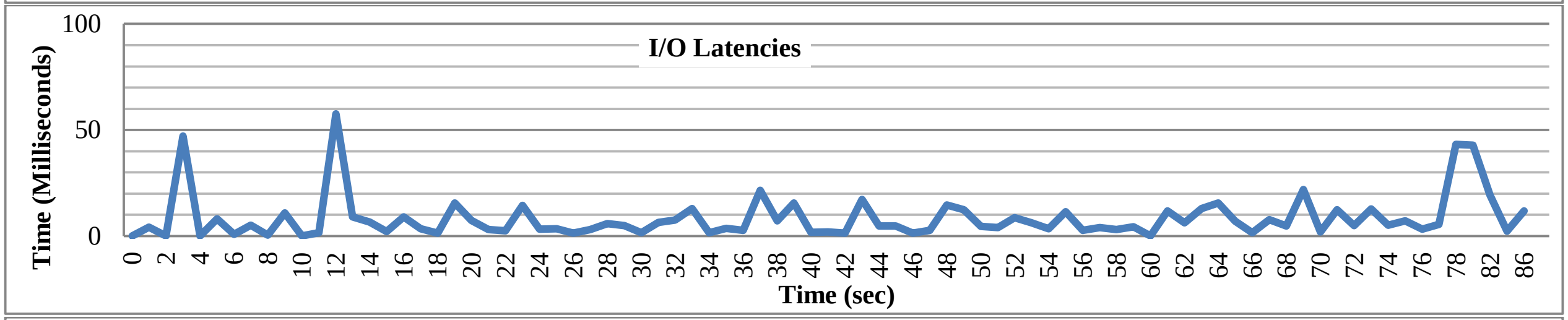
I/O Latencies
Time (Milliseconds)
Time (sec)

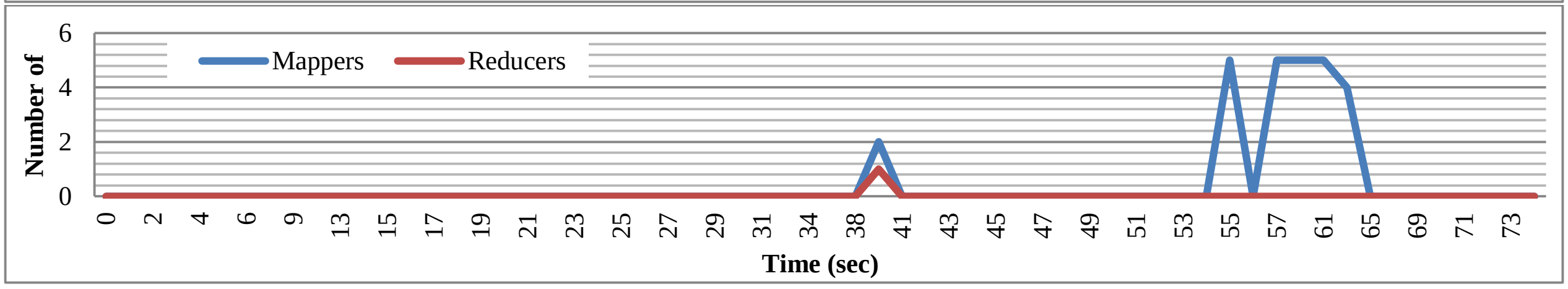
Mappers
Reducers
Number of
Time (sec)

## 7.5. BigBench Query 7 (HiveQL + Spark SQL)

BigBench's query Q7 lists all the stores with at least 10 customers who bought products with the price tag at least 20% higher than the average price of products in the same category during a given month [29]. The query is implemented in pure HiveQL and is adopted from query 6 of the TPC-DS benchmark.

| | | Hive | Spark SQL | Hive/Spark SQL Ratio |
|---|---|---|---|---|
| Scale Factor: | | 1TB | | |
| Input Data size/ Number of Tables: | | 70GB / 5 Tables | | |
| Average Runtime (minutes): | | 46.33 | 41.07 | 1.13 |
| Result table rows: | | 52 | | |
| Stages: | | MapReduce 39 | Spark 144 & 4474 Tasks | |
| Avg. CPU Utilization % | User % | 56.97% | 16.65% | 3.42 |
| | System % | 3.89% | 2.62% | 1.48 |
| | IOwait % | 0.40% | 21.28% | - |
| Memory Utilization % | | 94.33% | 93.78% | 1.01 |
| Avg. Kbytes Transmitted per Second | | 11650.07 | 3455.03 | 3.37 |
| Avg. Kbytes Received per Second | | 11654.28 | 3456.24 | 3.37 |
| Avg. Context Switches per Second | | 10251.24 | 8693.44 | 1.18 |
| Avg. Kbytes Read per Second | | 2739.21 | 6501.03 | - |
| Avg. Kbytes Written per Second | | 7190.15 | 3364.60 | 2.14 |
| Avg. Read Requests per Second | | 40.24 | 66.93 | - |
| Avg. Write Requests per Second | | 17.13 | 12.20 | 1.40 |
| Avg. I/O Latencies in Milliseconds | | 55.76 | 32.91 | 1.69 |

**Summary:** Hive (MR) is only 13% slower than Spark SQL. The Hive execution utilizes on average 57% of the CPU, whereas the Spark SQL uses on average 17% of the CPU. Both Hive and Spark SQL are memory bound utilizing on average 94% of the memory. However, Spark SQL spent on average around 21% on waiting for outstanding disk I/O requests (IOwait), which is much greater than the average for both Hive and Spark SQL. Additionally, Hive read data with on average 7 MB/sec and writes it with on average 4MB/sec, whereas Spark SQL reads with on average 6.3MB/sec and writes with on average 3.3MB/sec.

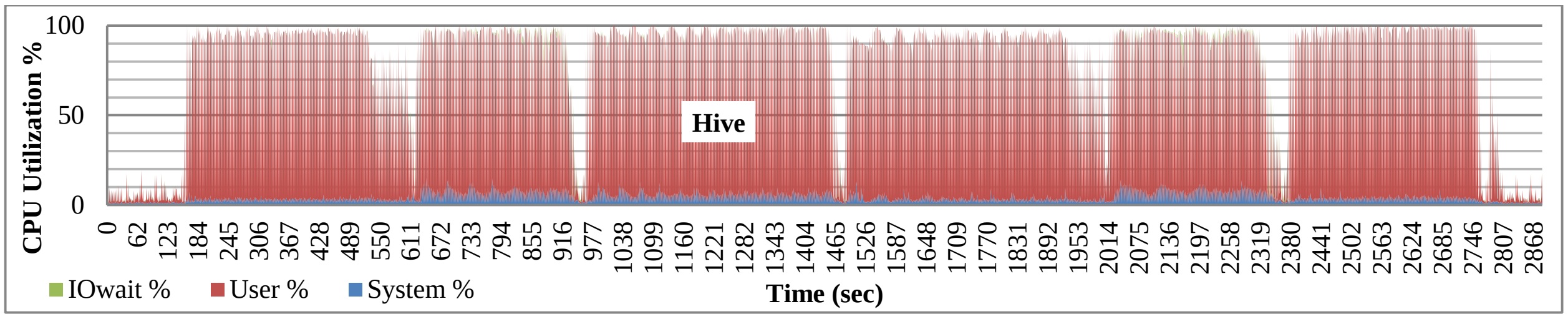

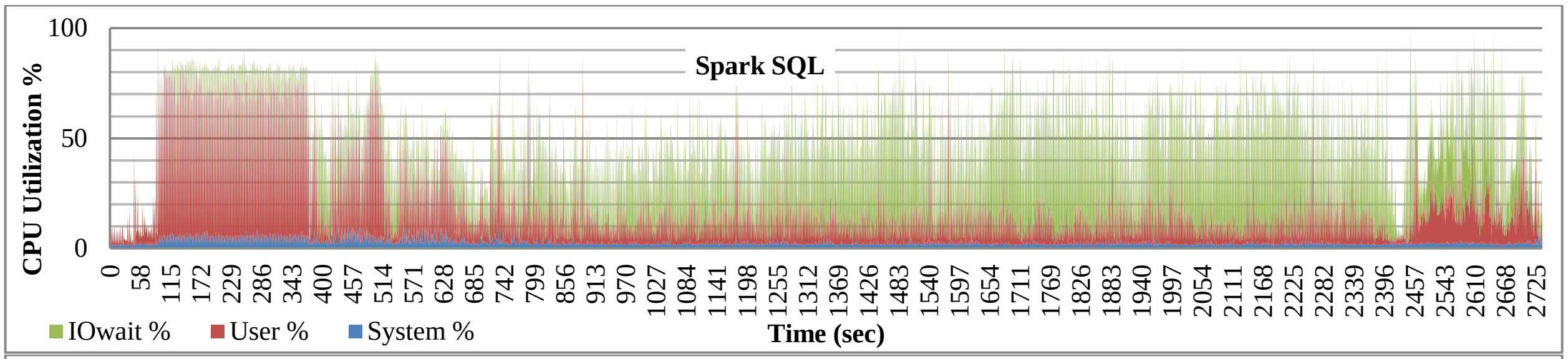
Spark SQL
CPU Utilization %
100
50
0
0 58 115 172 229 286 343 400 457 514 571 628 685 742 799 856 913 970 1027 1084 1141 1198 1255 1312 1369 1426 1483 1540 1597 1654 1711 1769 1826 1883 1940 1997 2054 2111 2168 2225 2282 2339 2396 2457 2543 2610 2668 2725
Time (sec)
IOwait %
User %
System %

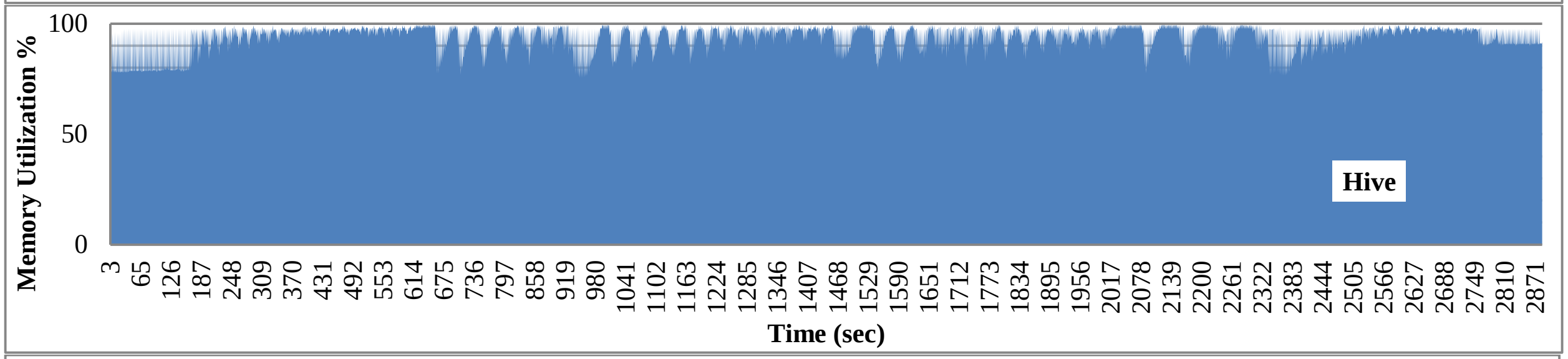
Hive
Memory Utilization %
100
50
0
3 65 126 187 248 309 370 431 492 553 614 675 736 797 858 919 980 1041 1102 1163 1224 1285 1346 1407 1468 1529 1590 1651 1712 1773 1834 1895 1956 2017 2078 2139 2200 2261 2322 2383 2444 2505 2566 2627 2688 2749 2810 2871
Time (sec)

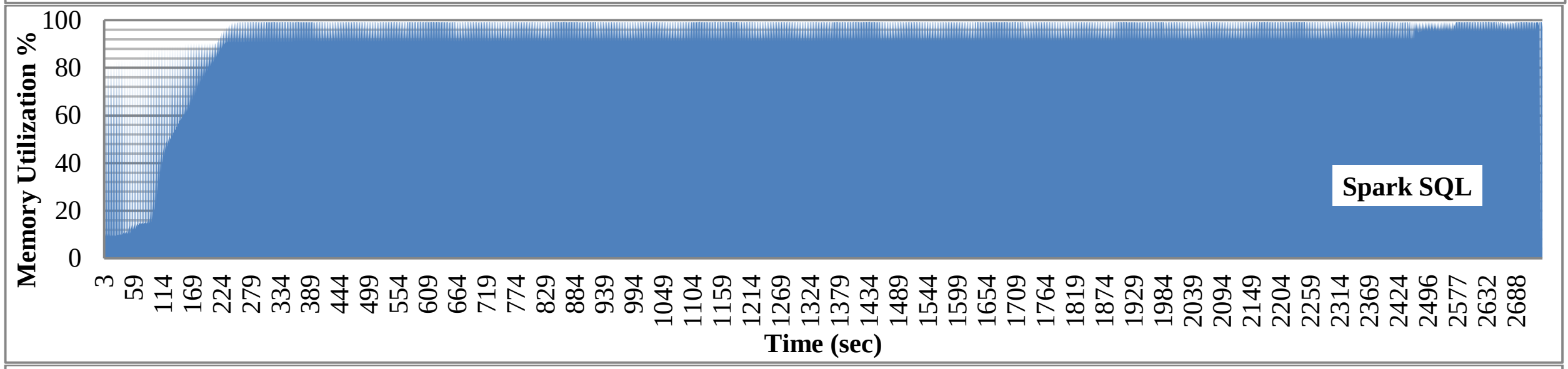
Spark SQL
Memory Utilization %
100
80
60
40
20
0
3 59 114 169 224 279 334 389 444 499 554 609 664 719 774 829 884 939 994 1049 1104 1159 1214 1269 1324 1379 1434 1489 1544 1599 1654 1709 1764 1819 1874 1929 1984 2039 2094 2149 2204 2259 2314 2369 2424 2496 2577 2632 2688
Time (sec)

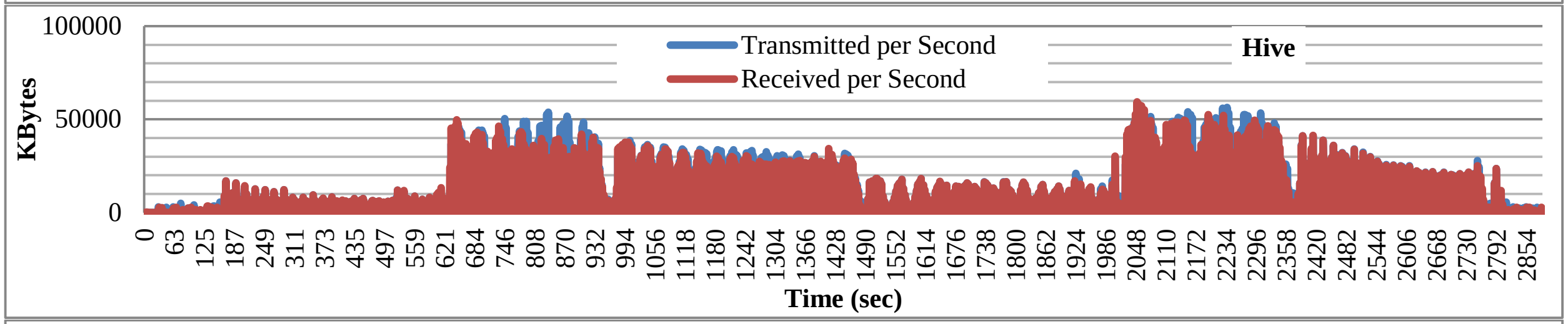
Hive
Transmitted per Second
Received per Second
KBytes
100000
50000
0
0 63 125 187 249 311 373 435 497 559 621 684 746 808 870 932 994 1056 1118 1180 1242 1304 1366 1428 1490 1552 1614 1676 1738 1800 1862 1924 1986 2048 2110 2172 2234 2296 2358 2420 2482 2544 2606 2668 2730 2792 2854
Time (sec)

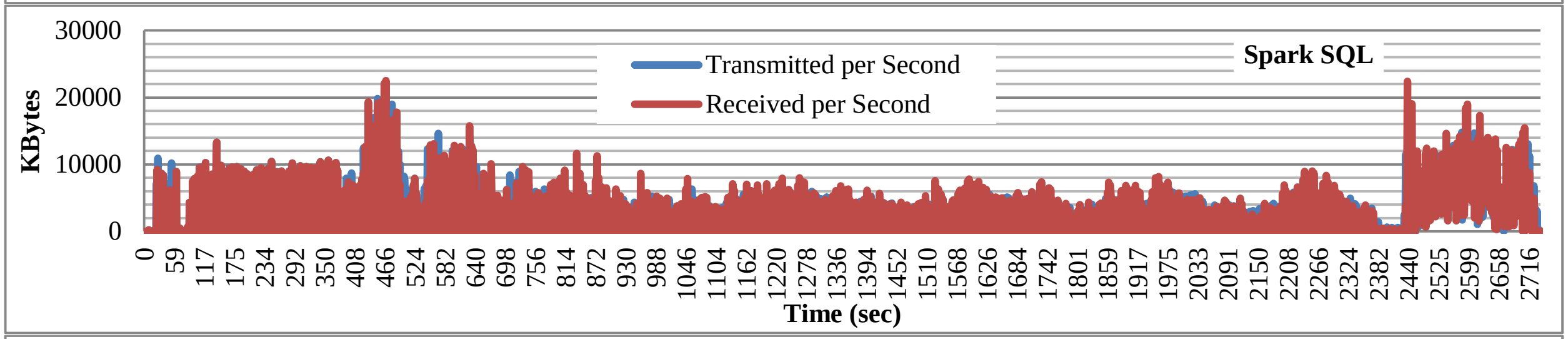
Spark SQL
Transmitted per Second
Received per Second
KBytes
30000
20000
10000
0
0 59 117 175 234 292 350 408 466 524 582 640 698 756 814 872 930 988 1046 1104 1162 1220 1278 1336 1394 1452 1510 1568 1626 1684 1742 1801 1859 1917 1975 2033 2091 2150 2208 2266 2324 2382 2440 2525 2599 2658 2716
Time (sec)

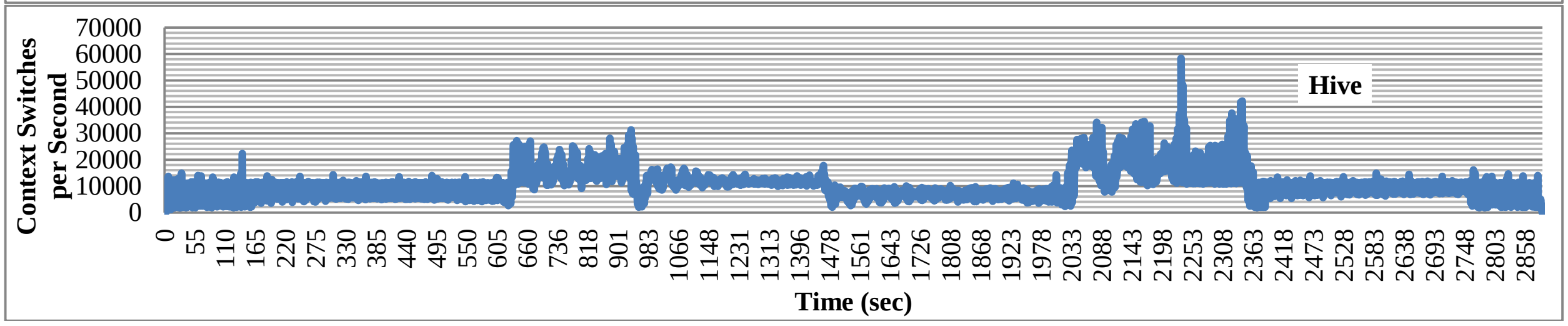
Hive
Context Switches per Second
70000
60000
50000
40000
30000
20000
10000
0
0 55 110 165 220 275 330 385 440 495 550 605 660 736 818 901 983 1066 1148 1231 1313 1396 1478 1561 1643 1726 1808 1868 1923 1978 2033 2088 2143 2198 2253 2308 2363 2418 2473 2528 2583 2638 2693 2748 2803 2858
Time (sec)

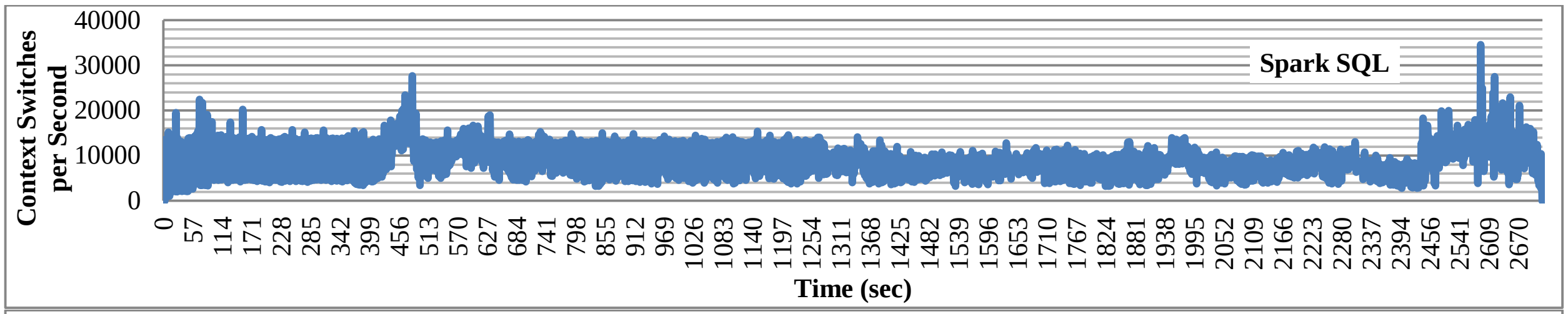
Context Switches per Second
40000
30000
20000
10000
0
Spark SQL
0 57 114 171 228 285 342 399 456 513 570 627 684 741 798 855 912 969 1026 1083 1140 1197 1254 1311 1368 1425 1482 1539 1596 1653 1710 1767 1824 1881 1938 1995 2052 2109 2166 2223 2280 2337 2394 2456 2541 2609 2670
Time (sec)

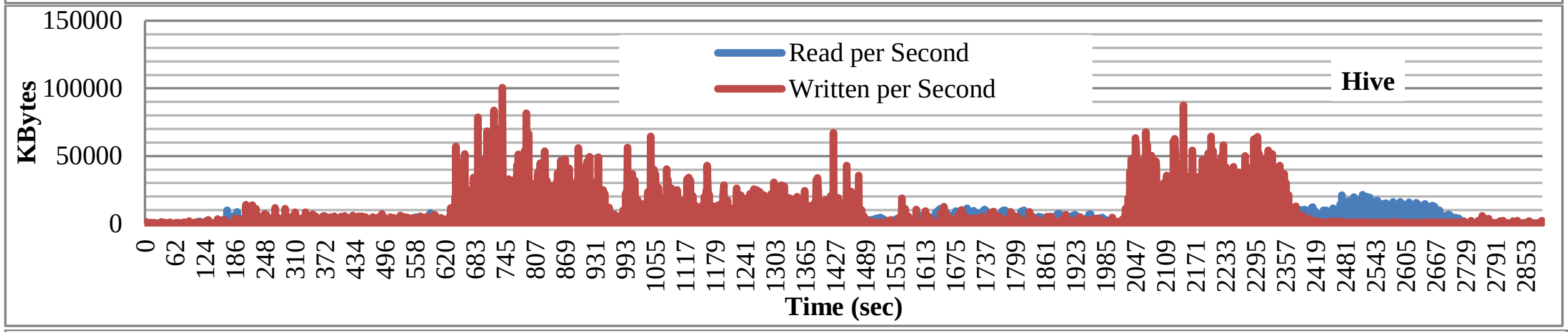
KBytes
150000
100000
50000
0
Read per Second
Written per Second
Hive
0 62 124 186 248 310 372 434 496 558 620 683 745 807 869 931 993 1055 1117 1179 1241 1303 1365 1427 1489 1551 1613 1675 1737 1799 1861 1923 1985 2047 2109 2171 2233 2295 2357 2419 2481 2543 2605 2667 2729 2791 2853
Time (sec)

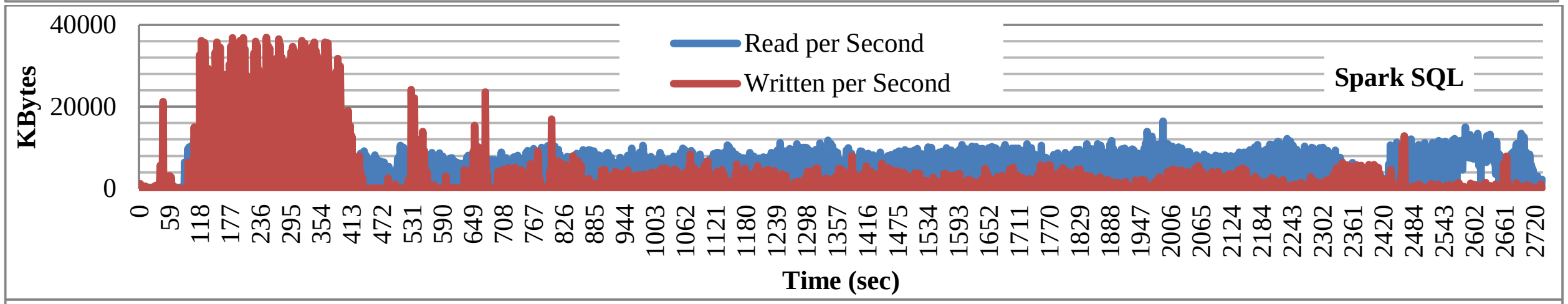
KBytes
40000
20000
0
Read per Second
Written per Second
Spark SQL
0 59 118 177 236 295 354 413 472 531 590 649 708 767 826 885 944 1003 1062 1121 1180 1239 1298 1357 1416 1475 1534 1593 1652 1711 1770 1829 1888 1947 2006 2065 2124 2184 2243 2302 2361 2420 2484 2543 2602 2661 2720
Time (sec)

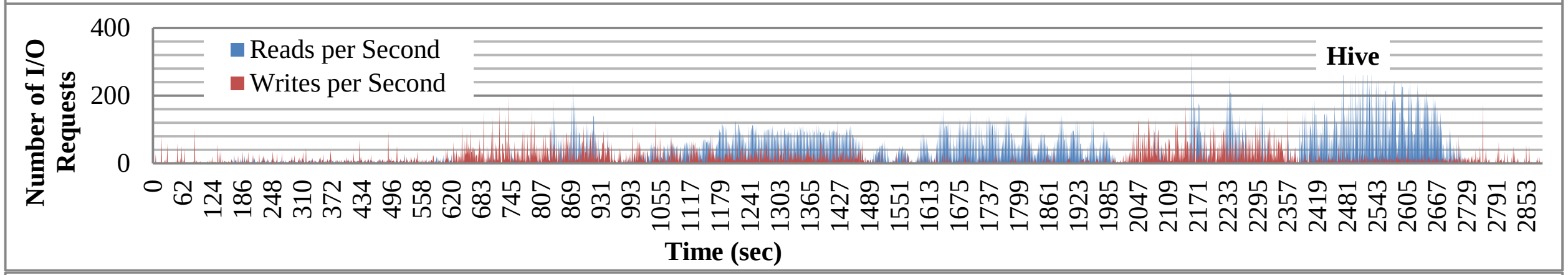
Number of I/O Requests
400
200
0
Reads per Second
Writes per Second
Hive
0 62 124 186 248 310 372 434 496 558 620 683 745 807 869 931 993 1055 1117 1179 1241 1303 1365 1427 1489 1551 1613 1675 1737 1799 1861 1923 1985 2047 2109 2171 2233 2295 2357 2419 2481 2543 2605 2667 2729 2791 2853
Time (sec)

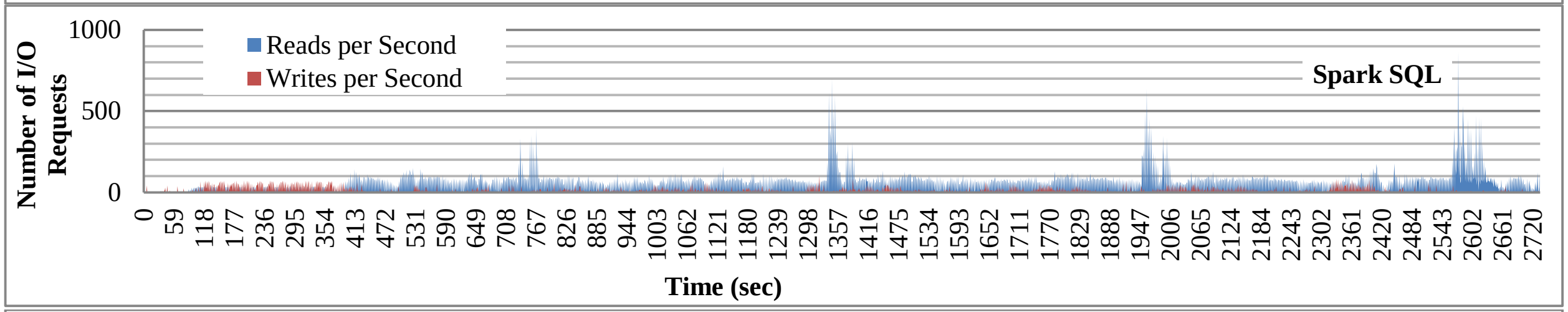
Number of I/O Requests
1000
500
0
Reads per Second
Writes per Second
Spark SQL
0 59 118 177 236 295 354 413 472 531 590 649 708 767 826 885 944 1003 1062 1121 1180 1239 1298 1357 1416 1475 1534 1593 1652 1711 1770 1829 1888 1947 2006 2065 2124 2184 2243 2302 2361 2420 2484 2543 2602 2661 2720
Time (sec)

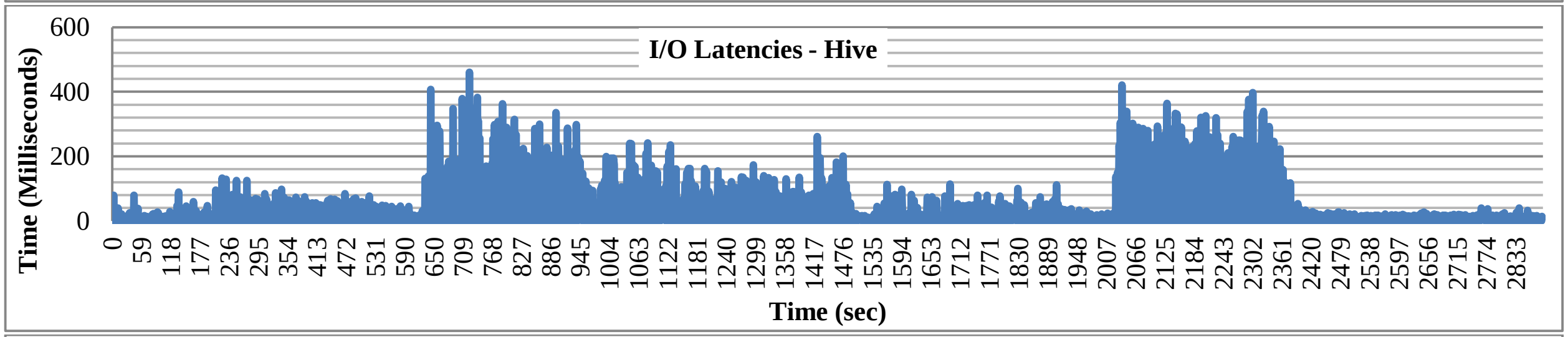
Time (Milliseconds)
600
400
200
0
I/O Latencies - Hive
0 59 118 177 236 295 354 413 472 531 590 650 709 768 827 886 945 1004 1063 1122 1181 1240 1299 1358 1417 1476 1535 1594 1653 1712 1771 1830 1889 1948 2007 2066 2125 2184 2243 2302 2361 2420 2479 2538 2597 2656 2715 2774 2833
Time (sec)

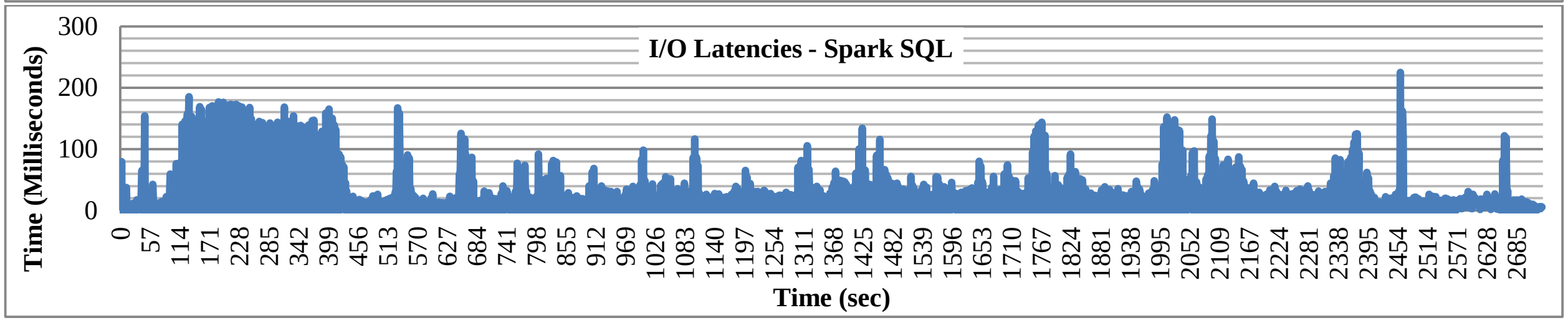
Time (Milliseconds)
300
200
100
0
I/O Latencies - Spark SQL
0 57 114 171 228 285 342 399 456 513 570 627 684 741 798 855 912 969 1026 1083 1140 1197 1254 1311 1368 1425 1482 1539 1596 1653 1710 1767 1824 1881 1938 1995 2052 2109 2167 2224 2281 2338 2395 2454 2514 2571 2628 2685
Time (sec)

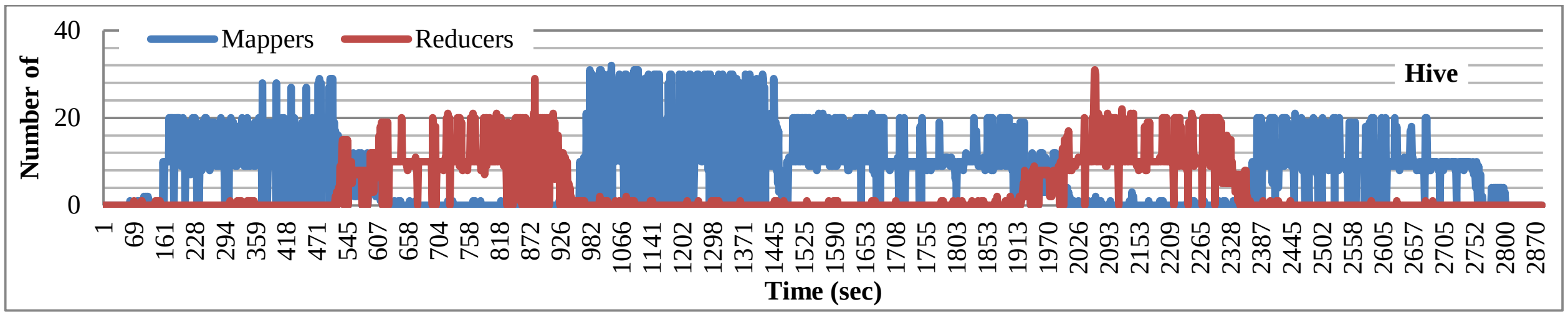

## 7.6. BigBench Query 9 (HiveQL + Spark SQL)

BigBench's query Q9 calculates the total sales for different types of customers (e.g. based on marital status, education status), sales price and different combinations of state and sales profit [29]. The query is implemented in pure HiveQL and was adopted from query 48 of the TPC-DS benchmark.

| | | Hive | Spark SQL | Hive/ Spark SQL Ratio |
|---|---|---|---|---|
| Scale Factor: | | 1TB | | |
| Input Data size/ Number of Tables: | | 69GB / 5 Tables | | |
| Result table rows: | | 1 | | |
| Stages: | | MapReduce 7 | Spark 135 & 3065 Tasks | |
| Average Runtime (minutes): | | 17.72 | 2.82 | 6.28 |
| Avg. CPU Utilization % | User % | 60.34% | 27.87% | 2.17 |
| | System % | 3.44% | 2.22% | 1.55 |
| | IOwait % | 0.38% | 4.09% | - |
| Memory Utilization % | | 78.87% | 61.27% | 1.29 |
| Avg. Kbytes Transmitted per Second | | 7512.13 | 7690.59 | - |
| Avg. Kbytes Received per Second | | 7514.87 | 7691.04 | - |
| Avg. Context Switches per Second | | 19757.83 | 7284.11 | 2.71 |
| Avg. Kbytes Read per Second | | 2741.72 | 13174.12 | - |
| Avg. Kbytes Written per Second | | 4098.95 | 1043.45 | 3.93 |
| Avg. Read Requests per Second | | 9.76 | 48.91 | - |
| Avg. Write Requests per Second | | 10.84 | 3.62 | 2.99 |
| Avg. I/O Latencies in Milliseconds | | 41.67 | 27.32 | 1.53 |

**Summary:** Hive is 6 times slower than Spark SQL. The Hive execution is CPU (on average 60%) and memory (78%) bound, whereas the Spark SQL execution consumes on average 28% CPU and 61% memory. Additionally, Hive read data with on average 2.6 MB/sec and writes it with on average 4MB/sec, whereas Spark SQL reads with on average 13MB/sec and writes with on average 1MB/sec.

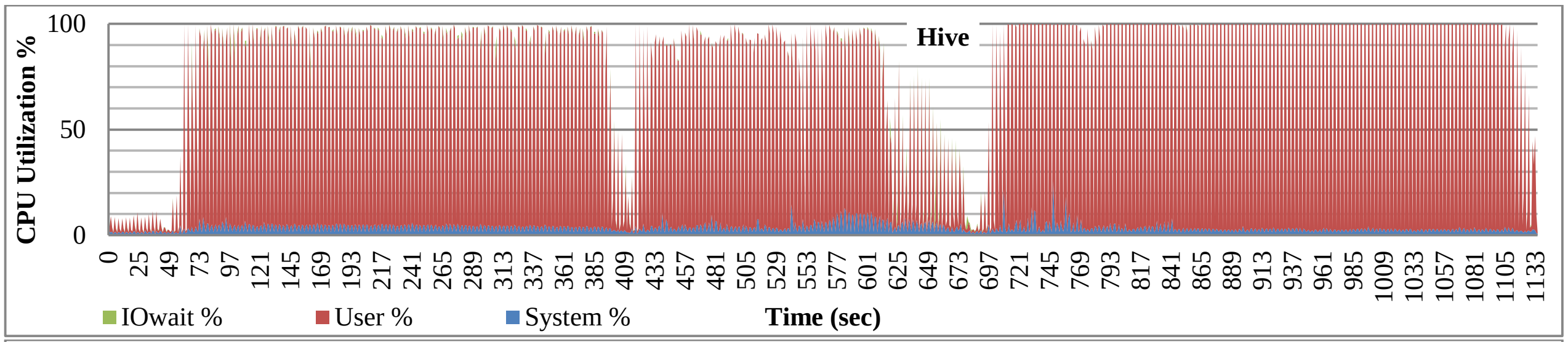

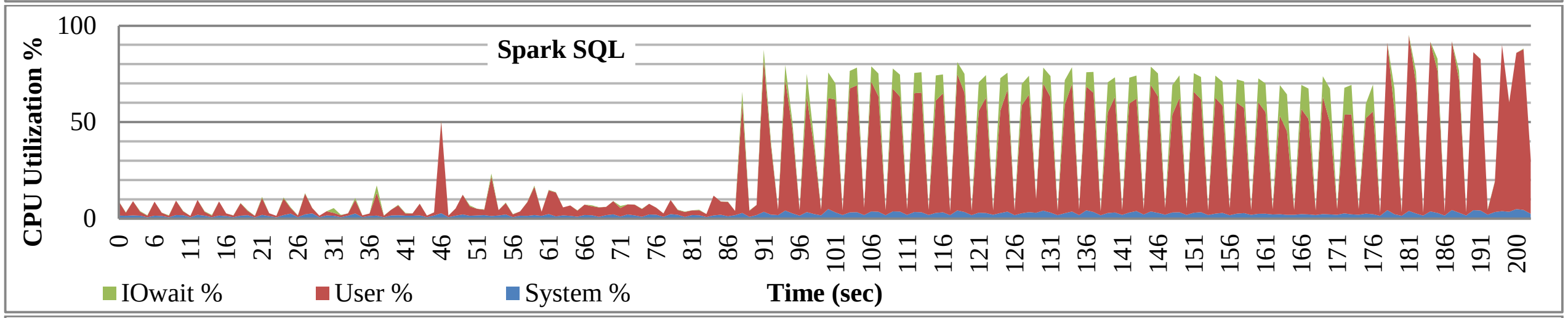

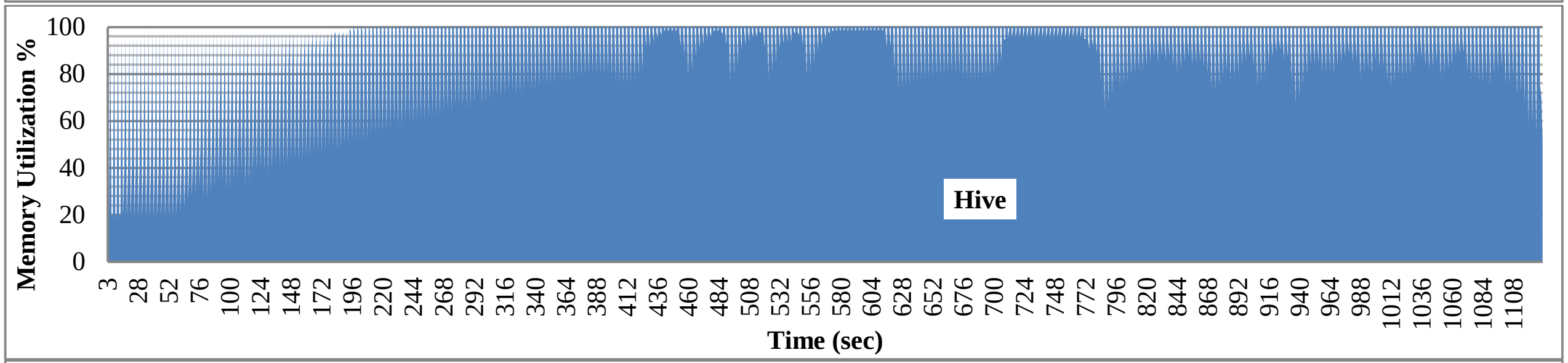
Hive
Memory Utilization %
Time (sec)

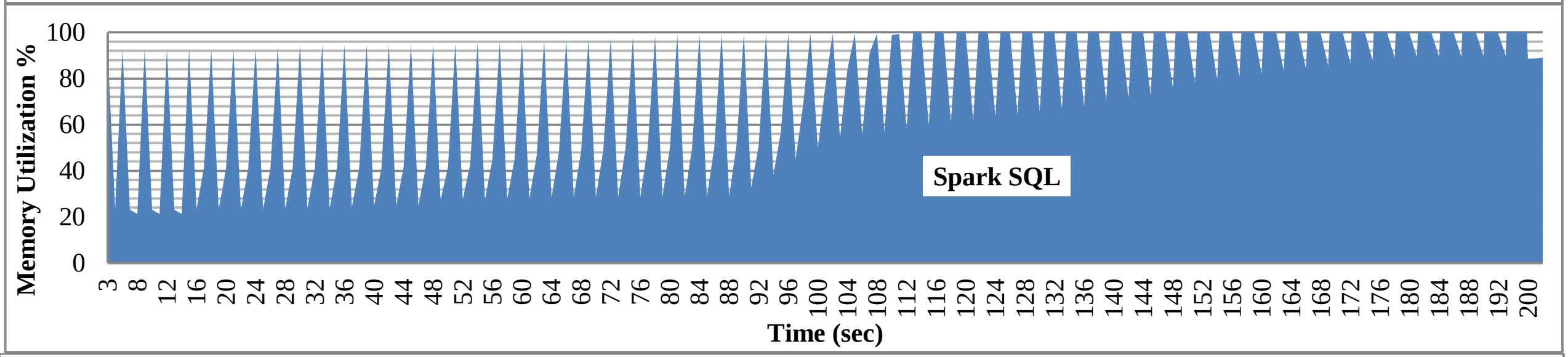
Spark SQL
Memory Utilization %
Time (sec)

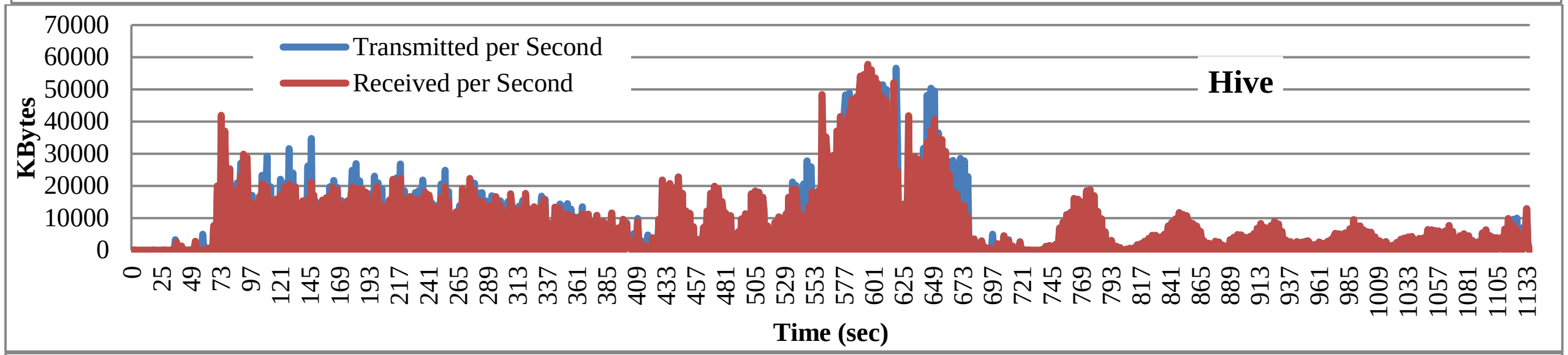
Hive
KBytes
Time (sec)
Transmitted per Second
Received per Second

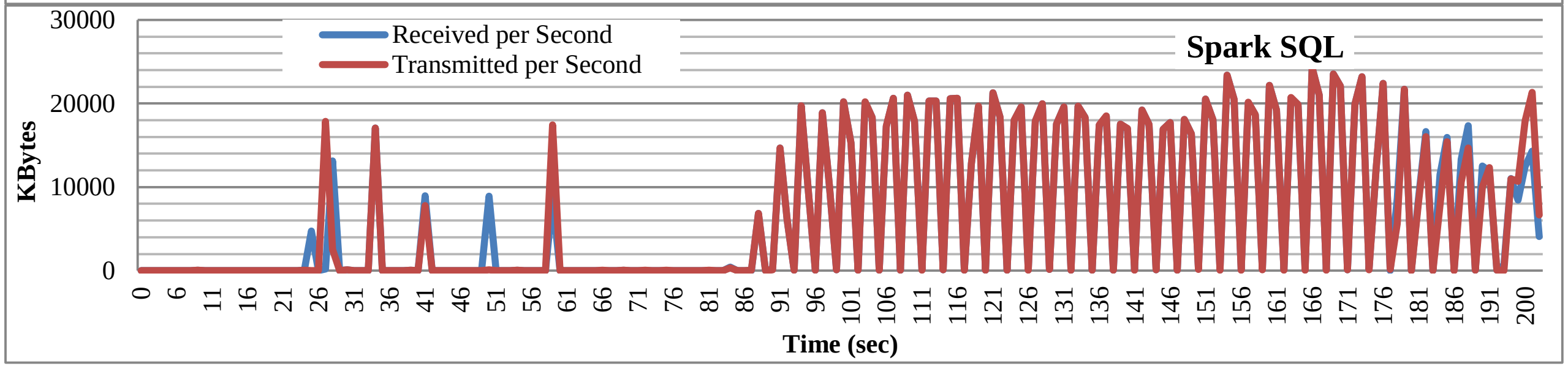
Spark SQL
KBytes
Time (sec)
Received per Second
Transmitted per Second

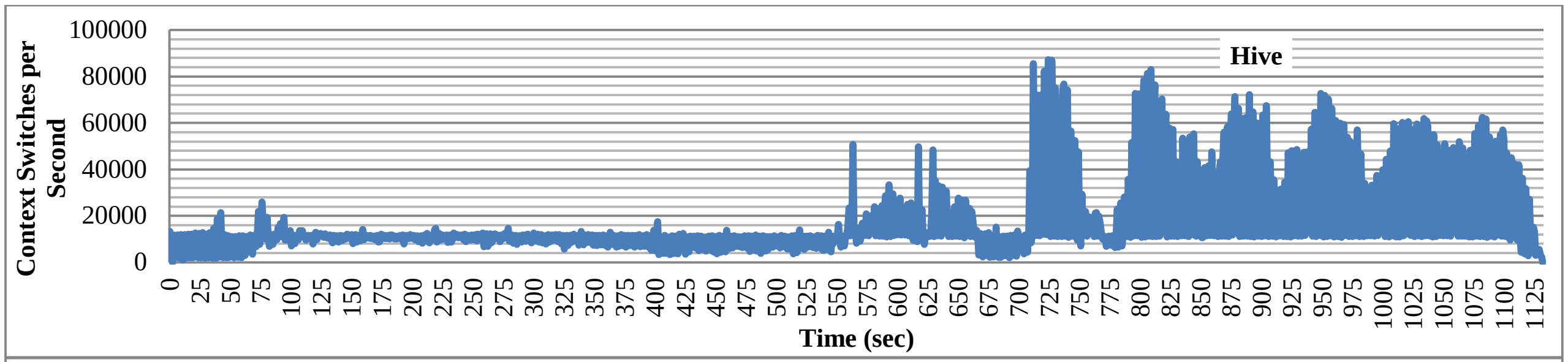
Hive
Context Switches per Second
Time (sec)

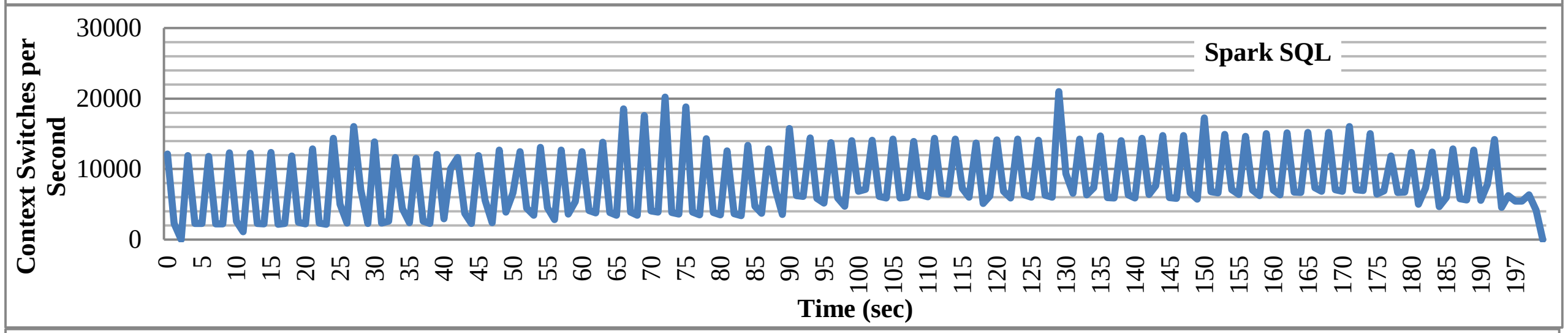
Spark SQL
Context Switches per Second
Time (sec)

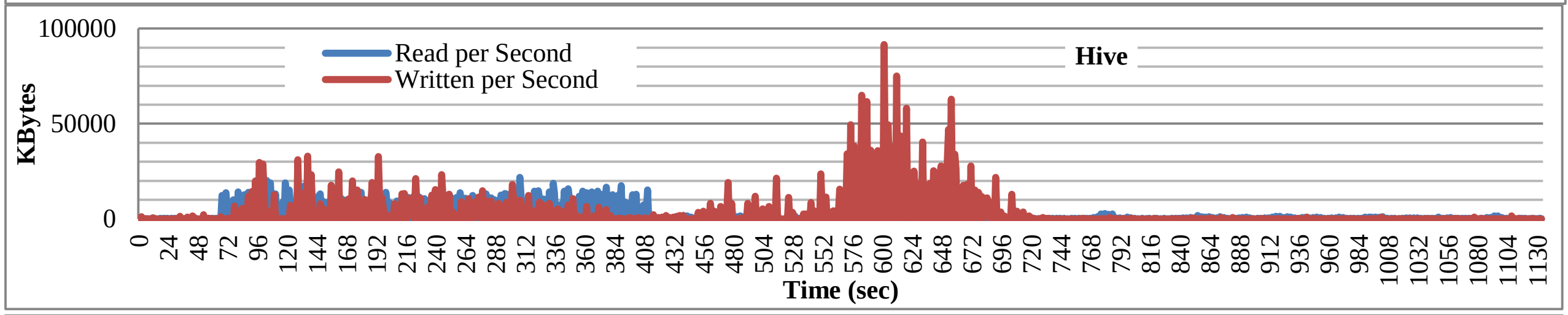
Read per Second
Written per Second
Hive
KBytes
Time (sec)

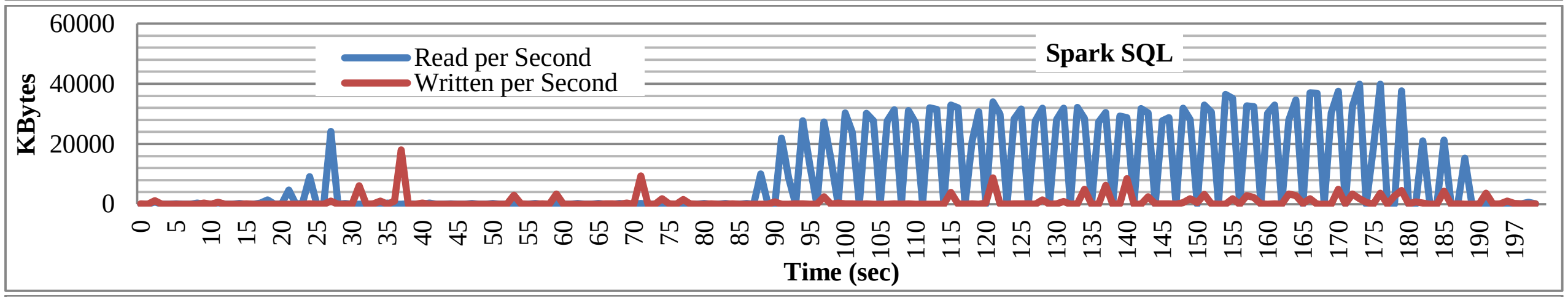
Read per Second
Written per Second
Spark SQL
KBytes
Time (sec)

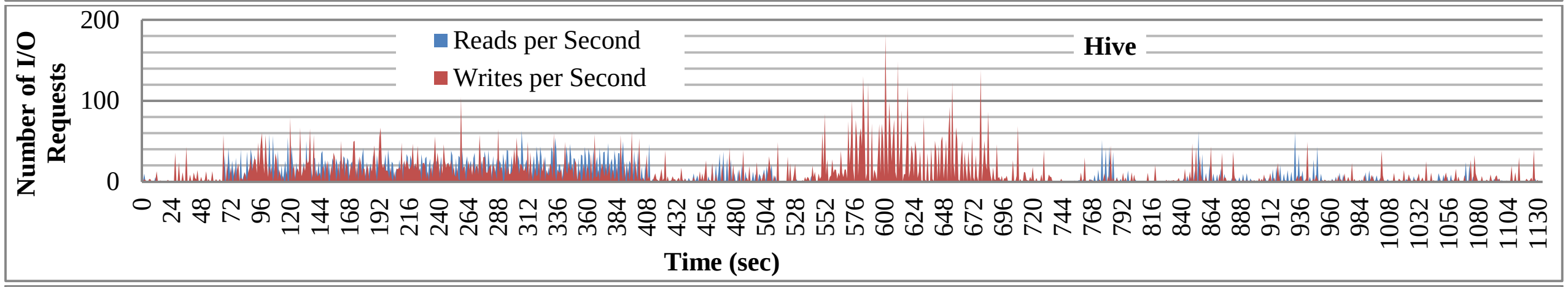
Reads per Second
Writes per Second
Hive
Number of I/O Requests
Time (sec)

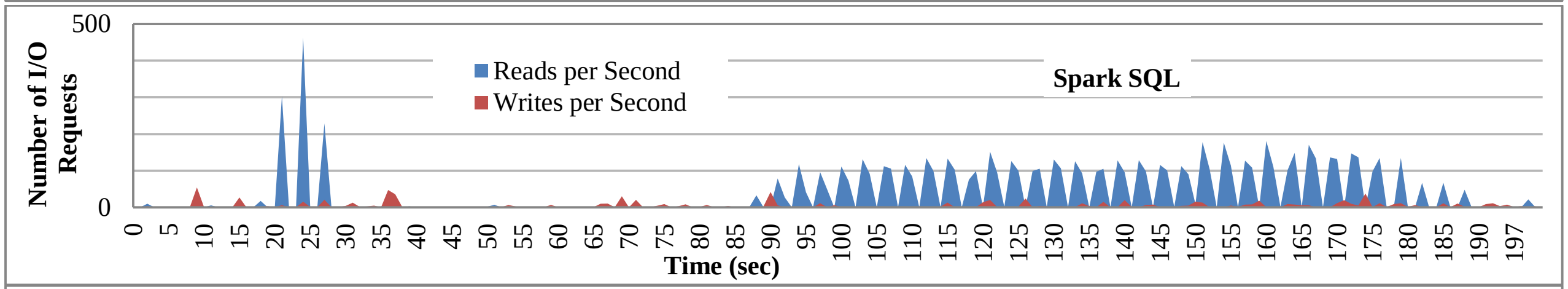
Reads per Second
Writes per Second
Spark SQL
Number of I/O Requests
Time (sec)

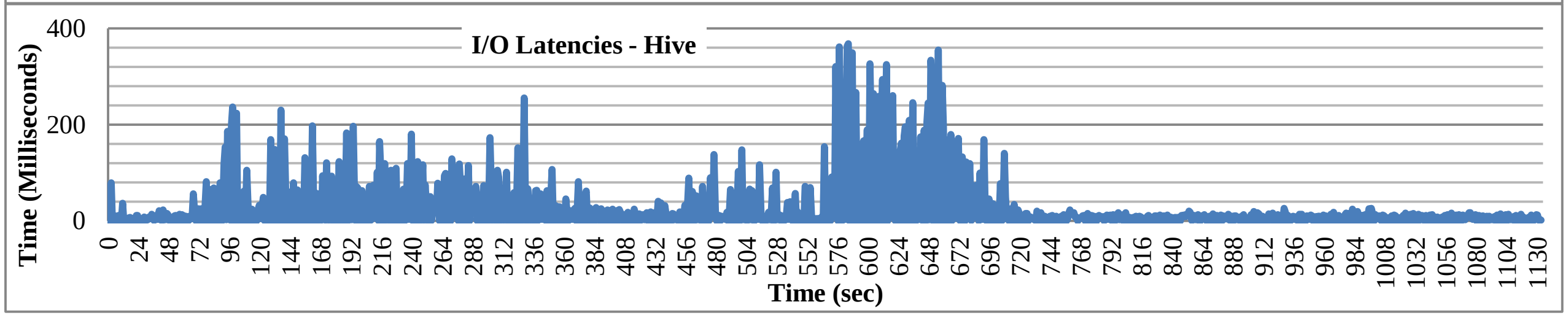
I/O Latencies - Hive
Time (Milliseconds)
Time (sec)

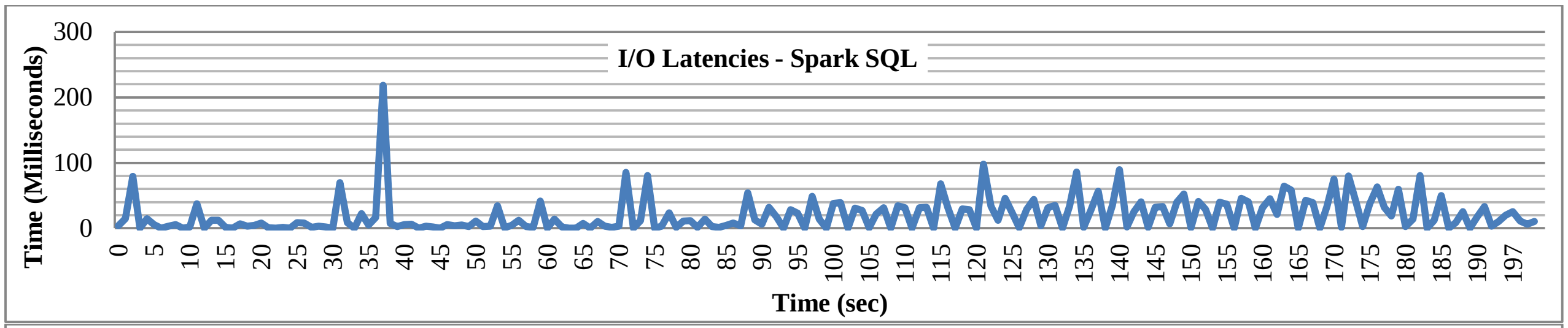

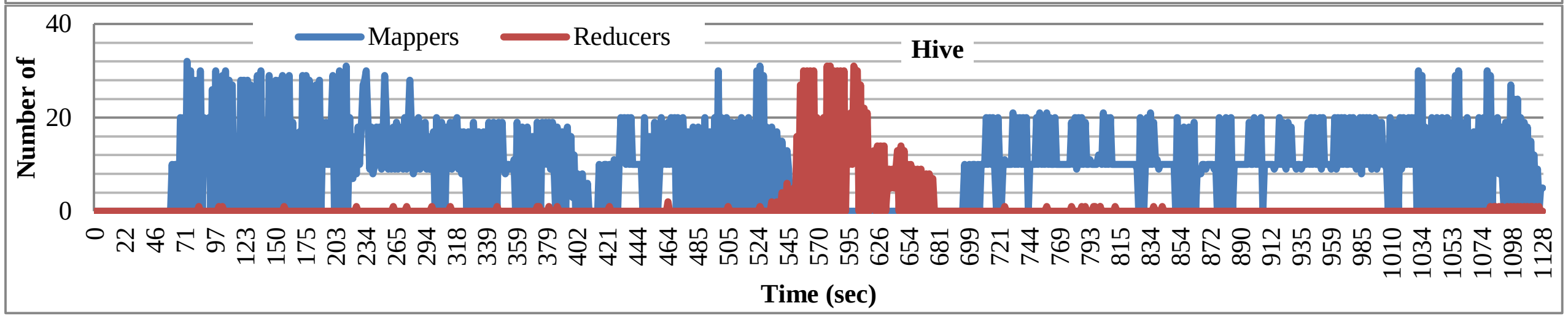

## 7.7. BigBench Query 24 (HiveQL + Spark SQL)

BigBench's query Q24 measures the effect of competitors' prices on products' in-store and online sales for a given product [29] (Compute the cross-price elasticity of demand for a given product). The query is implemented in pure HiveQL.

| | | Hive | Spark SQL | Spark SQL/Hive Ratio |
|---|---|---|---|---|
| Scale Factor: | | 1TB | | |
| Input Data size/ Number of Tables: | | 99GB / 4 Tables | | |
| Result table rows: | | 8 | | |
| Stages: | | MapReduce 39 | Spark 42 & 6996 Tasks | |
| Average Runtime (minutes): | | 14.75 | 77.05 | 5.22 |
| Avg. CPU Utilization % | User % | 48.92% | 17.52% | - |
| | System % | 2.01% | 1.61% | - |
| | IOwait % | 0.48% | 11.21% | 23.35 |
| Memory Utilization % | | 43.60% | 82.84% | 1.90 |
| Avg. Kbytes Transmitted per Second | | 3123.24 | 4373.39 | 1.40 |
| Avg. Kbytes Received per Second | | 3122.92 | 4374.41 | 1.40 |
| Avg. Context Switches per Second | | 7077.10 | 8821.01 | 1.25 |
| Avg. Kbytes Read per Second | | 7148.77 | 7810.38 | 1.09 |
| Avg. Kbytes Written per Second | | 169.46 | 3762.42 | 22.20 |
| Avg. Read Requests per Second | | 22.28 | 64.38 | 2.89 |
| Avg. Write Requests per Second | | 4.71 | 8.29 | 1.76 |
| Avg. I/O Latencies in Milliseconds | | 21.38 | 27.66 | 1.29 |

**Summary:** Overall Hive (MapReduce) is around 81% faster than Spark SQL. The query execution on Hive utilizes on average 49% of the CPU, whereas the Spark SQL uses on average 18% of the CPU. However, for Spark SQL around 11% of the time is spent on waiting for outstanding disk I/O requests (IOwait), which is much greater than the average for both Hive and Spark SQL. Also Spark SQL execution is memory bound utilizing 83% of the memory in comparison to the 44% utilization for Hive.

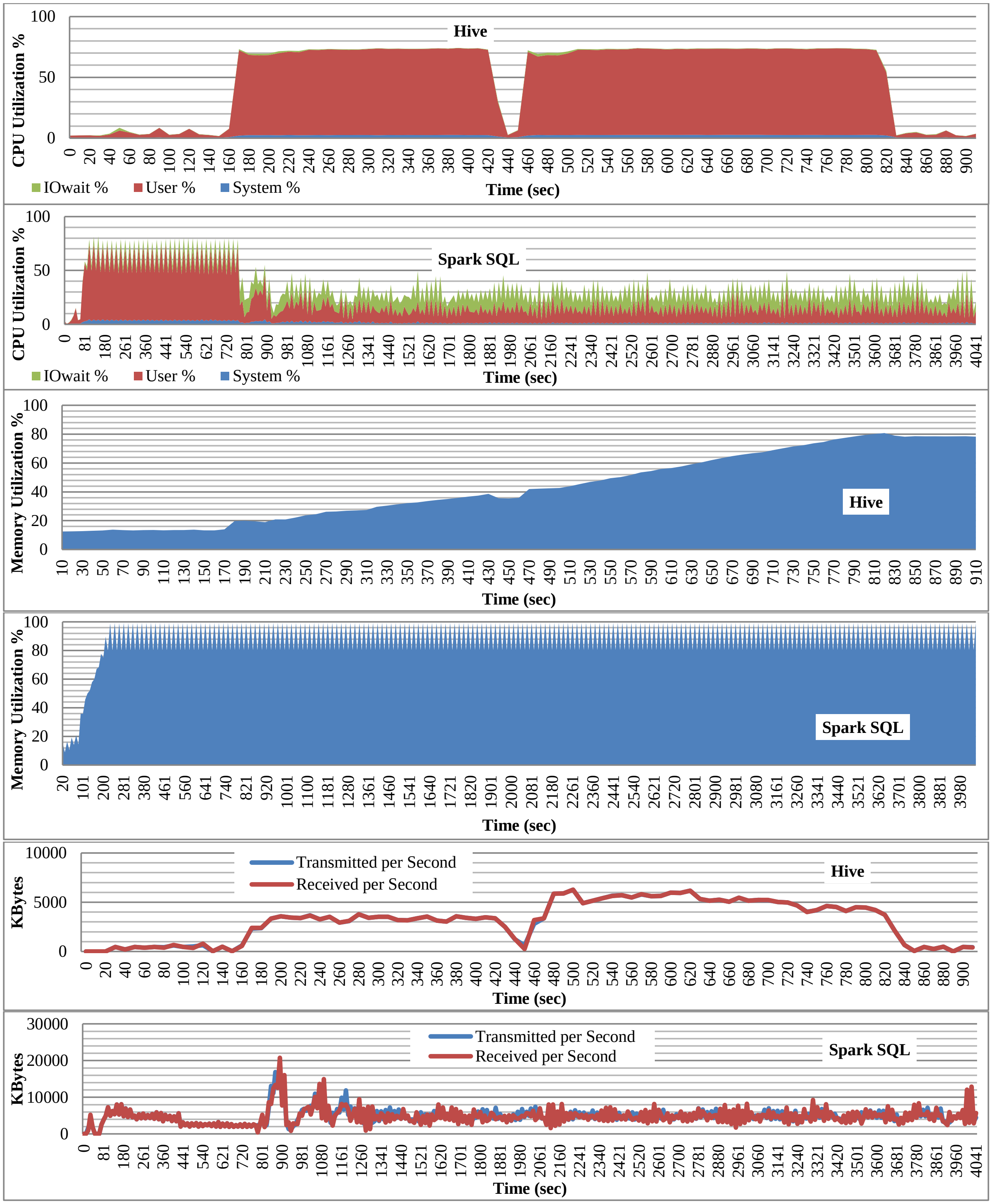

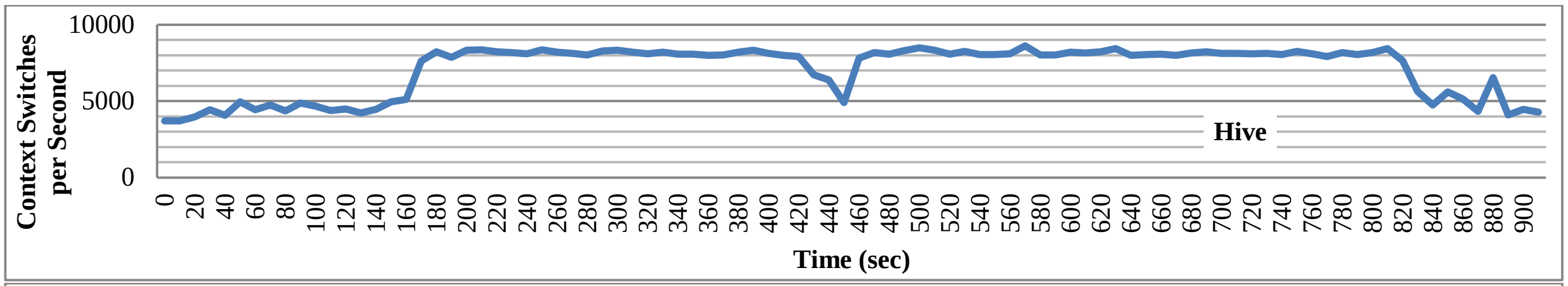
Context Switches per Second
Hive
Time (sec)

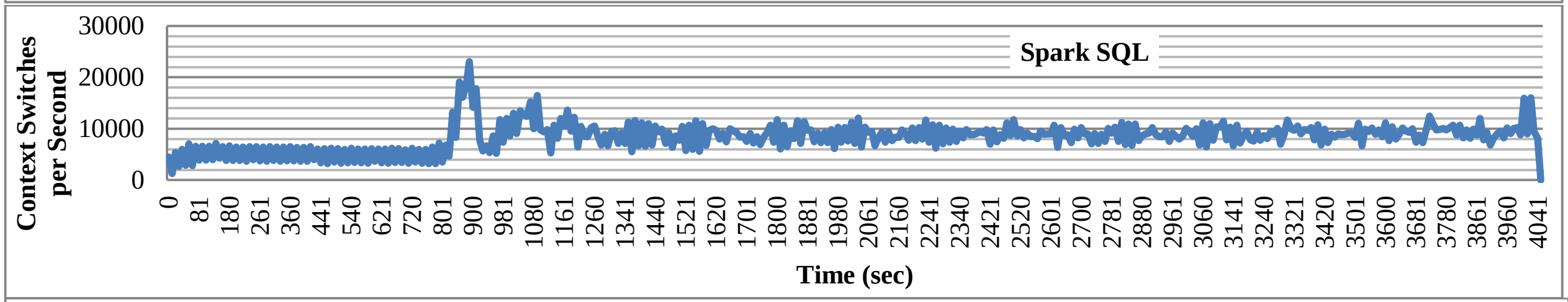
Context Switches per Second
Spark SQL
Time (sec)

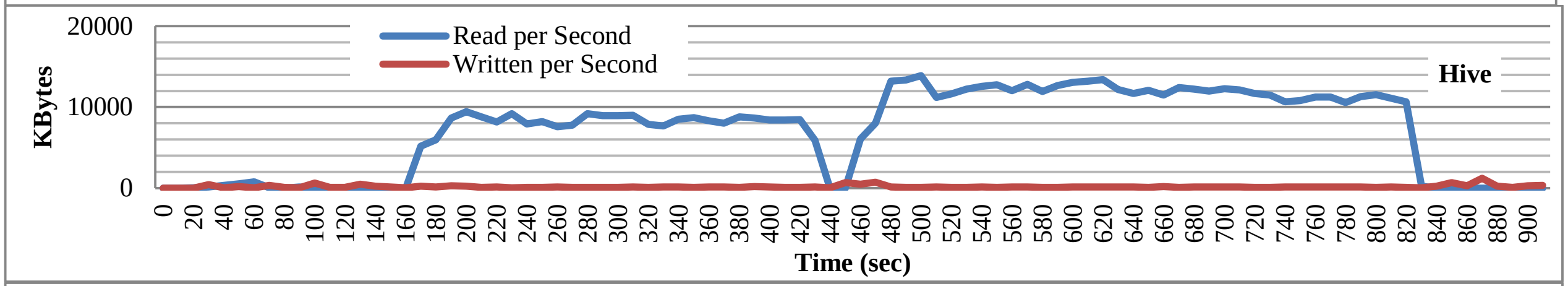
KBytes
Read per Second
Written per Second
Hive
Time (sec)

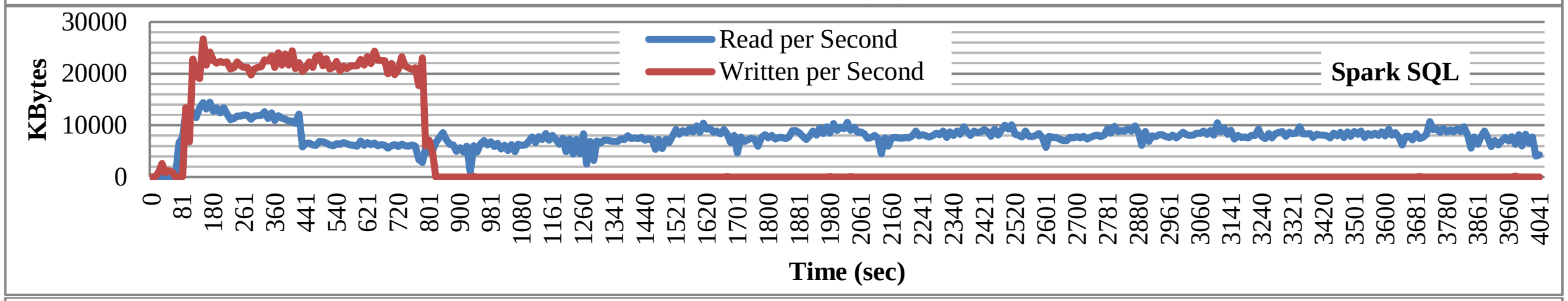
KBytes
Read per Second
Written per Second
Spark SQL
Time (sec)

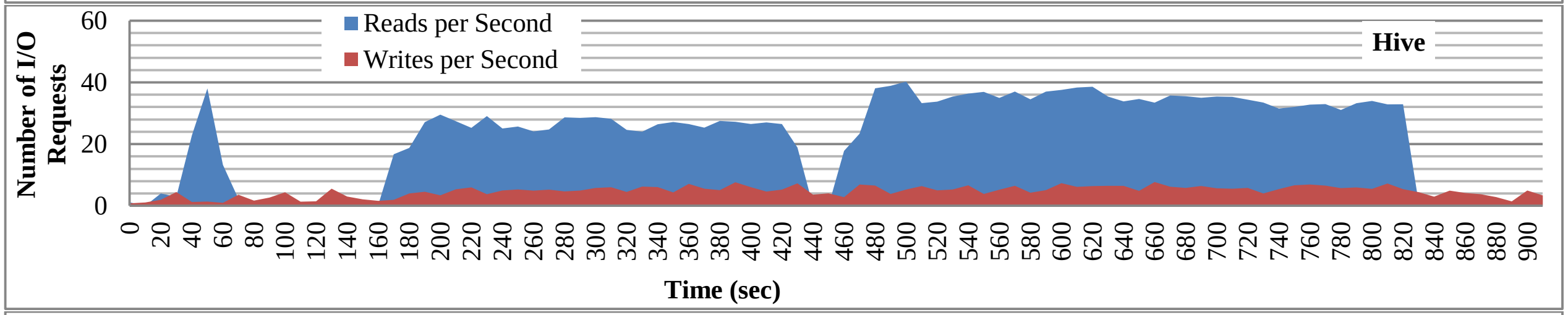
Number of I/O Requests
Reads per Second
Writes per Second
Hive
Time (sec)

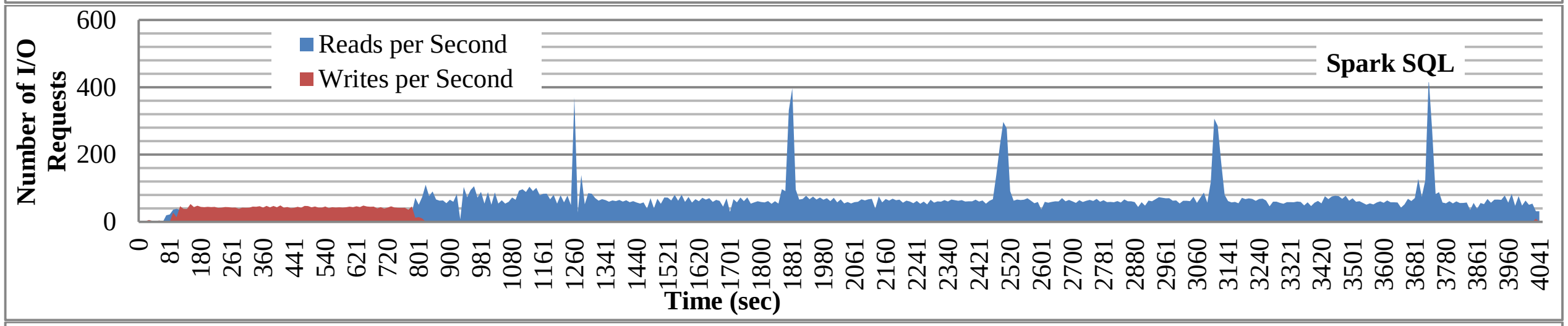
Number of I/O Requests
Reads per Second
Writes per Second
Spark SQL
Time (sec)

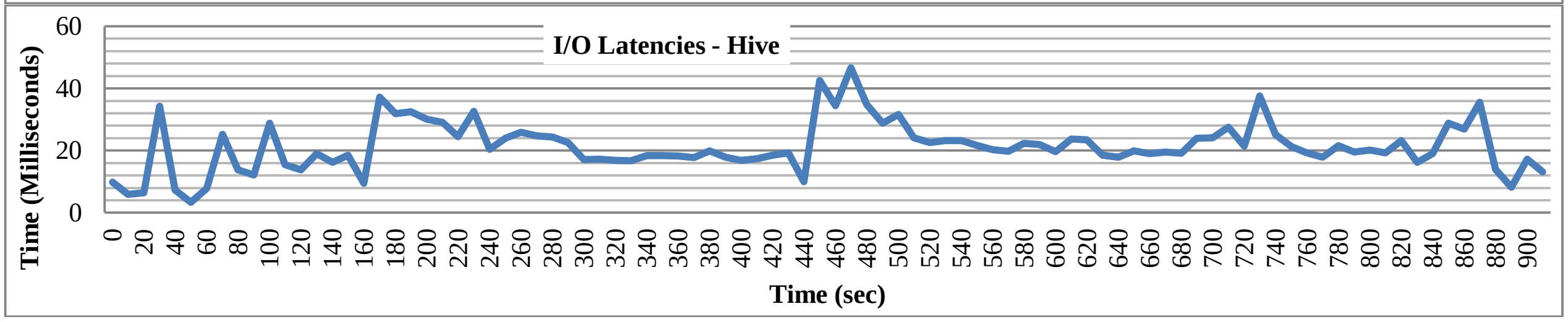
Time (Milliseconds)
I/O Latencies - Hive
Time (sec)

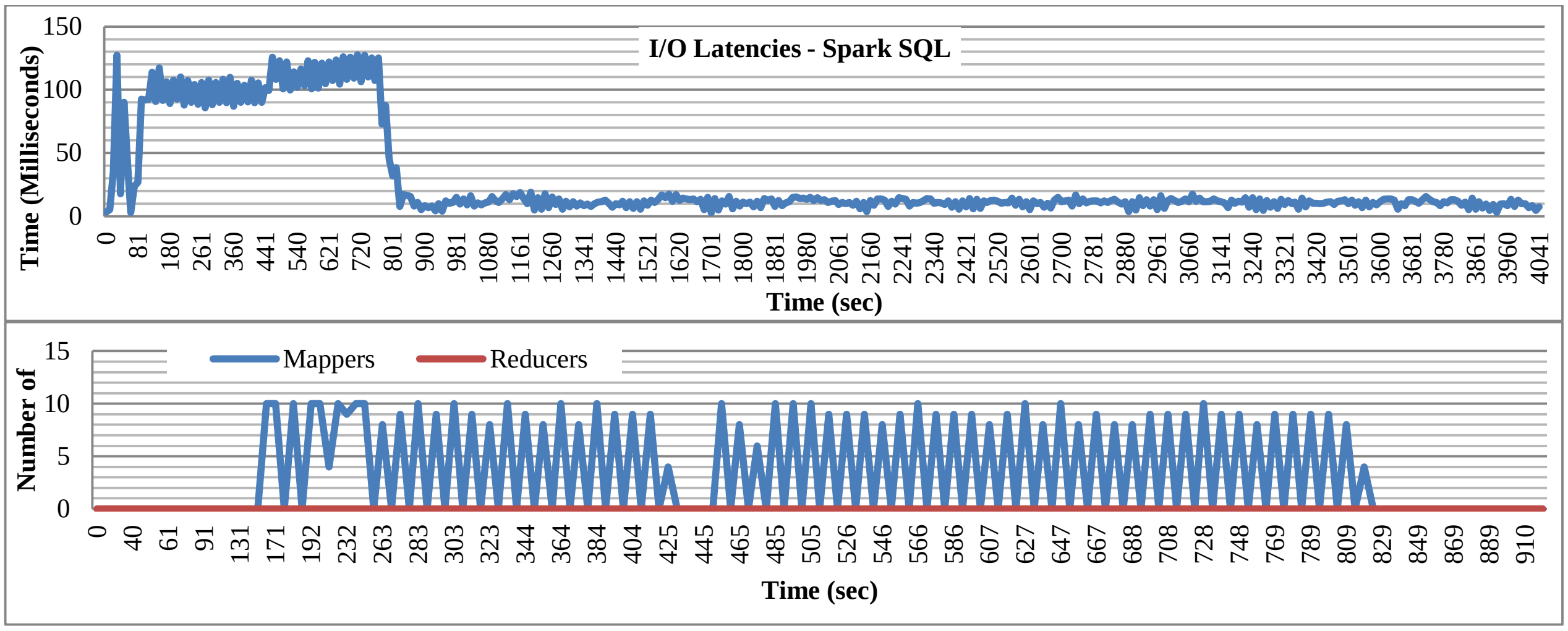

## 8. Lessons Learned

This report presented our first attempt to use the BigBench benchmark to evaluate the data scaling capabilities of a Hadoop cluster on both MapReduce/Hive and Spark SQL. Furthermore, multiple issues and fixes were presented as part of our initiative to execute BigBench on Spark. Our experiments showed that the group of 14 pure HiveQL queries can be successfully executed on Spark SQL.

The Spark SQL performance greatly varied among the type of queries and the data sizes on which they were executed. On one hand, a group of HiveQL queries (Q6, Q11, Q12, Q13, Q14, Q15 and Q17) performed best, with Q9 being around 6.3 times faster than Hive with the increase of the data size. On the other hand, we observed a group of queries (Q7, Q16, Q21, Q22 and Q23) performed worst, with Q24 being 5.2 times slower on Spark SQL than on Hive with the increase of the data size. The reason for this is the reported join issue [43] in the current Spark SQL version.

In terms of resource utilization, our analysis showed that Spark SQL:

- Utilized less CPU, whereas it showed higher I/O wait than Hive.
- Read more data from disk, whereas it wrote less data than Hive.
- Utilized less memory than Hive.
- Sent less data over the network than Hive.

In the future, we plan to rerun the BigBench queries on the latest version of Spark SQL, where the join issue should be fixed and offer more stable experience. Also we plan running the remaining groups of BigBench queries using other components from the Spark framework.

## Acknowledgements

This work has benefited from valuable discussions in the SPEC Research Group's Big Data Working Group. We would like to thank Tilmann Rabl (University of Toronto), John Poelman (IBM), Bhaskar Gowda (Intel), Yi Yao (Intel), Marten Rosselli, Karsten Tolle, Roberto V. Zicari and Raik Niemann of the Frankfurt Big Data Lab for their valuable feedback. We would like to thank the Fields Institute for supporting our visit to the Sixth Workshop on Big Data Benchmarking at the University of Toronto.

# Appendix

| System Information | | Description |
|---|---|---|
| Manufacturer: | Dell Inc. | |
| Product Name: | PowerEdge T420 | |
| BIOS: | 1.5.1 | Release Date: 03/08/2013 |
| Memory | | |
| Total Memory: | 32 GB | |
| DIMMs: | 10 | |
| Configured Clock Speed: | 1333 MHz | Part Number: M393B5273CH0-YH9 |
| Size: | 4096 MB | |
| CPU | | |
| Model Name: | Intel(R) Xeon(R) CPU E5-2420 0 @ 1.90GHz | |
| Architecture: | x86_64 | |
| CPU(s): | 24 | |
| On-line CPU(s) list: | 0-23 | |
| Thread(s) per core: | 2 | |
| Core(s) per socket: | 6 | |
| Socket(s): | 2 | |
| CPU MHz: | 1200.000 | |
| L1d cache: | 32K | |
| L1i cache: | 32K | |
| L2 cache: | 256K | |
| L3 cache: | 15360K | |
| NUMA node0 CPU(s): | 0,2,4,6,8,10,12,14,16,18,20,22 | |
| NUMA node1 CPU(s): | 1,3,5,7,9,11,13,15,17,19,21,23 | |
| NIC | | |
| Settings for em1: | Speed: 1000Mb/s | |
| Ethernet controller: | Broadcom Corporation NetXtreme BCM5720 Gigabit Ethernet PCIe | |
| Storage | | |
| Storage Controller: | LSI Logic / Symbios Logic MegaRAID SAS 2008 [Falcon] (rev 03) | 08:00.0 RAID bus controller |
| Drive / Name | Formatted Size | Model |
| Disk 1/ sda1 | 931.5 GB | Western Digital, WD1003FBYX RE4-1TB, SATA3, 3.5 in, 7200RPM, 64MB Cache |

Table 19: Master Node

| System Information | | Description |
|---|---|---|
| Manufacturer: | Dell Inc. | |
| Product Name: | PowerEdge T420 | |
| BIOS: | 2.1.2 | Release Date: 01/20/2014 |
| Memory | | |
| Total Memory: | 32 GB | |
| DIMMs: | 4 | |
| Configured Clock Speed: | 1600 MHz | Part Number: M393B2G70DB0-YK0 |
| Size: | 16384 MB | |
| CPU | | |
| Model Name: | Intel(R) Xeon(R) CPU E5-2420 v2 @ 2.20GHz | |
| Architecture: | x86_64 | |
| CPU(s): | 12 | |
| On-line CPU(s) list: | 0-11 | |
| Thread(s) per core: | 2 | |
| Core(s) per socket: | 6 | |
| Socket(s): | 1 | |
| CPU MHz: | 2200.000 | |
| L1d cache: | 32K | |
| L1i cache: | 32K | |
| L2 cache: | 256K | |
| L3 cache: | 15360K | |
| NUMA node0 CPU(s): | 0-11 | |
| NIC | | |
| Settings for em1: | Speed: 1000Mb/s | |
| Ethernet controller: | Broadcom Corporation NetXtreme BCM5720 Gigabit Ethernet PCIe | |
| Storage | | |
| Storage Controller: | Intel Corporation C600/X79 series chipset SATA RAID Controller (rev 05) | 00:1f.2 RAID bus controller |
| Drive / Name | Formatted Size | Model |
| Disk 1/ sda1 | 931.5 GB | Dell- 1TB, SATA3, 3.5 in, 7200RPM, 64MB Cache |
| Disk 2/ sdb1 | 931.5 GB | WD Blue Desktop WD10EZEX - 1TB, SATA3, 3.5 in, 7200RPM, 64MB Cache |
| Disk 3/ sdc1 | 931.5 GB | |
| Disk 4/ sdd1 | 931.5 GB | |

Table 20: Data Node